\pdfoutput=1
\documentclass[%
reprint,
showpacs,
preprintnumbers,
showkeys,
amsmath,amssymb,
prd,
nofootinbib,
]{revtex4-1}

\newcommand{\TITLE}{Eikonal model analysis of elastic proton-proton collisions at 52.8~GeV and 8~TeV}

\RequirePackage[T1]{fontenc}

\usepackage{amssymb,amsmath,latexsym}
\RequirePackage{graphicx}
\RequirePackage{mathptmx}      % use Times fonts if available on your TeX system
\RequirePackage{flushend}
\RequirePackage[colorlinks,citecolor=blue,urlcolor=blue,linkcolor=blue]{hyperref}

\usepackage{sidecap}
\usepackage{ifthen}
\usepackage{keycommand}
\usepackage[toc,page]{appendix}
\usepackage[labelfont={small,bf},textfont=small]{caption}
\usepackage[labelfont={small,bf},textfont=small]{subcaption}

\usepackage{placeins} % It gives a \FloatBarrier command that you can use to prevent floats to appear beyond some point in your document.
\usepackage{multirow}

\usepackage{enumitem}
\setlist{  
  listparindent=\parindent,
  parsep=0pt,
}

\makeatletter \let\cl@chapter\relax \makeatother % workaround to fix cleveref and svjour incompatibilities http://phaseportrait.blogspot.fr/2007/08/cleveref-incompatibilities-with.html (need to be before amsmath)
\usepackage{cleveref}  % Basically, cleveref must be loaded last! (it manipulates ref)
\newcommand{\refp}[1]{(\ref{#1})}

\crefname{equation}{eq.}{eqs.}
\crefname{section}{sect.}{sects.}
\crefname{chapter}{chapter}{chapters}
\crefname{table}{table}{tables}
\crefname{figure}{fig.}{figs.}
\crefname{appsec}{appendix}{appendices}
\crefname{appchap}{appendix}{appendices}

\renewcommand\Re{\operatorname{Re}}
\renewcommand\Im{\operatorname{Im}}
\DeclareMathOperator{\e}{e}
\makeatletter
\def\blfootnote{\xdef\@thefnmark{}\@footnotetext}
\makeatother

\usepackage{xifthen}
\usepackage{keycommand}
\newcommand{\abs}[1]{\ensuremath{\left| {#1} \right|}}
\newcommand{\ampl}[2][]{\ensuremath{F_{#1}^{\text{#2}}(s,t)}}
\newcommand{\modulus}[2][]{\ensuremath{\abs{\ampl[#1]{#2}}}}
\newcommand{\phase}[1][]{\ensuremath{\zeta^\text{N}_{#1}(s,t)}}

\newcommand{\PROF}[1]{\ensuremath{D^{\text{#1}}}}

\newcommand{\dcss}[3][]{
\ifthenelse{\equal{#1}{}}
{\ensuremath{      \frac{\text{d}\sigma^{#2}{#3}}{\text{d}t}}}
{\ensuremath{\left.\frac{\text{d}\sigma^{#2}{#3}}{\text{d}t}\right|_{#1}}}
}

\newcommand{\dcs}[2][]{\dcss[#1]{#2}{}}

\newcommand{\weight}[0]{w}
\newkeycommand{\meanb}[n=1,etype=,j=,weight=]{\ensuremath{\langle \ifthenelse{\equal{\commandkey{n}}{1}}{b}{b^{\commandkey{n}}} \rangle^{\text{\commandkey{etype}}\ifthenelse{\equal{\commandkey{weight}}{}}{}{,{\text{\weight}=\commandkey{weight}}}}_{\commandkey{j}} }}
\newkeycommand{\bmax}[j=]{ \ensuremath{b^{\text{max}}\ifcommandkey{j}{_{\commandkey{j}}}{} }}

\newkeycommand{\PROB}[etype=,j=]{ \ensuremath{\ifcommandkey{etype}{P^{\text{\commandkey{etype}}}}{P}_{\ifcommandkey{j}{\commandkey{j}}{}} }}

\newkeycommand{\CS}[etype=,j=]{ \ensuremath{\ifcommandkey{etype}{\sigma^{\text{\commandkey{etype}}}}{\sigma}_{\ifcommandkey{j}{\commandkey{j}}{}} }}

\newkeycommand{\dbt}[j=]{ \ensuremath{ d_{b\ifthenelse{ \equal{\commandkey{j}}{} }{}{,\commandkey{j}}}(t) }}

\graphicspath{{figures/}} 

\DeclareCaptionJustification{justified}{\justifying}
\begin{document}
\title{\TITLE}
\author{Ji\v{r}\'{\i} Proch\'{a}zka}
\email{jiri.prochazka@fzu.cz}
\author{Vojt\v{e}ch Kundr\'{a}t}
\email{kundrat@fzu.cz}
%\affiliation{Institute of Physics of the AS CR, v.v.i., 18221 Prague 8, Czech Republic} 
\affiliation{The Czech Academy of Sciences, Institute of Physics, 18221 Prague 8, Czech Republic}


\begin{abstract}
Under the influence of standardly used description of Coulomb-hadronic interference proposed by West and Yennie the protons have been interpreted as transparent objects; elastic events have been interpreted as more central than inelastic ones. It will be shown that using eikonal model the protons may be interpreted in agreement with usual ontological conception; elastic processes being more peripheral than inelastic ones. The corresponding results (differing fundamentally from the suggested hitherto models) will be presented by analyzing the most ample elastic data set measured at the ISR energy of 52.8~GeV and the LHC energy of 8~TeV. Detailed analysis of measured differential cross section will be performed and possibility of peripheral behavior on the basis of eikonal model will be presented. The impact of recently established electromagnetic form factors on determination of quantities specifying hadron interaction determined from the fits of experimental elastic data will be analyzed. The influence of some other assumptions on proton characteristics derived from elastic hadronic amplitude determined on the basis of experimental data will be studied, too.
\end{abstract}


\maketitle

%%%%%%%%%%%%%%%%%%%%%%
\section{\label{sec:introduction}Introduction}
\begin{figure*}%[!htb]
\centering
\begin{subfigure}[b]{0.48\textwidth}
\includegraphics*[width=\textwidth]{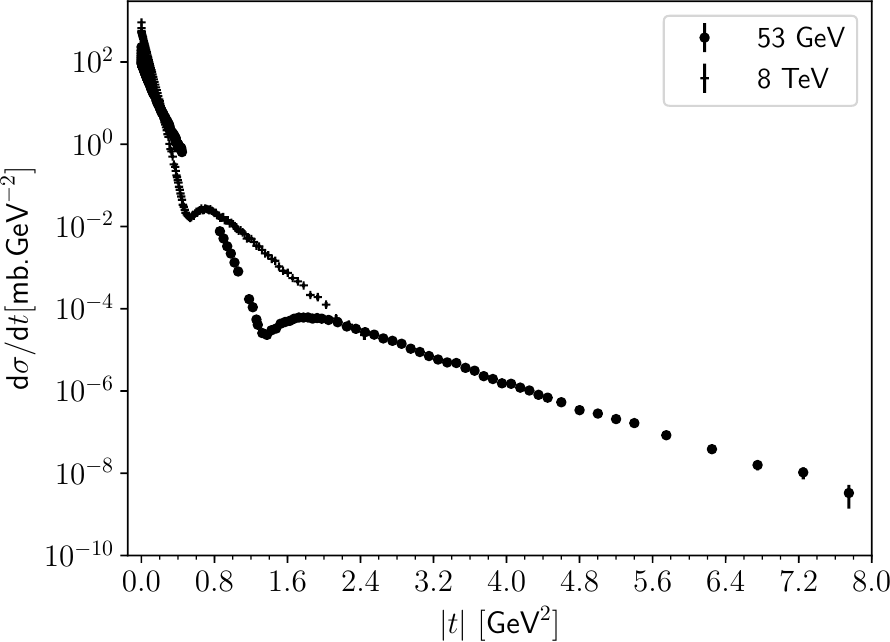}
\caption{$|t|$ values up to 8~GeV$^2$}
\end{subfigure}
\quad%add desired spacing between images, e. g. ~, \quad, \qquad etc.
\begin{subfigure}[b]{0.48\textwidth}
\includegraphics*[width=\textwidth]{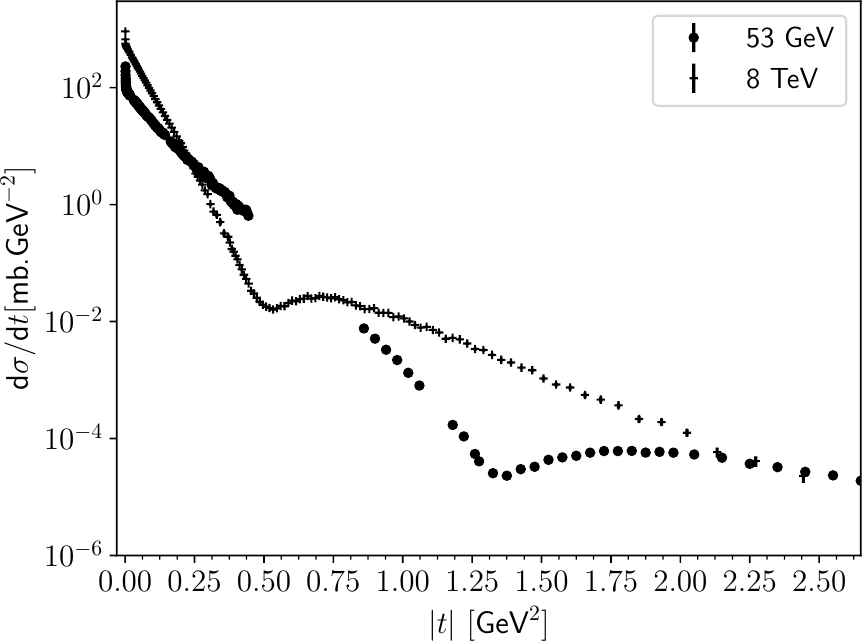}
%\vspace{0.05cm}
\caption{$|t|$ values up to $\approx2.5$~GeV$^2$}
\end{subfigure}
\begin{minipage}[t]{.9\textwidth}
		\caption{\label{fig:dsdt_data_pp53gev_pp8000gev}Comparison of measured elastic pp differential cross section at 52.8~GeV and 8~TeV (see \cref{sec:pp53gev_data,sec:pp8000gev_data} for details). Measured $|t|$-dependences of differential cross sections at these two energies are significantly different but they have similar structure: peak at very low values of $|t|$ and dip-bump at higher values of $|t|$.}
\end{minipage}
\end{figure*}

%%%%%%%%%%%%%%%%%%%%%%
Elastic differential cross section $\text{d}\sigma/\text{d}t$ represents basic experimental characteristic established in elastic collisions of hadrons. If the influence of spins is not considered, the $t$ (four momentum transfer squared) dependence exhibits a very similar structure in all cases of elastic scattering of charged hadrons at contemporary high energies: there is a peak at very low values of $|t|$, followed by a (nearly) exponential region and then there is a dip-bump or shoulder structure at even higher values of $|t|$ practically for all colliding hadrons \cite{Carter1986}. \Cref{fig:dsdt_data_pp53gev_pp8000gev} shows comparison of measured $\text{d}\sigma/\text{d}t$ for pp scattering at the ISR energy of 52.8~GeV and much higher LHC energy of 8~TeV as examples.

The elastic differential cross section  is standardly defined using elastic scattering amplitude $\ampl{}$ as (common units $\hbar = c = 1$ used)
\begin{equation}
\frac{\text{d} \sigma}{\text{d}t} = \frac{\pi}{s p^2} \modulus{}^2
\label{eq:difamp_gen}
\end{equation}
where $s$ is the square of the total center-of-mass energy and $p$ is the value of momentum of one incident hadron in the center-of-mass system; $t = - 4p^2 \sin^2 \frac{\theta}{2}$ where $\theta$ is scattering angle.

Two fundamental interactions have been commonly used for description of measured elastic differential cross section of two charged hadrons: the long-ranged Coulomb (electromagnetic) interaction and much stronger but short-ranged hadronic interaction. While the former one is assumed to be well known from QED (except electromagnetic form factors), the determination of the latter one is more complicated. %Well known theory of strong interactions, QCD, is not applicable in this case. 
The simultaneous action of both the interactions already in the case of elastic collisions represents, therefore, quite delicate problem even thought some theoretical approaches exist. The contemporary situation was recently summarized, e.g., in an Italian strategic document \cite{Andreazza2015} (see sect.~7.5 concerning total, elastic and diffractive cross sections therein):
%by the CSN1, the Italian INFN committee which manages experiments at high energy particle accelerators (see sect.~7.5 concerning total, elastic and diffractive cross sections in \cite{Andreazza2015}):
%It was pointed out in a document describing its understanding of the current scientific scenarios and proposing a strategy for the next 10 years in the context of a 20-year global vision for this field of research (see sect.~7.5 concerning total, elastic and diffractive cross sections in \cite{Andreazza2015}): 

\noindent
%\begin{center}
%\begin{minipage}{.9\columnwidth}
\textit{"Several theoretical models have been developed during the last decades to interpret the experimental results. Unfortunately, the perturbative QCD approach cannot be used in this context since most of the processes contributing to the total cross section are characterised by low momentum transfer. Some of the models are still based on Regge theory, while others prefer using optical or eikonal approaches. Moreover, so-called QCD-inspired models are trying to connect the concepts of Pomeron trajectories and proton opacity to the QCD description of elementary interactions between quarks and gluons. At the moment, no model manages to describe qualitatively and quantitatively the large amount of data available; they all have merits and shortcomings. Typically, they successfully describe the experimental results in a certain kinematic range but completely fail in other ones."}
%\end{minipage}
%\end{center}

\noindent
Indeed, only some phenomenological models have been applied to in interpreting experimental data represented by elastic differential cross sections until now. Theoretical description of elastic scattering is more complicated that it may seem at first glance, even if it is kinematically the simplest process. 

It is evident that result of collision of two particles may depend strongly on sizes and internal structures of colliding particles. The characteristics of their individual collisions may depend also on the values of their impact parameter. Impact parameter analysis of experimental collision data may, therefore, shed light on the dynamics of hadronic collisions and spatial characteristics of colliding particles which can be hardly obtained in different way. From the contemporary approaches (see also above) only the eikonal one has allowed to obtain some knowledge in this direction. This is also one of the main reasons why possibilities of this approach (applied to experimental data) will be discussed in more details in this paper.

One of the first attempts to describe elastic Coulomb-hadronic scattering of two charged hadrons was done by West and Yennie (WY) in 1968 \cite{WY1968} who proposed a formula for corresponding complete elastic scattering amplitude $\ampl{C+N}$. However, the formula has been derived under very simplified and limited conditions. It has been assumed to be valid only in region of very low values of $|t|$. It has been further assumed at any high collision energy (value of $\sqrt{s}$) that modulus of hadronic amplitude is purely exponential function of $t$ and that the ratio of the real to imaginary part of hadronic amplitude is $t$-independent; both these a priori very strong assumptions have been assumed to be valid at \emph{all} kinematically allowed values of $t$ (including region beyond assumed validity of the formula and even outside the region of measured $t$-values). The approach of WY has not allowed to study shape of $t$-dependence of hadronic amplitude on the basis of experimental data as the shape has been, without any justification, fixed by the used assumptions.

Under the influence of the approach of WY the measured elastic pp (or $\bar{\text{p}}$p) differential cross section at given energy has been commonly divided into two regions. Region of very low values of $|t|$ has been commonly analyzed since the era of ISR with the help of the simplified formula of WY for determination of 3 free parameters: total hadronic cross section $\CS[etype={tot,N}]$, quantity $\rho$ (the ratio of the real to imaginary parts of hadronic amplitude in forward direction) and diffractive slope $B$ at the given energy. In the second region of higher values of $|t|$ (including the dip-bump structure) any Coulomb effect has been commonly a priori neglected. This region cannot be described with the help of purely exponential modulus of hadronic amplitude as assumed in the WY approach. Different models of hadronic amplitude based on assumptions inconsistent with the WY approach have been used in many analyses of measured data in this region. The hadronic models have been often constrained by the values of parameters $\CS[etype={tot,N}]$, $\rho$ and $B$ determined by the simplified formula of WY (see, e.g., \cite{Fagundes2011_ijmpa26}). Full measured region of data have been, therefore, described in dual and inconsistent way which can be hardly denoted as satisfactory. The values of $\CS[etype={tot,N}]$ and $\rho$ determined with the help of the WY approach at different energies have been often used also in connection with dispersion relations (see, e.g., \cite{COMPETE2002}). These approaches have been, therefore, based on all the problems and limitations involved in the WY approach; neither one of these models has provided consistent description of elastic differential cross section at \emph{all} values of~$t$. 

In order to avoid the before mentioned discrepancies another more convenient approach based on the eikonal model has been proposed; see \cite{Kundrat1994_unpolarized}. In this case the used complete eikonal elastic scattering amplitude \ampl{C+N} describes the influence of both Coulomb and hadronic scattering with the help of only one formula in the whole measured region of momentum transfers in a unique and consistent way. It is based on additivity of individual eikonals of the Coulomb and hadron interactions. The formula for complete amplitude \ampl{C+N} has been derived with the aim not to impose any strong assumptions concerning $t$-dependence of elastic hadronic amplitude \ampl{N}.
%At difference to the WY approach, consisting in determining the relative phase between the Coulomb and elastic hadronic amplitudes, the eikonal model approach has consisted in calculating the complete elastic amplitude and has specified directly the complete elastic differential cross section $\frac{\text{d} \sigma^{\text{C+N}}}{\text{d}t}$ according to \refp{eq:difamp_gen} in the whole measured region of momentum transfers. 
As the Coulomb scattering amplitude has been assumed to be known from the QED (apart from form factors), the only task has consisted in determining the elastic hadronic amplitude \ampl{N}. 

The complex function \ampl{N} cannot be, however, derived from the mere experimental data with the help of \cref{eq:difamp_gen} if only hadronic interaction is taken into account; only its modulus in the measured region of $t$-values may be determined and then extrapolated outside this region. Coulomb-hadronic interference has been standardly used to explain observed peak in measured differential cross section at very small values of $|t|$ (see \cref{fig:dsdt_data_pp53gev_pp8000gev,eq:difamp_gen}) and also to constrain not only the modulus but also the $t$-dependence of phase of hadronic amplitude. However, even in this case the elastic hadronic amplitude \ampl{N} has been uniquelly established if and only if some additional assumptions have been applied to. It has been almost generally assumed (under the influence of the WY approach) that imaginary part of hadronic amplitude has been dominant in a broad region of $t$ around $t=0$ and that it has vanished around the region of diffractive minimum. It has concerned nearly all contemporary models of elastic (hadronic) scattering including the most recently published papers, see e.g., \cite{Franca1979,Carvalho1997,Carvalho2005,Menon2005,Menon2007,Avila2006,Avila2007,Avila2008,Silva2007,Ferreira1997,Ferreira1998,Campos2010,Fagundes2011_epjc71,Fagundes2011_ijmpa26,Fagundes2012,Fagundes2016,Menon2013_125001,Fagundes2013_065005,Shoshi2002,Ryskin2007,Lipari2013,Kohara2014,Arriola2017,Albacete2017,Broniowski2018,Alkin2018,Goncalves2019,Martynov2019,Csorgo2019} (in some papers the $t$-dependence of the phase of \ampl{N} has not been discussed at all). These additional assumptions constraining hadronic amplitude \ampl{N} have never been sufficiently reasoned (in some cases not even clearly mentioned). They have led to some unusual physical properties of protons - to large proton transparency in 'head-on' collisions at ISR energies \cite{Miettinen1975,Groot1973}, i.e., to the maximum probability of elastic processes at $b=0$ (denoted as centrality of elastic collisions), which has been regarded by some authors as a 'puzzle' \cite{Giacomelli1979}. One can also hardly understand why single diffractive production process \mbox{pp $\rightarrow$ p(n$\pi^+$)} should be peripheral \cite{Giovannini1979,Chou1985}, when its differential cross section is very similar to the elastic pp differential cross section (also having dip-bump structure). One would expect, in agreement to usual ideas corresponding to collisions of two matter objects, that collision processes in which at least one proton survives should be all peripheral.
 %and differ significantly from non-diffractive (inelastic) ones. It contradicts also the standard realistic picture of matter particles; elastic processes being expected to exhibit peripheral behavior (elastic collisions corresponding in average to higher impact parameters then inelastic ones).
%of both these kinds of diffractive processes are very similar and differ significantly from non-diffractive (inelastic) ones. 

It has been shown already in 1981 in \cite{Kundrat1981} that the central interpretation has followed as direct consequence of the mentioned $t$-dependence of the dominant imaginary part of $\ampl{N}$, or equivalently as a consequence of the amplitude phase very slowly changing with rising $|t|$ in broad region of $t$ close to $t=0$. Such $t$-dependence has never been theoretically justified (up to our knowledge) in the literature. 

It has been found already in \cite{Kundrat1981} that the high-energy elastic hadronic scattering may be described with the help of eikonal model as a fully peripheral process if the phase has been allowed to change rather quickly with changing $t$ (for more details see \cite{Kundrat1987,KL1989,Kundrat1990,Kundrat1992,Kundrat1994_unpolarized,KL1996}); or if the imaginary part of the elastic hadronic amplitude decreases quickly and vanishes at $|t| \sim 0.1$ GeV$^2$ (at 52.8~GeV). Such a $t$-dependence of peripheral hadronic phase has been found numerically by means of minimization technique.

The centrality of elastic collisions has been recently "rediscovered" by some authors under the term \emph{hollowness}, see, e.g., \cite{Arriola2017,Troshin2017,Albacete2017,Broniowski2018}. The result has followed mainly from the requirement of the dominance of the imaginary part of \ampl{N} in quite broad interval of $t$ around $t=0$ in the given models, see also useful comments related to the hollowness in \cite{Petrov2018}. The fact that $t$-dependence of the phase if \ampl{N} matters for determination of characteristics of collisions in impact parameter space has been recently pointed out in \cite{Dremin2019}. The question of peripheral behaviour of elastic collisions has been recognized as an interesting question in \cite{Kohara2017}. It has been summarized in \cite{Petrov2018_sizes} that protons cannot be taken as point-like particles during collisions and that it is necessary to take into account sizes of colliding particles in description of the physical process. In other words, some even basic questions and problems concerning description of elastic pp scattering have not been satisfactorily understood and solved up to know.

We have revisited possibilities of basically all older as well as more recent approaches (models) trying to describe elastic scattering data, especially within the eikonal model framework (see also \cite{Prochazka2018_phd_thesis}). This widely used theoretical framework at high energies is the only one which allows to take into account dependence of (elastic) collisions on impact parameter and Coulomb-hadronic interference. It is necessary to take into account impact parameter $b$ of colliding particles in analysis of experimental data of elastic collisions in order not to mix collisions at different values of $b$. This in turn may provide important information about shapes, dimensions and other characteristics of colliding particles which can be hardly obtained in different way. %These studies have been motivated by fundamental question concerning behavior of elastic scattering in the impact parameter space. 
The main aims of this paper may be summarized as follows:

\begin{itemize}
\item{to expose and improve consistent mathematical and physical analysis of elastic pp scattering at high energies and at \emph{all} measured values of momentum transfers within the eikonal model framework proposed originally in \cite{Kundrat1994_unpolarized} (i.e., to solve the problem related to the mentioned dual and inconsistent description of data introduced (directly or indirectly) by the usage of the simplified formula of WY);}
\item{and to study different interpretation possibilities of elastic pp collisions in the impact parameter space and the assumptions leading to the given behavior; with focus on showing possibility of peripheral description of elastic scattering as this solution of the collision process is not sufficiently known although its existence has been published several times.}
\end{itemize}

Both the tasks will be demonstrated in the case of pp elastic scattering at the ISR energy of 52.8~GeV and the LHC energy of 8~TeV (see \cref{fig:dsdt_data_pp53gev_pp8000gev}). Our goal consisted also in comparison of determined quantities characterizing pp collisions on the basis of experimental data at the two \emph{very different} energies. This approach requires to use the formalism of impact parameter representation of scattering amplitudes. And, of course, full description of Coulomb scattering, i.e., also the knowledge of $t$-dependence of form factors in the broadest possible region of $t$ variable. For this reason impact of choice of different form factors on determination of hadronic quantities will be also studied. The eikonal model interference formula for complete amplitude enabling to describe the contemporary influence of both the Coulomb and hadronic scattering of charged hadrons at any (high) energy and at any measured $t$ value will be updated and improved. It will take into account both the Coulomb interactions described via electric and magnetic form factors. It will be shown that it can substitute the commonly used simplified amplitude of WY for the analysis of corresponding differential cross section data. The eikonal model allows, therefore, consistent and more reliable description of data.

This paper is structured as follows. The simplified description of Coulomb and hadron interference proposed by WY, which influenced directly or indirectly many contemporary models of elastic (hadronic) scattering, is summarized in \cref{sec:WY}. Electromagnetic form factors needed in description of elastic pp collisions are discussed in \cref{sec:ff_elcmag}. There are several formulas or parameterizations which have been used recently by several (group of) authors for determination of $t$-dependences of electric and magnetic form factors. However, it seems that there is no comparative study between them showing how much the $t$-dependences differ (if at all). In \cref{sec:ff_elcmag} new plots comparing several alternatives available in the literature are, therefore, shown. The influence of Coulomb interaction described with the help of \emph{both} the electric and magnetic form factors in elastic scattering of charged hadrons in the eikonal model approach is for the first time analyzed in detail in \cref{sec:eikonal} (originally only electric form factors have been used in \cite{Kundrat1994_unpolarized}). Assumptions concerning parameterizations of elastic hadronic amplitude in contemporary models which are commonly applied to experimental data and leading to central behavior of elastic hadron collisions are discussed in \cref{sec:ampl_n} together with assumptions leading to peripheral behavior. Both the fundamentally different alternatives are fitted to experimental data of elastic pp scattering at the ISR energy of 52.8~GeV and much higher LHC energy of 8~TeV in \cref{sec:data_analysis} under different assumptions. The results are then compared and further discussed in greater detail than it was done in the past. E.g., different peripheral alternatives differing in value of mean impact parameter corresponding to elastic scattering are shown in \cref{sec:data_analysis} for the first time. The impact of choice of form factor on the determined results will be also discussed in \cref{sec:data_analysis}; it represents another new result. Concluding remarks are then given in \cref{sec:conclusion}.
The corresponding formalism of impact parameter representation of the elastic hadron scattering amplitude (valid at any $s$ and $t$) at \emph{finite} energies used in \cref{sec:data_analysis} is summarized briefly in appendix~\ref{sec:ampl_b}.  
%Appendix~\ref{sec:wy_study} is devoted to some aspects of the relative phase $\alpha\phi(s,t)$ in the WY approach. Analytical expression \refp{eq:phiWY} will be compared to corresponding numerical calculation obtained with the help of \cref{eq:phaseWY}. It will be shown in this section, too, that the integral formula \refp{eq:phaseWY} is consistent with $t$-independent hadronic phase only and that it cannot be, therefore, used for arbitrary $t$-dependence of \ampl{N}. 
Appendix~\ref{sec:wy_study} is devoted to new calculations explicitly demonstrating that the approach of WY is consistent with $t$-independent hadronic phase only; and that the approach cannot be, therefore, used for general analysis of experimental data with arbitrary $t$-dependence of \ampl{N}.

\section{\label{sec:WY}Simplified description of Coulomb and hadron interference by West and Yennie}

According to Bethe \cite{Bethe1958} (1958) the complete elastic scattering amplitude $\ampl{C+N}$ of two charged hadrons (neglecting spins)\footnote{Taking into account also spins represent much more delicate problems from both theoretical as well as experimental point of view. A theoretical attempt how to take into account spins with the help of helicity amplitudes may be found in, e.g., \cite{Buttimore1978}.} has been commonly decomposed into the sum of the Coulomb scattering amplitude $\ampl{C}$ and the hadronic amplitude $\ampl{N}$ bound mutually with the help of relative phase $\alpha\phi(s,t)$ 
\begin{equation}
\ampl{C+N} = \ampl{C}\e^{\text{i}\alpha\phi(s,t)}+\ampl{N};
\label{eq:FCNbethe} 
\end{equation}
$\alpha=1/137.036$ being the fine structure constant. The $t$-dependence of the relative phase factor $\alpha\phi(s,t)$ has been determined on various levels of sophistication. The dependence having been commonly accepted in the past was proposed by West and Yennie (WY) \cite{WY1968} (1968) within the framework of Feynman diagram technique (one-photon exchange) in the case of charged point-like particles and for $s \gg m^2$ ($m$ standing for nucleon mass) as
\begin{equation}
\alpha\phi(s,t) = \mp \alpha\left[\ln\left (\frac{-t}{s}\right) + \int_{-4p^2}^0 \frac{\text{d}t'}{\abs{t-t'}} \left(1-\frac{F^\text{N}(s,t')}{F^\text{N}(s,t)}\right)\right].
\label{eq:phaseWY}
\end{equation}
The upper (lower) sign corresponds to the scattering of particles with the same (opposite) electric charges.

Formula \refp{eq:phaseWY} containing the integration over all admissible values of four-momentum transfer squared $t'$ seemed to be complicated when it was proposed. It has been simplified for practical use to perform the analytical integration. The $t$-dependencies of modulus and phase of the hadronic amplitude \ampl{N} defined as 
\begin{equation} 
\ampl{N} = \text{i}\modulus{N} \e^{-\text{i}\phase}
\label{eq:modphas}
\end{equation}
have been strongly limited. It has been assumed: 
\begin{itemize}
		\item[(i)]{the modulus \modulus{N} has had purely exponential $t$-dependence at \emph{all} kinematically allowed $t$ values;}
		\item[(ii)]{the phase $\phase$ has been $t$-independent for \emph{all} kinematically allowed $t$ values (see \cite{WY1968,Amaldi1973_231}, for more details see \cite{KL1989,KL1996}).}
\end{itemize}
As introduced in \cite{KL2005} some other high energy approximations and simplifications were added, too (see also \cite{Prochazka2018_phd_thesis}).

For the relative phase between the Coulomb and elastic hadronic amplitude the following simplified expression has been then obtained:
\begin{equation}
\alpha \phi(s,t) = \mp \alpha \left [\ln{\left(\frac{-B(s)t}{2}\right)}+\gamma \right]
\label{eq:phiWY}
\end{equation}
where $\gamma=0.577215$ is Euler constant and $B(s)$ is the value of diffractive slope $B(s,t)$ at $t=0$ generally defined as
\begin{equation}
B(s,t) = \frac{\text{d}}{\text{d} t} \left[ \ln \dcs{\text{N}}(s,t)\right]
= \frac{2}{\modulus{N}} \frac{\text{d}}{\text{d}t}\modulus{N} \; .
\label{eq:slope}
\end{equation}
The $t$-independence of $B(t)$ is equivalent to the requirement of purely exponential $t$-dependence of \modulus{N}. 

One may further define quantity $\rho(s,t)$ as ratio of the real to imaginary parts of elastic hadronic amplitude 
\begin{equation}
\rho(s,t) = \frac{\Re \ampl{N}}{\Im \ampl{N}}.
\label{eq:rho}
\end{equation}
It follows from \cref{eq:modphas,eq:rho} that
\begin{equation}
\tan{\phase} = \rho(s,t) \; , 
\label{eq:tanzeta}
\end{equation}
i.e., the assumption concerning $t$-independence of hadronic phase $\phase$ is equivalent to assumption of quantity $\rho(s,t)$ being $t$-independent.

The complete elastic scattering amplitude $\ampl{C+N}$ has been then written as
\begin{equation} 
\begin{split}
\ampl[\text{WY}]{C+N}  =&  \pm \frac{\alpha s}{t}G_1(t)G_2(t) \e^{\text{i}\alpha \phi(s,t)} \\
&+ \frac{\CS[etype={tot,N}](s)}{4\pi}p\sqrt{s}(\rho(s)+\text{i})\e^{B(s)t/2}.
\label{eq:simplifiedWY}
\end{split}
\end{equation}
Here the first term corresponds to the Coulomb scattering amplitude (relative phase included) while the second term represents the elastic hadronic amplitude in which the quantity $\CS[etype={tot,N}](s)$ is the total cross section given by optical theorem
\begin{equation}
  \CS[etype={tot,N}](s) = \frac{4 \pi}{p\sqrt{s}} \Im F^{\text{N}}(s,t=0)
  \label{eq:optical_theorem}
\end{equation}
and the quantity $\rho(s)$ is value of the assumed $t$-independent quantity $\rho(s,t)$. The two quantities $G_1(t)$ and $G_2(t)$ stand for the electric form factors taken commonly in standard dipole form (see, e.g., \cite{Block1985}) as
\begin{equation}
G_{\text{E}}^{\text{D}}(t) = \left(1 - \frac{t}{\Lambda^2}\right)^{-2}
\label{eq:G_E_dipol}
\end{equation}
where $\Lambda^2 = 0.71\; \text{GeV}^2$. The electric form factors as Fourier-Bessel (FB) transformation of electric charge distribution of colliding hadrons have been put into formula~\refp{eq:simplifiedWY} by hand.

The Coulomb differential cross section (including form factors) has been, therefore, taken as
\begin{equation}
\frac{\text{d}\CS[etype=C](s,t)}{\text{d}t} = \frac{\pi s}{p^2} \frac{\alpha^2}{t^2} G_1^2(t)G_2^2(t),
\label{eq:dcs_c_qed}
\end{equation}
i.e., diverging at $t\!=\!0\,$. In high energy limit the Coulomb differential cross section~\refp{eq:dcs_c_qed} may be further simplified to known form
\begin{equation}
\frac{\text{d}\CS[etype=C](s,t)}{\text{d}t} = \frac{4\pi\alpha^2}{t^2} G_1^2(t)G_2^2(t).
\label{eq:dcs_ampl_c_high_energy_limit}
\end{equation}

As to \cref{eq:simplifiedWY,eq:phiWY} they were derived also by Locher~\cite{Locher1967} one year earlier than \cref{eq:phaseWY} proposed by WY~\cite{WY1968}. Locher assumed from the very beginning the validity of both the mentioned assumptions (\textit{i}) and (\textit{ii}) limiting the general $t$-dependence of the elastic hadronic amplitude \ampl{N}. He, therefore, avoided the misleading idea that WY integral formula \refp{eq:phaseWY} may be correctly used for determination of the relative phase for any $t$-dependent elastic hadronic amplitude \ampl{N}. The high-energy approximations used in the given approach might be regarded as acceptable at that time when nothing was known about actual structure of elastic differential cross section data. However, the questions have arisen when experimental data have shown not to be in agreement with the mentioned assumptions (for details see \cite{KL1996,Kundrat2001}).

\Cref{eq:simplifiedWY,eq:phiWY,eq:difamp_gen} have been used practically for the analysis of all hitherto elastic scattering data of charged hadrons in the forward region, i.e., for $|t|\lesssim 0.05$ GeV$^2$ (see, e.g., \cite{Amaldi1973_231,Block1985}, \cite{Amaldi1973,Bartenev1973,Amaldi1976,Baksay1978,Nagy1979,Fajardo1981,Favart1981,Amos1983,Carboni1985,Bernard1987,Bozzo1984,Bozzo1985,Augier1993,Block2006}); contrary to the fact that both the mentioned theoretical assumptions (\textit{i}) and (\textit{ii}) justifying the correctness of both \cref{eq:phiWY,eq:simplifiedWY} have not been fulfilled in the analyzed experimental data. At higher values of $|t|$ the influence of Coulomb scattering has been then fully neglected and the elastic scattering of charged hadrons has been described only with the help of the elastic hadronic amplitude being constructed on a phenomenological basis with completely different $t$-dependence. Such type of fundamentally inconsistent description of elastic scattering by two different approaches in diverse regions of $t$ has been pointed out and further analyzed in, e.g., \cite{KL1996,Kundrat2000,Deile:2010mv_Kundrat,elba2010_klkp,KL2007}. 

The elastic hadronic amplitude \ampl{N} has been then transformed into impact parameter representation of elastic scattering amplitude $h_{\text{el}}(s,b)$ introduced with the help of FB transform:
\begin{equation} 
h_{\text{el}}(s,b) =\frac{1}{4p\sqrt{s}}\int\limits_{-\infty}^0 \ampl{N} J_0(b\sqrt{-t})\text{d}t;
\label{eq:hel_standard} 
\end{equation}
$J_0(x)$ being the Bessel function of the zeroth order. The elastic scattering amplitude $h_{\text{el}}(s,b)$ %or the elastic profile $|h_{\text{el}}(s,b)|^2$ 
%$\;\;\;{\bf\underline{and}}\;\;\;$ the elastic profile $|h_{\text{el}}(s,b)|^2$ has $\;\;\;{\bf\underline{have}}\;\;\;$
has been then required to fulfill the unitarity equation 
\begin{equation} 
\Im h_{\text{el}}(s,b) = |h_{\text{el}}(s,b)|^2+g_{\text{inel}}(s,b)
\label{eq:unitarity_standard}
\end{equation}
with the inelastic impact parameter profile $g_{\text{inel}}(s,b)$ being defined similarly as the FB transform of the inelastic overlap function $G_{\text{inel}}(s,t)$ fulfilling the unitarity relation \cite{Hove1963,Hove1964} (see also \cite{Islam1968})
\begin{equation} 
\Im \ampl{N} = \frac{p}{4\pi\sqrt{s}} \int \text{d}
\Omega' F^{\text{N}^*}(s,t') F^{\text{N}}(s,t'') + G_{\text{inel}}(s,t),
\label{eq:preunitarity}
\end{equation}
being valid at any $s$ and kinematically allowed value of $t$. The function $G_{\text{inel}}(s,t)$ represents summation of all possible inelastic states including integration over all remaining kinematical variables specifying corresponding production amplitude; $\text{d}\Omega' = \sin \vartheta'\text{d}\vartheta' \text{d}\Phi'$, $t = -4p^2 \sin^2{\frac{\vartheta}{2}}$, $t' = -4p^2 \sin^2{\frac{\vartheta'}{2}}$, $t'' = -4p^2 \sin^2{\frac{\vartheta''}{2}}$ and $\cos \vartheta'' = \cos \vartheta \cos \vartheta' +\sin \vartheta \sin \vartheta' \cos \Phi'$. Variables $\vartheta$, $\vartheta'$ and $\vartheta''$ are angles connected with the variables $t$, $t'$ and $t''$ in the center of mass system.

Formulas \refp{eq:hel_standard} and \refp{eq:unitarity_standard} have represented the starting basis practically in all phenomenological model analyses at finite energies where the impact parameter representation of elastic hadronic scattering amplitudes has been made use of, in spite of the fact that the formulas have been derived at asymptotic energies only (see, e.g., \cite{Miettinen1975,Groot1973,Amaldi1976_385,Amaldi1980,Castaldi1985,Barone2002,Ayres1976,Bailly1987}).

%%%%%%%%%%%%%%%%%%%%%%%%%%%%%%%%%%%
\section{\label{sec:ff_elcmag}Electromagnetic proton form factors determined from elastic ep scattering}
%%%%%%%%%%%%%%%%%%%%%%%%%%%%%%%%%%%

The proton cannot be taken as point-like object, which represents a modification of the simple Coulomb interaction as its charge is distributed in a larger space. The shape of this distribution and its influence on the corresponding interactions is commonly characterized by elastic electromagnetic form factors. The corresponding differential cross section $\frac{\text{d} \sigma}{\text{d}\Omega}$ (in one-photon exchange) in the laboratory frame has been described by Rosenbluth formula (see \cite{Rosenbluth1950,Thomas2001,Jain2006,Arrington2007_nucl_el_ff}) which has been rewritten later by Sachs~\cite{Sachs1962} in the form 
\begin{equation}
  \begin{split}
\left( \frac{\text{d} \sigma}{\text{d} \Omega} \right)_{ep} =
\left( \frac{\text{d} \sigma}{\text{d} \Omega} \right)_{\text{Mott}} \;\; 
&\left\{ \frac{1}{1 + \tau}
\left[ G_{\text{E}}^2(Q^2) \; + \; \tau G_{\text{M}}^2(Q^2) \right] \right.  \\
&+ \left. 2 \tau G_{\text{M}}^2(Q^2) 
\tan^2 \left(\frac{\theta}{2}\right) \right\}
\end{split}
\label{eq:ros1}
\end{equation}
where
\begin{equation} 
Q^2 = 4\; E E' \sin^2 \left( \frac{\theta}{2} \right), 
\label{eq:ros2}
\end{equation}
\begin{equation} 
\tau = \frac{Q^2}{4 m^2}
\label{eq:ros2b}
\end{equation}
and $E$ and $E'$ are the incident and final electron energies, respectively, which are bound due to the conservation of the total four-momentum by relation
\begin{equation}
E' \; = \; \frac{E}{1 + \frac{2 E}{m} \sin^2(\frac{\theta}{2})};
\label{eq:ros3}
\end{equation}
$\theta$ is the scattering angle of the electron in the laboratory frame. $G_{\text{E}}$ and $G_{\text{M}}$ stand for electric and magnetic form factor. The expression
\begin{equation}
\left(  \frac{\text{d} \sigma}{\text{d} \Omega} (E, \theta) \right)_{\text{Mott}} =
\frac{\alpha^2}{4 E^2 \;{\sin^4 (\frac{\theta}{2}})} \;
\frac{E'}{E} \; \cos^2 \left(\frac{\theta}{2}\right)
\label{eq:mott}
\end{equation}
is the Mott formula~\cite{Mott1930} (in one-photon exchange approximation) for the differential cross section describing the elastic scattering of Dirac electron with point-like and spinless charged particle of proton mass $m$ at incident energy $E$ in the same frame (see, e.g., \cite{Thomas2001}). 

The formula~\refp{eq:ros1} contains electric form factor $G_{\text{E}}(Q^2)$ and magnetic form factor $G_{\text{M}}(Q^2)$ which depend only on the square of exchanged momentum transfer
\begin{equation} 
t = -Q^2
\label{eq:ros6}
\end{equation}
and which should satisfy the initial conditions
\begin{equation}
 G_{\text{E}}(0) \;= \; {{G_\text{M}(0)}/{\mu_p}} \;=\;1;
\label{eq:G_EM_at_t0}
\end{equation}
here $\mu_p \approx 2.793$ is the proton magnetic moment divided by nuclear magneton.
%here $\mu_p \approx 2.793 \mu_N$ is the proton magnetic moment ($\mu_N$ being nuclear magneton).
Variable $\tau$ may be expressed also as (see \cref{eq:ros2b,eq:ros6})
\begin{equation}
\tau = \frac{-t}{4 m^2} \, .
\label{eq:t2tau}
\end{equation}

From early measurements of the elastic ep scattering at lower energies it has been also deduced that electric $G_{\text{E}}(t)$ proton form factor can be described by the dipole formula~\refp{eq:G_E_dipol} and the magnetic one by
\begin{equation}
G_{\text{M}}^{\text{D}}(t) \approx  \mu_p G_{\text{E}}^{\text{D}}(t) \; .
\label{eq:G_M_dipol}
\end{equation}

Borkowski et al.~\cite{Borkowski1974,Borkowski1975} analyzed elastic ep scattering data at several energies with the help of Rosenbluth differential cross section formula~\refp{eq:ros1} where the $t$-dependencies of both the electric and magnetic form factors have been parametrized by the formulas 
\begin{align}
G_{\text{E}}^{\text{B}}(t) &= \sum_{j=1}^4 \frac{g_k^{\text{E}}}{w_k^{\text{E}} - t}, \;\;\;  \label{eq:G_E_Borkowski} \\
G_{\text{M}}^{\text{B}}(t) &= \mu_p \; \sum_{j=1}^4 \frac{g_k^{\text{M}}}{w_k^{\text{M}} - t} \label{eq:G_M_Borkowski}
\end{align}
inspired by the vector dominance model. The original values of the parameters $g_k^{\text{E,M}}$ and $w_k^{\text{E,M}}$ (being different for both the electric and magnetic form factors) may be found in \cite{Borkowski1975}; the corresponding electric and magnetic form factors may be denoted as $G_{\text{E}}^{\text{BO}}(t)$ and $G_{\text{M}}^{\text{BO}}(t)$. Different shapes of electromagnetic form factor parametrizations have been proposed by Arrington et al.~\cite{Arrington2005,Arrington2007} (denoted as $G_{\text{E}}^{\text{AR}}(t)$ and $G_{\text{M}}^{\text{AR}}(t)$) and Kelly \cite{Kelly2004} which has been applied by Puckett \cite{Pucket2010} (denoted as $G_{\text{E}}^{\text{PU}}(t)$ and $G_{\text{M}}^{\text{PU}}(t)$), too.  

Extending the measurements of the proton electric and magnetic form factors to higher values of $|t|$ has offered a chance for a better description of the influence of electromagnetic proton structure in the elastic pp collisions at high energies. However, this approach may be considered as fully entitled assuming that the electric and magnetic form factors determined from an analysis of elastic ep scattering are identical with the form factors involved in a description of pp elastic scattering (which should be tested in the future).

\begin{figure}
\centering
%\begin{minipage}[t]{.49\textwidth}
\begin{minipage}[t]{1.\columnwidth}
\includegraphics[width=\textwidth,keepaspectratio]{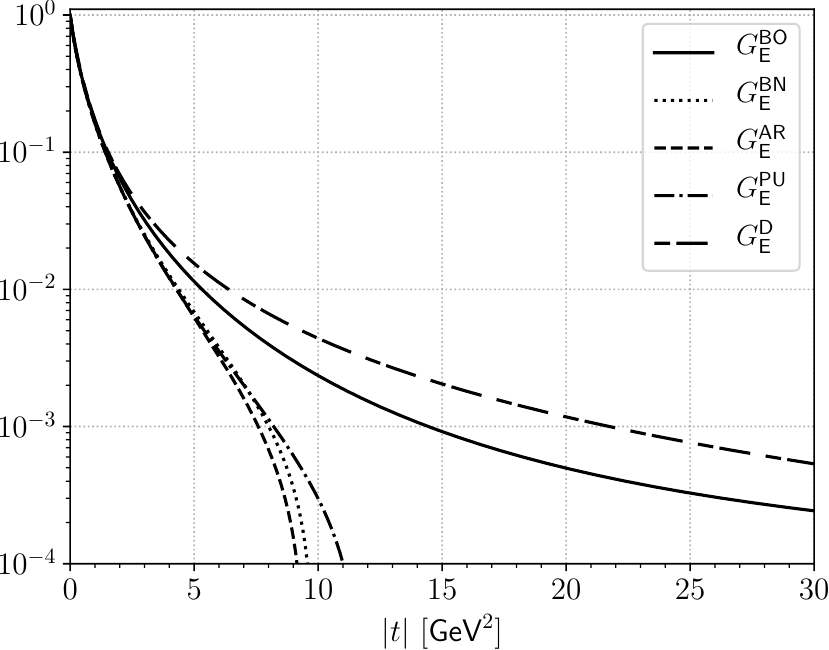} 
\caption{\label{fig:ff_elc} Proton electric form factors $G_{\text{E}}(t)$. 
}
\end{minipage}%
\hfill
%\begin{minipage}[t]{.49\textwidth}
\begin{minipage}[t]{1.\columnwidth}
  \centering
\includegraphics[width=\textwidth,keepaspectratio]{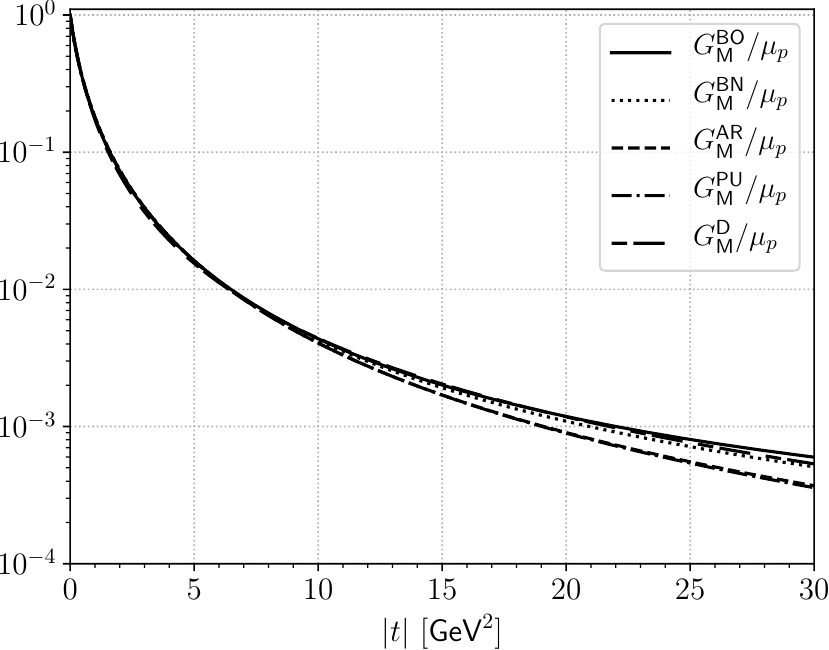} 
\caption{\label{fig:ff_mag}
Proton magnetic form factors $G_{\text{M}}(t)/\mu_p$ having very similar $t$-dependences. 
}
\end{minipage}%
\end{figure}

The relatively recent determination of $t$-dependent electric and magnetic form factors has been done by Arrington et al.~\cite{Arrington2007} (see also \cite{Arrington2007_nucl_el_ff,Arrington2005}) in the relatively broad region of $-t \in (0.007, 5.85)$ GeV$^2$. In this region we may express (refit) the form factors using the parameterizations of Borkowski given by \cref{eq:G_E_Borkowski,eq:G_M_Borkowski}.
%published from the historical reasons in the form of ratios of the measured electric and magnetic form factors to their dipole forms, e.g., as $G_{\text{E}}(t)/G_{\text{E}}^{\text{D}}(t)$ and $G_{\text{M}}(t)/({\mu_p}G_{\text{E}}^{\text{D}}(t))$ 
The refitted parameters are in \cref{tab:Bork_new_par_values}; the corresponding electric and magnetic form factors may be denoted as $G_{\text{E}}^{\text{BN}}(t)$ and $G_{\text{M}}^{\text{BN}}(t)$; they we will be used for data analysis in \cref{sec:data_analysis}. The mentioned electric and magnetic form factors (in different parameterizations) $G_{\text{E,M}}^{\text{AR}}(t)$, $G_{\text{E,M}}^{\text{PU}}(t)$, $G_{\text{E,M}}^{\text{BO}}(t)$, $G_{\text{E,M}}^{\text{BN}}(t)$ and $G_{\text{E,M}}^{\text{D}}(t)$ are shown in \cref{fig:ff_elc,fig:ff_mag}. 

%%%%%%%%%%%%%%%%%%%%%%%%%%%%%%%%%%%%%%%%%%%%%%%%%%%%%%%%
\begin{table}
\centering
\begin{tabular}{cccccc}  
\hline  \hline
k                 & 1       & 2      & 3       & 4       \\
\hline
$g_k^{\text{E}} $ & 0.1344  & 5.014  & -7.922  & 2.747   \\
$w_k^{\text{E}} $ & 0.2398  & 1.135  &  1.530  & 2.284   \\
$g_k^{\text{M}} $ & 0.2987  & 27.73  & -28.15  & 0.1274  \\
$w_k^{\text{M}} $ & 0.3276  & 1.253  &  1.276  & 6.361   \\
\hline\hline   
\end{tabular}
%\begin{minipage}[t]{.8\textwidth}
\begin{minipage}[t]{1.\columnwidth}
\caption{\label{tab:Bork_new_par_values} 
New values of refitted parameters specifying electromagnetic proton form factors in Borkowski's parameterization, see \cref{eq:G_E_Borkowski,eq:G_M_Borkowski,eq:Geff}. The parameters are expressed in units of GeV$^{2}$.
%The values of parameters specifying the new Borkowski's et al.~electromagnetic proton form factors $G_{\text{E}}^{\text{BN}}(t)$ and $G_{\text{M}}^{\text{BN}}(t)/\mu_p$ taken from \cite{Arrington2007}; here the parameters have been expressed in units of GeV$^{2}$.
}  
\end{minipage}
\end{table} 
%%%%%%%%%%%%%%%%%%%%

The effective electromagnetic form factor squared
\begin{equation}
G_{e\!f\!f}^2 (t) = \frac{1}{1 + \tau} \left[ G_{\text{E}}^2(t) +  \tau \; G_{\text{M}}^2(t) \right] \;,
\label{eq:Geff}
\end{equation}
appearing in \cref{eq:ros1} has been introduced in \cite{Block1996} for analysis of elastic pp scattering as the term in \cref{eq:ros1} proportional to $\tan^2 \left(\frac{\theta}{2}\right)$ can be neglected in linear $\alpha$ approximation (one-photon exchange) \cite{Arrington2007}.
One may define effective electric form factor squared as 
\begin{equation}
G_{\text{E},e\!f\!f}^2 (t) = \frac{1}{1 + \tau}  G_{\text{E}}^2(t) 
\label{eq:G_E_eff}
\end{equation}
and effective magnetic form factor as
\begin{equation}
G_{\text{M},e\!f\!f}^2 (t) = \frac{\tau}{1 + \tau} \; G_{\text{M}}^2(t)  \, .
\label{eq:G_M_eff}
\end{equation}

\begin{figure}
\center
%\includegraphics[width=0.6\textwidth,keepaspectratio]{ff_xix_ff_eff.pdf}
%\begin{minipage}[t]{0.60\textwidth}
\begin{minipage}[t]{1.\columnwidth}
\includegraphics[width=1.\textwidth,keepaspectratio]{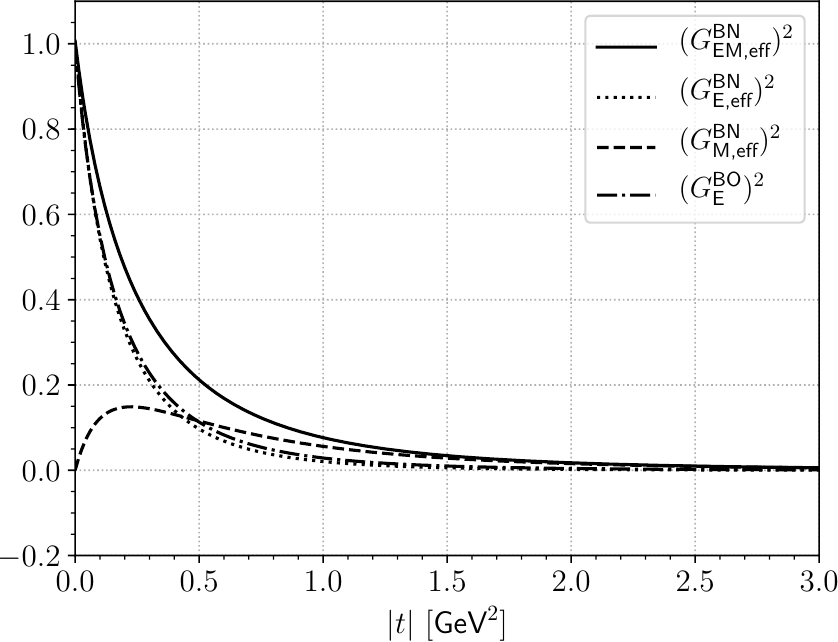}
\caption{\label{fig:ff_eff}
Effective form factors corresponding to $G_{\text{E}}^{\text{BN}}(t)$ and $G_{\text{M}}^{\text{BN}}(t)$ (see \cref{eq:Geff,eq:G_E_eff,eq:G_M_eff}) and compared to $(G_{\text{E}}^{\text{BO}}(t))^2$.}
\end{minipage}
\end{figure}

The graphs of the effective electric form factor $G_{\text{E},e\!f\!f}^2(t)$, the effective magnetic form factor $G_{\text{M},e\!f\!f}^2(t)$ and effective electromagnetic form factor $G_{e\!f\!f}^2(t)$ corresponding to the $G_{\text{E}}^{\text{BN}}(t)$ and $G_{\text{M}}^{\text{BN}}(t)$ (i.e., Borkowski's parameterization with the newly determined values of free parameters) are shown in \cref{fig:ff_eff}. For the comparison also the electric form factor $(G_{\text{E}}^{\text{BO}}(t))^2$ used in \cite{Kundrat1994_unpolarized} is shown. 

\Cref{fig:ff_eff} shows that the $t$-dependence of the effective electromagnetic form factor $G_{e\!f\!f}^2 (t)$ in \cref{eq:Geff} is different from that one appearing in original Borkowski's et al.~parameterization \cref{eq:G_E_Borkowski} which has been used in analysis of experimental elastic pp data in \cite{Kundrat1994_unpolarized}. One may ask what may be the difference in the result if also magnetic form factor is included. In next section it will be, therefore, shown how to generalize the approach in \cite{Kundrat1994_unpolarized} to take into account either the effective electric or the effective electromagnetic form factor in the eikonal model description of elastic pp collisions.

%%%%%%%%%%%%%%
%%%%%%%%%%%%

%\FloatBarrier

%%%%%%%%%%%%%%%%%%%%%%%%%%%%%%%%%%%%%%%%%%%%%%%%%%%%%%%%%%%%%%%%%%%%%%
\section{\label{sec:eikonal}Eikonal model description of Coulomb and hadron interference}
%%%%%%%%%%%%%%%%%%%%%%%%%%%%%%%%%%%%%%%%%%%%%%%%%%%%%%%%%%%%%%%%%%%%%%
%%%%%%%%%%%%%%%%%%%%%%%%%%%
\subsection{\label{sec:ampl_cn}Eikonal complete amplitude with effective electromagnetic form factors}

Instead of the limited approach of WY (see \cref{sec:introduction}) it is necessary to give the preference to a more suitable eikonal approach concerning description of Coulomb-hadronic interference, based on impact parameter representation which has been proved to be mathematically consistent and valid at any $s$ and $t$ \cite{Islam1976}. In the eikonal model the complete elastic scattering amplitude \ampl{C+N} has been introduced as the function of common eikonal being equal to the sum of individual (Coulomb and hadronic) eikonals \cite{Franco1966,Franco1973}. This approach has been used by Cahn \cite{Cahn1982} who has rederived the West and Yennie simplified formula~\refp{eq:simplifiedWY} using several approximations similar to the ones used by WY. %an approximation the concept of an angularly hadronic amplitude.

However, the eikonal model approach can be used in a more general way as it has been shown in \cite{Kundrat1994_unpolarized}. The complete elastic scattering amplitude in this approach may be written as %\cite{Kundrat1994_unpolarized} 
\begin{equation}
%\ampl{C+N} = \pm \frac{\alpha s}{t} G_1(t)G_2(t)  + \ampl{N}[1 \mp \text{i} \alpha \bar{G}(s,t)],
\ampl{C+N} = \pm \frac{\alpha s}{t} G^2_{\text{eff}}(t) + \ampl{N}[1 \mp \text{i} \alpha \bar{G}(s,t)],
\label{eq:kl1}
\end{equation}
where
\begin{equation}
\begin{split}
\bar{G}(s,t) = \int\limits_{t_{\text{min}}}^0 \text{d}t'
%&\left\{ \ln \left( \frac{t'}{t} \right) \frac{\text{d}}{\text{d}t'} \left[{G_1(t')G_2(t')}\right] \right.\\
&\left\{ \ln \left( \frac{t'}{t} \right) \frac{\text{d}}{\text{d}t'} \left[{ G^2_{\text{eff}}(t') }\right] \right.\\
&- \left. \frac{1}{2\pi} \left[ \frac{F^{\text{N}}(s,t')}{\ampl{N}} - 1 \right]
I(t,t')\right\} 
%+ \ln \left( \frac{t}{t_{\text{min}}} \right){G_1(t_{\text{min}}) G_2(t_{\text{min}})}
,
\end{split}
\label{eq:kl2}
\end{equation}
%\begin{equation} 
%\begin{split}
%  G(s,t) =& \int\limits^{0}_{t_{\text{min}}} \text{d}t' \left \{\ln \left (\frac{t'}{t} \right ) \frac{\text{d}}{\text{d}t'}[f_1(t')f_2(t')] \right.  \\
%&-\left.\frac{1}{2\pi}\left[ \frac{F^{\text{N}}(s,t')}{F^{\text{N}}(s,t)}-1\right ]I(t,t') \right \}
%  \label{eq:KLampG} 
%\end{split}
%\end{equation}
%%
and
\begin{equation}
I(t,t')=\int\limits_0^{2\pi}\text{d}{\Phi''}
%\frac{G_1(t'')G_2(t'')}{t''};
\frac{G^2_{\text{eff}}(t'')}{t''};
\label{eq:KLampI}
\end{equation}
%%%
%\begin{equation}
%I(t,t') =\; \int\limits_0^{2\pi}\text{d}{\Phi''}\;
%\frac{G_{e\!f\!f}^2(t'')}{t''}\;, \;\;
%t''=t+t'+2\sqrt{tt'}\cos{\Phi''}\;.
%\label{eq:kl4}
%\end{equation}
%%%
%\begin{equation}
%  I(t,t') = \int\limits_{0}^{2\pi} \text{d} \Phi
%  ''\frac{f_{1}(t'')f_{2}(t'')}{t''}.
%  \label{eq:KLampI}
%\end{equation}
%%
%here $G_1(t)$ and $G_2(t)$ are form factors 
here $G^2_{\text{eff}}$ is effective form factor squared given by~\refp{eq:Geff}
reflecting the electromagnetic structure of colliding charged hadrons %, $\alpha = 1/137.036$ is the fine structure constant 
and $t'' = t+t'+2\sqrt{tt'}\cos{\Phi''}$. The lowest value of $t$ is limited by kinematical limit $t_{\text{min}}= -s+4m^2$ where $m$ is rest mass of hadron in the case of elastic hadron-hadron scattering.

The upper (lower) sign in \cref{eq:kl1} corresponds to the scattering of particles with the same (opposite) electric charges. %The last term in \cref{eq:kl2}

Comparing the $t$-dependence of the complete eikonal scattering amplitude given by \cref{eq:kl1} with the standardly used complete WY scattering amplitude~\refp{eq:simplifiedWY} one may see the substantial difference between these two approaches. Instead of calculating the relative phase between the Coulomb and elastic hadron components the shape of the whole complete elastic amplitude has been derived in the eikonal model approach. More detailed analysis \cite{Kundrat2000,Kundrat2002} shows then that the function $\bar{G}(s,t)$ represents the convolution between the Coulomb and hadronic amplitudes, which is in general a complex function.

At difference to the previous approaches one complete amplitude $\ampl{C+N}$ describes the influence of both the Coulomb and elastic hadron collisions at any finite $s$ in the whole interval of $t\in\langle t_{\text{min}},0 \rangle$ up to the terms linear in $\alpha$. Formulas \refp{eq:kl1}, \refp{eq:kl2} and \refp{eq:KLampI} may be used in two ways: either for establishing the elastic hadronic amplitude $\ampl{N}$ from the analysis of measured corresponding differential cross section data provided the hadronic amplitude is conveniently parametrized as it has been done in \cite{Kundrat1994_unpolarized}. Or for a consistent inclusion of the influence of Coulomb scattering if the elastic hadronic amplitude is phenomenologically established as it has been done in \cite{Kaspar2011} in the case of predictions of pp elastic differential cross sections at the LHC.

The use of electromagnetic form factors reflects the influence of both the electric and magnetic charge structures of colliding nucleons.
Only the electric form factors given by \cref{eq:G_E_Borkowski} have been used originally in \cite{Kundrat1994_unpolarized} to calculate \ampl{C+N} according to \cref{eq:kl1} for analysis of experimental data. It has enabled to include in the elastic scattering the influence of electric space structure of colliding protons. Such an approach can be generalized by taking into account also the influence of the proton magnetic form factor, i.e., the interaction of magnetic moment of the proton with Coulomb field of the other colliding proton. 

The influence of the magnetic form factors in the case of elastic pp scattering at high energies have been theoretically studied by Block \cite{Block1996,Block2006}. However, this approach has been based on the application of standard WY complete elastic amplitude containing originally only the dipole electric proton form factors given by \cref{eq:G_E_dipol} which have been replaced by effective electromagnetic form factor \refp{eq:Geff} containing also dipole magnetic form factor~\refp{eq:G_M_dipol}. Such an approach, however, contains many limitations and deficiencies as it has been discussed in \cref{sec:introduction}. 

Unlike the approach of WY (see \cref{sec:introduction}) the electromagnetic form factors form the part of Coulomb amplitude from the very beginning in the eikonal model. Due to the integration over all kinematically allowed region of $t'$ in \cref{eq:kl2} the $t'$-dependence of effective electromagnetic form factors should describe the charge distributions in the largest possible interval of momentum transfers $t'$. For some suitable $t$-dependent parameterizations of electromagnetic proton form factor the integral $I(t,t')$ may be analytically calculated (see \cref{sec:integral_I}) which helps in numerical calculations in application of the eikonal model to experimental data. The elaborated approach then enables to study either the influence of individual effective electric or magnetic form factor or the common influence of both of them.

%%%%%%%%%%%%%%%%%%%%%%%%%%%%%%%%%%%%%%%%%%%%%%%%%%%%%%%%%%%%%%%%%%%
\subsection{\label{sec:integral_I}Analytical expression of integral $I(t,t')$}
%%%%%%%%%%%%%%%%%%%%%%%%%%%%%%%%%%%%%%%%%%%%%%%%%%%%%%%%%%%%%%%%%%%

It has been mentioned in \cref{sec:ampl_cn} that the integral involving the electromagnetic proton form factors~\refp{eq:KLampI} may be calculated analytically for conveniently parameterized form factors. It is sufficient to integrate only over a finite region of momentum transfers in formula \refp{eq:kl1} since the whole integral is multiplied by the elastic hadronic amplitude $\ampl{N}$ the modulus of which decreases very strongly at high $|t|$ (see \cref{fig:dsdt_data_pp53gev_pp8000gev}). 
%approximately like $|t|^{-4}$ - see, e.g., \cite{Donnachie1992}. 
The used limited integration region of momentum transfers allows us to use some simpler formulas for the ep form factors enabling us much simpler analytical calculation. 

%The analytical calculation of the integral $I(t,t')$ given by \cref{eq:KLampI} containing Borkowski's et al.~parameterizations of electric proton form factors $G_{\text{E}}^{\text{B}}(t)$ entering into formula \refp{eq:kl1} has been already described in \cite{Kundrat1994_unpolarized}. Such analytical calculations may be generalized to take into account the effective electromagnetic form factor given by \cref{eq:Geff} and Borkowski's et al.~parameterizations \refp{eq:G_E_Borkowski} and \refp{eq:G_M_Borkowski} of $G_{\text{E}}$ and $G_{\text{M}}$. 

In \cite{Kundrat1994_unpolarized} the integral $I(t,t')$ %given by \cref{eq:KLampI} 
was analytically calculated only for \emph{electric} form factor parameterized according to \refp{eq:G_E_Borkowski}. The same analytical formulas for this integral have been also used in \cite{Kaspar2011}. The integral may be analytically calculated also for \emph{effective electromagnetic} form factor given by \cref{eq:Geff} if the corresponding electric and magnetic form factors are given by \cref{eq:G_E_Borkowski,eq:G_M_Borkowski}. Due to the fact that both the effective form factors have more complicated $t$-dependences the corresponding formulas will be also a little bit more complicated than that ones in \cite{Kundrat1994_unpolarized}.

The analytical calculation of the new form of the integral $I(t,t')$ in \cref{eq:KLampI} has been calculated with the program Mathematica \cite{Wolfram1991} and equals to the sum of two contributions coming from the electric and magnetic form factors which contain now some kinematical factors ($r_p=-\tau/t=1/(4m^2)$)   
\newcommand{\Pelc}[1]{P^{\text{E}}_{#1}}
\newcommand{\Pmag}[1]{P^{\text{M}}_{#1}}
\begin{equation}
\begin{split}
I(t,t')\; =\;- &\left[ \sum_{j,k=1}^4 g_j^{\text{E}} g_k^{\text{E}} \; 
W_{jk}^{\text{E}}(t,t')\;I_{jk}^{\text{E}}(t,t') \right. \\
\; &+ \left. \; r_p {\mu_p}^2 \sum_{m,n=1}^4 g_m^{\text{M}} g_n^{\text{M}} \; 
W_{mn}^{\text{M}}(t,t')\;I_{mn}^{\text{M}}(t,t') \right].
\end{split}
\label{eq:I}
\end{equation}
The contribution of electric form factor in this equation is given as follows. For $j \ne k$ it holds
\begin{equation}
\begin{split}
I_{jk}^{\text{E}}(t,t') = 
2 \pi &\left[\frac{(U-1)^3}{\sqrt{U}(U-R)(U-\Pelc{j})(U-\Pelc{k})} \right. \\
          +& \frac{(R-1)^3}{\sqrt{R}(R-U)(R-\Pelc{j})(R-\Pelc{k})}  \\
          +& \frac{(\Pelc{j}-1)^3}{\sqrt{\Pelc{j}}(\Pelc{j}-U)(\Pelc{j}-R)(\Pelc{j}-\Pelc{k})} \\
          +& \left. \frac{(\Pelc{k}-1)^3}{\sqrt{\Pelc{k}}(\Pelc{k}-U)(\Pelc{k}-R)(\Pelc{k}-\Pelc{j})} \right]\; , 
\label{eq:Ielc_jk}
\end{split}
\end{equation}
while for $j=k$ one has
\begin{equation}
\begin{split}
I_{jj}^{\text{E}}(t,t') = 
2 \pi &\left[\frac{(U-1)^3}{\sqrt{U}(U-R) (U-\Pelc{j})^2} \right.\\
&+ \frac{(R-1)^3}{\sqrt{R}(R-U) (R-\Pelc{j})^2}  \\
&+  \frac{(\Pelc{j}-1)^2}{2 (U-\Pelc{j})^2 (R - \Pelc{j})^2 (\Pelc{j})^{3/2}} \\
& \;\;\; \left[U\left(R+5 R \Pelc{j}-3 \Pelc{j}(\Pelc{j}+1)\right) + \right. \\
& \;\;\; \left. \Pelc{j}\left(-3 R (\Pelc{j}+1) + \Pelc{j} (5+\Pelc{j})\right)\right]
                \Bigg]\; .
\label{eq:Ielc_jj}
\end{split}
\end{equation}
The quantities $U$, $R$ and $\Pelc{j}$ are the functions of $t$ and $t'$ variables defined as
\begin{align}
U        =& \frac{(\sqrt{-t}+\sqrt{-t'})^2}{(\sqrt{-t}-\sqrt{-t'})^2}, \label{eq:Uelc}\\
R        =& \frac{1 + r_p(\sqrt{-t} + \sqrt{-t'})^2}{1 + r_p (\sqrt{-t}-\sqrt{-t'})^2}, \label{eq:Relc} \\
\Pelc{j} =& \frac{w_j^{\text{E}} + (\sqrt {-t} + \sqrt{-t'})^2}{w_j^{\text{E}} + (\sqrt {-t} - \sqrt{-t'})^2}.
\label{eq:Pjelc}
\end{align}
Similarly the quantity $W_{jk}^{\text{E}}$ is also the function of $t$ and $t'$ variables and equals
\begin{equation}
\begin{split}
W_{jk}^{\text{E}}(t, t') =  
&\left[ [w_j^{\text{E}} + (\sqrt{-t}-\sqrt{-t'})^2]
[w_k^{\text{E}} + (\sqrt{-t}-\sqrt{-t'})^2] \right.\\
&\left. [\sqrt{-t} - \sqrt{-t'}]^2   
[1 + r_p(\sqrt{-t} - \sqrt{-t'})^2]  \right]^{-1}\; .
\label{eq:Welc}
\end{split}
\end{equation}
The contribution of magnetic form factor is represented by the second term in \cref{eq:I}. The integral for $m \ne n$ equals to
\begin{equation}
\begin{split}
I_{mn}^{\text{M}}(t,t') = 
2 \pi &\left[ \frac{(\Pmag{m}-1)^2}{\sqrt{\Pmag{m}}(\Pmag{m}-R)(\Pmag{m}-\Pmag{n})} \right.\\
&+ \frac{(R-1)^2}{\sqrt{R}(R-\Pmag{m})(R-\Pmag{n})} \\
&+ \left. \frac{(\Pmag{n} - 1)^2}{\sqrt{\Pmag{n}}(\Pmag{n} - R)(\Pmag{n} - \Pmag{m})} \right]
\end{split}
\label{eq:Imag_mn}
\end{equation}
and for $m=n$ it equals
\begin{equation}
\begin{split}
I_{mm}^{\text{M}}(t,t')  =&  
2 \pi \left[\frac{(R-1)^2}{\sqrt{R}(R-\Pmag{m})^2} \right. \\
&+ \left. \frac{(\Pmag{m}-1)\left[\Pmag{m}(\Pmag{m}+3)-R(3\Pmag{m}+1)\right]}{2 \; (\Pmag{m})^{3/2}(R-\Pmag{m})^2}\right]\; .
\end{split}
\label{eq:Imag_mm}
\end{equation}
The quantities $\Pmag{m}$ and $W_{mn}^{\text{M}}$ are the functions of $t$ and $t'$ variables and equal
\begin{equation}
\Pmag{m} = \frac{w_m^{\text{M}}+(\sqrt{-t}+\sqrt{-t'})^2}{{w_m^{\text{M}}}+(\sqrt{-t}-\sqrt{-t'})^2}
\label{eq:Pmag}
\end{equation}
and
\begin{equation}
\begin{split}
W_{mn}^{\text{M}} =  &\left[ [w_m^{\text{M}}+(\sqrt{-t}-\sqrt{-t'})^2]
[w_n^{\text{M}}+(\sqrt{-t}-\sqrt{-t'})^2] \right. \\ 
&\;\left. [1 + r_p(\sqrt{-t} - \sqrt{-t'})^2]  \right]^{-1}\; .
\end{split}
\label{eq:Wmag}
\end{equation}

Then the complete elastic scattering amplitude in the eikonal model describing the common influence of Coulomb and hadron scattering in one-photon exchange approach which is valid up to the terms linear in $\alpha$ is generally given by \cref{eq:kl1,eq:kl2,eq:KLampI} with the quantity $I(t,t')$ given by \cref{eq:I,eq:Ielc_jk,eq:Ielc_jj,eq:Uelc,eq:Relc,eq:Pjelc,eq:Welc,eq:Imag_mn,eq:Imag_mm,eq:Pmag,eq:Wmag}. 
%This form of the complete elastic scattering amplitude, enabling to study influence of different form factors, will be used for the analysis of pp elastic scattering data in \cref{sec:data_analysis}. 
This newly derived form of the complete elastic scattering amplitude, enabling to study influence of different form factors, will be used for the analysis of pp elastic scattering data at given energy at all measured values of $t$ in a consistent way in \cref{sec:data_analysis}.

%%%%%%%%%%%%%%%%%%%%%%%%%%%%%%%%%%%%%%%%%%%%%%%%%%%%%%%%%%%%%%%%%%
\section{\label{sec:ampl_n}Elastic hadronic amplitude in many contemporary models of elastic scattering}
%%%%%%%%%%%%%%%%%%%%%%%%%%%%%%%%%%%%%%%%%%%%%%%%%%%%%%%%%%%%%%%%%%

For the description of hadron interactions, mainly in the case of deep inelastic scattering processes, quantum chromodynamics has been commonly made use of. However, in the case of elastic and other diffractive processes there has not been any significant progress in spite of enormous effort having been produced. The point is that the perturbative methods, being principally involved in QCD descriptions of hard processes, may be hardly applied to in the case of soft diffractive processes, see the introduction.

This has been especially the case of elastic hadronic amplitude describing the scattering of charged nucleons where differential cross section data have been obtained with relatively large statistics. The observed dip-bump (or shoulder) structure of high-energy data has been usually described with the help of a complex hadronic amplitude \ampl{N} having the dominant imaginary part in a broader region of lower $|t|$ and vanishing in the region of diffractive minimum. The real part (very small in the region of low deviations) has been introduced to obtain a non-zero value at the diffractive minimum.

This currently accepted dominance of the imaginary part of the hadron elastic amplitude has seemed to be supported by the theorems derived at asymptotic energies and has been introduced on the basis of some \emph{a priori} assumptions (being accepted by most physicists) \cite{Hove1963_252,Hove1963_76,Logunov1963,Logunov1965,Cornille1972_plb40,Cornille1972_npb48,Martin1973,Auberson1971}. However, it has been shown \cite{KL1989,KL1996,Kundrat1985,Kundrat1997} that the experimental data, e.g., for pp and $\bar{\text{p}}$p elastic hadron scattering at the ISR energies, have behaved according to these theorems at most only in a very narrow interval of $t$ close to $t=0$ where the dominance of imaginary part may exist while fundamental deviations may appear in a greater interval. Consequently, the application of the mentioned assumptions to elastic hadron scattering at present energies in a broad interval of momentum transfers can be hardly justified.

The mentioned standard properties of hadronic amplitude might seem, of course, to be justified for the authors of the first papers analyzing the elastic pp scattering at the ISR energies \cite{Miettinen1975,Groot1973,Amaldi1976_385,Amaldi1980,Castaldi1985}; consequently, they obtained the central profile function of elastic hadron scattering $\PROF{el}(s,b)$ in the impact parameter space, represented by a Gaussian function narrower than that obtained for inelastic one. All consequences have been denoted as reliable results, even if the colliding protons have had to behave as transparent objects in elastic collisions.

Similar amplitude characteristics have been used as a starting point of many analyses concerning the elastic pp and $\bar{\text{p}}$p scattering at different energies; see, e.g., \cite{Carvalho1997,Carvalho2005,Menon2005,Menon2007,Avila2006,Avila2007,Avila2008,Silva2007,Ferreira1997,Ferreira1998,Campos2010,Fagundes2011_epjc71,Fagundes2011_ijmpa26,Kohara2014,Arriola2017,Albacete2017,Broniowski2018,Alkin2018,Goncalves2019,Martynov2019,Csorgo2019}. In \cite{Franca1979} the elastic hadronic amplitude has been taken as smoothly energy dependent and purely imaginary. Then the imaginary part has been parameterized by a sum of $n$ ($n \leq 5$) differently weighted exponentials in $t$:
\begin{equation}
\Im \ampl{N} \; = \; \Im F^{\text{N}}(s,0) \sum_{j=1}^{n} {\alpha_j(s) \e^{- \beta_j(s) |t|}}.
\label{eq:br1}
\end{equation}
The role of the real part has been admitted only as a small partial fraction of corresponding imaginary part, i.e., the number of its contributing terms has been smaller than $n$ (as, e.g., in \cite{Menon2005,Menon2007,Silva2007}); or specified with the help of derivative dispersion relations as in \cite{Carvalho2005,Avila2006,Avila2007,Avila2008,Campos2010}. Also additional linear logarithmic $t$-dependencies of all quantities $\beta_j(s)$ and quadratic logarithmic $t$-dependencies of all quantities $\alpha_j(s)$ have been introduced in order to better reproduce the corresponding differential cross section. Similar behavior of elastic hadronic amplitude has been also used in papers \cite{Ferreira1997,Ferreira1998} where  the model of stochastic vacuum to the pp and $\bar{\text{p}}$p elastic scattering has been applied to. The individual free parameters specifying the quantities $\alpha_j(s)$ and $\beta_j(s)$ have been determined by fitting measured differential elastic cross section.

However, as the FB transform of \ampl{N} (see \cref{eq:hel_standard,eq:imp7}) is additive and as it holds (see, e.g., formula (6.631.4) in \cite{Gradshteyn1980})
\begin{equation}
\int\limits_{0}^{\infty}\! \sqrt {-t}\; \text{d} \sqrt{-t} \;
 \e^{-\beta_j(s)|t|} J_0(b \sqrt{-t}) =
\frac{1}{2 \beta_j(s)} \e^{- b^2/(4 \beta_j(s))},
\label{eq:br2}
\end{equation}
final elastic impact parameter profile $\PROF{el}(s,b)$ is given by superposition of different central Gaussian functions with the maximum at $b=0$; their shapes being chosen as central from the very beginning. It is already the choice of the parameterization of \ampl{N} which predetermined the result independently of actual values of the free parameters.  

Similar weak $t$-dependence of hadronic phase $\phase$ in quite broad interval of lower $|t|$ values and imaginary part of \ampl{N} being equal to zero in the dip region has been used in majority of contemporary published papers practically without any deeper reasoning - see, e.g., \cite{Shoshi2002,Ryskin2007} and discussion of some other phenomenological models in \cite{Kaspar2011} (see also fig.~14 in \cite{totem2015}). It means that in all cases the elastic collisions have been taken as central from the very beginning.

%Published phenomenological models of elastic pp scattering at high energies have typically elastic hadron amplitude with dominant imaginary part vanishing in a neighbourhood of diffractive minimum and leading to a central image of elastic hadron pp scattering (see, e.g., \cite{Kaspar2011} and fig.~14 in \cite{totem2015}). 
Similar conclusions can be obtained with the "QCD-inspired" model of Block et al.~\cite{Block2015}. Their eikonal model has been based on the idea that the interactions between hadrons are described in terms of interactions between their constituents: quarks, gluons and with allowance for soft interactions at small values of $|t|$. As to each sort of interaction corresponds one type of eikonal which is given by the product of corresponding constituent cross section with overlap function being defined in terms of the relevant distributions in the proton (by their convolutions). Assuming that the hadron matter distributions are similar as the distributions of their electric charge then all the distributions of the hadronic components have been central which finely have led to central distributions of elastic hadron scattering. It is possible to expect that if the peripheral character of individual hadronic components is allowed then the resulting hadron interaction might be peripheral, too.

The existence of minimum (dip) in the differential cross section observed in all elastic pp collisions at high energies (see, e.g., \cite{Carter1986}) does not require zero value of the imaginary part of the amplitude; \emph{only the sum of the squares of both the real and imaginary parts should be minimal in this region}. The mentioned requirement that the imaginary part should vanish in this region represents much stronger and more limiting condition that the theory and experiment require.

It has not been respected at all, either, that a very different behavior of pp collisions may be derived with the help of a non-dominant imaginary part in a broad region of momentum transfers as it has been shown already earlier in 1981 \cite{Kundrat1981}. In such a case a peripheral behavior of elastic processes may be derived. It has been shown then in \cite{KL1989,KL1996,Kundrat1987,Kundrat1990,Kundrat1992,Kundrat1994_unpolarized} that one may obtain a peripheral picture of elastic hadron scattering for pp collisions at the ISR energy of 52.8~GeV and for $\bar{\text{p}}$p scattering at the energy of 541~GeV if the hadronic phase $\phase$ changes rather rapidly (see the second term in \cref{eq:msel}). In peripheral case the imaginary part of the amplitude goes to zero at value of $|t| \sim 0.1$ GeV$^2$ at 52.8~GeV and 8 TeV, as it will be shown in \cref{sec:data_analysis}. It means that the imaginary part of the elastic hadronic amplitude may be dominant only in a very narrow region of momentum transfers near the forward direction; the given behavior of the hadronic phase $\phase$ being still in a full agreement with the assertions of the mentioned asymptotic theorems.

To interpret elastic hadron collisions as peripheral has gained significant support in the analysis of elastic hadron processes between light nuclei. Franco and Yin \cite{Franco1985,Franco1986} studied the elastic scattering of $\alpha$ particles on various targets \mbox{($^1H$, $^2H$, $^3He$, $^4He$)}. They tried to reproduce the momentum transfer distribution of elastic collisions of two objects composed of individual nucleons using the Glauber model approach \cite{Glauber1967,Glauber1959}. As an input they used the 'elementary' nucleon - nucleon elastic scattering amplitude, assuming (in the first approximation) to be the same for all possible combinations of nucleons involved in the scattering. The data were selected in order to have practically the same energy per one nucleon. They obtained an agreement with experimental data in all considered types of scattering if they introduced the strongly $t$-dependent elementary elastic hadronic phase of the form $\phase = \frac{\Tilde{\gamma} \;t}{2} +\text{const}$ with $|\Tilde{\gamma}| > 10$~GeV$^{-2}$. Such a simple $t$-dependence of the phase together with a purely exponential $t$-dependence of the corresponding modulus was chosen in order to perform analytically all the needed multiple integrals involved in the Glauber model approach. Their elementary nucleon - nucleon elastic scattering amplitude had imaginary part vanishing at $|t| \le 0.1$~GeV$^2$, which corresponded to the result obtained in \cite{Kundrat1981,KL1989,KL1996}. The technique similar to the Glauber approximation has been also used by Franco \cite{Franco1973} in re-deriving the WY integral formula for the relative phase~\refp{eq:phaseWY} appearing in the Coulomb-hadronic interference.

%If the spins of elastically colliding hadrons are not taken into account then according to \cref{eq:difamp_gen} the elastic hadron differential cross section $\frac{\text{d}\sigma}{\text{d}t}$ is determined only by the square of the modulus $\modulus{N}$ of elastic hadronic amplitude. As its phase $\phase$ does not enter into this equation a conveniently parameterized $t$-dependence of the modulus alone can be used for fitting the experimental differential cross section data.

If one assumes that the measured elastic differential cross section is given by hadronic interaction (Coulomb effects neglected), i.e., $\frac{\text{d}\sigma}{\text{d}t}$ = $\frac{\text{d}\sigma^{\text{N}}}{\text{d}t}$, then according to \cref{eq:difamp_gen} the measured differential cross section is determined only by the square of the modulus $\modulus{N}$ of elastic hadronic amplitude; phase $\phase$ does not enter into the calculations. A conveniently parameterized $t$-dependence of the modulus alone can be used for fitting the experimental data.

On the other hand for the determination of both the real and imaginary parts of elastic hadronic amplitude the knowledge of its modulus is not sufficient; the behavior of $t$-dependent phase $\phase$ should be known, too. Performing the FB transform of both of these parts the behavior of all the profiles in the impact parameter space may be determined, see appendix~\ref{sec:ampl_b}. Thus the $t$-dependence of the phase $\phase$ specifies the behavior of elastic hadron scattering in the impact parameter space.

If in fitting procedure of some arbitrarily chosen parameterizations of both the imaginary and real parts have been used (see, e.g., \cite{Carvalho1997,Carvalho2005,Menon2005,Menon2007,Avila2006,Avila2007,Avila2008,Ferreira1997,Ferreira1998,Campos2010,Fagundes2011_ijmpa26,Arriola2017,Albacete2017,Broniowski2018,Alkin2018,Goncalves2019,Martynov2019,Csorgo2019}) then the dominance of the imaginary part of elastic hadronic amplitude in a much broader region of momentum transfers then needed has been often implicitly incorporated; it has led to the central image of elastic hadron collisions. For example, the hadronic phase has been strongly limited in the approach introduced in \cite{Silva2007,Fagundes2011_epjc71} where the parametrization ($a_i$, $b_i$, $c_j$ and $d_j$ being energy dependent free parameters)
\begin{equation}
\begin{split}
\ampl{N} \; \sim & \;    \left\{\left[\frac{\rho \CS[etype={tot,N}]}{4\pi} - \sum_{i=2}^{m}a_i \right] \e^{-b_1|t|} + \sum_{i=2}^{m} a_i \e^{-b_i|t|} \right\} \\
                 & \; + i\left\{\left[\frac{\CS[etype={tot,N}]}{4\pi} - \sum_{j=2}^{n}c_j \right] \e^{-d_1|t|} + \sum_{j=2}^{n} c_j \e^{-d_j|t|} \right\}
\end{split}
\label{eq:ampl_n_silva}
\end{equation}
has been suggested. The parameterization may look quite general, however, it has been constrained as the values of $\rho$ and $\CS[etype={tot,N}]$ have been taken inconsistently from the simplified WY approach which is strictly assuming purely exponential modulus \modulus{N} and constant phase $\phase$ at all kinematically allowed values of $t$, see \cref{sec:introduction}. The hadronic amplitude given by \cref{eq:ampl_n_silva} has been fitted to measured data with the help of \cref{eq:difamp_gen} (i.e., without considering any Coulomb effect) which means that the $t$-dependence of its phase has not been constrained by data at all (as mentioned above) and had to be chosen differently. Such data analysis can be, therefore, hardly denoted as unconstrained and model-independent. %The dominance of the imaginary part of elastic hadronic amplitude in quite broad interval of $t$ had been implicitly assumed which has led to the central character of elastic collisions as shown in \cite{Fagundes2011_epjc71}. 
The description of elastic scattering data on the basis of the parameterization \refp{eq:ampl_n_silva} performed in \cite{Fagundes2011_epjc71} has lead to central character of elastic collisions; the $t$-dependence of the corresponding hadronic phase has not been, unfortunately, discussed at all.

If the real and imaginary parts of elastic hadronic amplitude are parameterized with the help of some free parameters as, e.g., in \cref{eq:ampl_n_silva} then only the free parameters of the corresponding modulus may be determined by fitting them to experimental data (if no Coulomb-hadronic interference is taken into account). In such a case the corresponding $t$-dependence of hadronic phase is established if set of the free parameters of the phase is a subset of the free parameters specifying the modulus - or some other constrains need to be introduced. In the former case the $t$-dependence of the phase is strongly constrained from the very beginning by the \emph{choice} of the parameterizations of the real and imaginary parts. In the later case the additional constrains should be always clearly mentioned in the corresponding papers. If a model (parameterization) fits data it is not a proof that all assumptions and consequences of the model are true. In both the cases one should, therefore, analyze the meaning and consequences of all the used assumptions.

In analysis of experimental data more general parameterizations of both the modulus $\modulus{N}$ and of the phase $\phase$ should be, therefore, allowed and preferred than those used in the mentioned models above and leading to centrality of elastic collisions. More attention should be devoted mainly to the determination (choice) of $t$-dependence of the hadronic phase and discussion of corresponding characteristics of collisions in dependence on impact parameter. 
The $t$-shape of hadronic amplitude may be also constrained by the well known requirements and theorems derived within the framework of axiomatic field theory (see, e.g., review \cite{Eden1971}). One may then ask whether it is possible to describe data under a given set of assumptions.
%It is evident that the chosen parameterizations of both the modulus and of the phase of elastic hadron scattering amplitude should fulfill the well known requirements and theorems derived within the framework of axiomatic field theory (see, e.g., review \cite{Eden1971}).
The results obtained under different assumptions should be then compared and discussed. Such an analysis of experimental data will be performed in next sections.

%%%%%%%%%%%%%%%%%%%%%%%%%%%%%%%%%%%%%%%%%%%%%%%%%%%%%%%%%%%%%
\section{\label{sec:data_analysis}Analysis of elastic pp scattering data}

%%%%%%%%%%%%%%%%%%%%%%%%%%%%%%%%%%%%%%%%%%%%%%%%%%%%%%%%%%%%%
\subsection{Fitting procedure}

It has been shown in \cref{sec:ff_elcmag} that the recent analyses of both electric and magnetic proton form factors showed some deviations from standardly used dipole formulas. One may see in \cref{fig:ff_eff} that the effective electromagnetic form factor has quite different values than the widely used electric one for analysis of pp experimental data represented by measured elastic differential cross section. Details and valuable comments concerning measurement of elastic pp scattering data at high energies may be found in chapter 1 and 2 in \cite{Prochazka2018_phd_thesis} and in papers quoted there. It is clear that the inclusion of magnetic form factor might have an impact also on the results of analysis of elastic pp scattering data at high energies. 

This was one of the reasons why we have performed new analysis of pp elastic scattering data (at the ISR~energy of 52.8~GeV and the LHC energy of 8~TeV) with the help of the eikonal model (see \cref{sec:eikonal}) similarly as it has been done in \cite{Kundrat1994_unpolarized} but now with the help of \emph{effective electric} form factor~\refp{eq:G_E_eff} and \emph{effective electromagnetic} form factor~\refp{eq:Geff}. Form factors $G_{\text{E}}^{\text{BN}}(t)$ and $G_{\text{M}}^{\text{BN}}(t)$ (i.e., Borkowski's et al.~parameterizations~\refp{eq:G_E_Borkowski} and \refp{eq:G_M_Borkowski} specified by parameters taken from \cref{tab:Bork_new_par_values}) have been used for this purpose. However, impact of the choice of form factor on determination of \ampl{N} and corresponding hadronic quantities has been found to be very small or negligible (see \cite{Prochazka2018_phd_thesis} for numerical details). In the following results corresponding to only effective electromagnetic form factors will be, therefore, shown.

Main unknown function in the eikonal interference formula given by \cref{eq:kl1,eq:kl2} is elastic hadronic amplitude \ampl{N}. It may be, therefore, parametrized and one may try to determine it from experimental data under different assumptions (constrains).
%For both the form factors a description of pp elastic collision data based on elastic hadronic amplitude corresponding to the widely used assumptions (namely to the dominance of its imaginary part in quite broad region around $t=0$) and leading to the central behavior of elastic collisions has been compared to the alternative peripheral description having different $t$-dependence of the phase. In the peripheral case some new possibilities corresponding to different values of elastic root-mean-square $\sqrt{\meanb[n=2,etype=el]}$ have been newly performed.
Conveniently parameterized elastic hadronic amplitude \ampl{N} has been fitted to the measured pp elastic differential cross section at given energy in broad interval of $t$ values including both peak at very low values of $|t|$ and dip-bump structure at higher values of $|t|$ with the help of \cref{eq:difamp_gen} and complete amplitude \ampl{C+N} given by \cref{eq:kl1,eq:kl2}. 

The integral $I(t,t')$ in \cref{eq:kl2} has been analytically calculated using \cref{eq:I,eq:Ielc_jk,eq:Ielc_jj,eq:Uelc,eq:Relc,eq:Pjelc,eq:Welc} and parameters from \cref{tab:Bork_new_par_values}, and compared to corresponding numerical integration \refp{eq:KLampI}. 
The result of numerical integration of the complete amplitude performed for the measured  $t$  values should be finite. The formulas (both analytical and numerical) for the integral $I(t,t')$ contain singularity at $t=t'$.
%as can be seen from, e.g., \cref{eq:Pjelc}. 
However, this singularity is canceled by the factor $\left[ \frac{F^{\text{N}}(s,t')}{\ampl{N}} - 1 \right]$ in \cref{eq:kl2}. The integration in \cref{eq:kl2} needs to be treated with care at $t'$ equal to $t$ and $0$ (for both numerically and analytically calculated function $I(t,t')$).\footnote{Similar difficulties may be identified also in the relative phase $\alpha\phi(s,t)$ of WY given by \cref{eq:phaseWY}.} The integrals in the regions $(t_{\text{min}}, t -\epsilon)$ and $(t+\epsilon, -\epsilon)$ where $\epsilon$ is small and positive, should be convergent. Also the integrand leading to the different improper integrals should be convergent in all the regions. Using the theorems valid for the values of improper integrals (see, e.g., \cite{Fichtengolc1962}) their values can be easily calculated in the limiting case when $\epsilon \rightarrow 0$. 

All the fits of experimental data under different assumptions have been performed by minimizing the corresponding $\chi^2$ function with the help of program MINUIT~\cite{James1975}. Quoted uncertainties of free parameters have been estimated with the help of HESSE procedure in MINUIT. Uncertainty $\sigma_{f}$ of a function $f$ depending on free parameters $x_i$ has been calculated with the help of
\begin{equation}
\sigma_{f} = \sqrt{\sum_i \left(\frac{\partial f(x)}{\partial x_i}\right)^2 (\sigma_{x_i})^2}
\label{eq:sigma_f}
\end{equation}
where $\sigma_{x_i}$ stands for uncertainty of the $i$-th parameter.

For each performed fit of data corresponding consistency tests were performed. Values of integrated cross section $\CS[etype={tot}]$ determined with the help of \cref{eq:optical_theorem} and $\CS[etype={el}]$ determined with the help of the first equation in \cref{eq:cs_el_integ_b} were compared to values obtained on the basis of integrating corresponding b-dependent profile functions $\PROF{X}(b)$ with the help of \cref{eq:integ_cs_prof}. Similarly, mean impact parameters $\sqrt{\meanb[n=2,etype=X]}$ calculated on the basis of \cref{eq:msel,eq:mstot,eq:msinel} have been compared to values obtained with the help of \cref{eq:integ_meanb}. In all cases good agreement was obtained. 

\subsection{\label{sec:analysis_parameterization}Parameterization of hadronic amplitude}

The analysis of experimental data with the help of \cref{eq:kl1,eq:kl2} requires a convenient parameterization of the complex elastic hadronic amplitude \ampl{N}, e.g., its modulus and its phase. The modulus may be parameterized very generally as
\begin{equation}
\begin{split}
  \modulus{N} \; = \;\;\; &(a_1 + a_2t)\e^{b_1t+b_2t^2+b_3t^3} \; \\
%                      +  &(c_1 + c_2t)\e^{d_1t+d_2t^2+d_3t^3} \;.
                      +  &(c_1 + c_2t)\e^{d_1t} \;.
\end{split}
  \label{eq:ampl_n_mod_param} 
\end{equation}
The integration limit $t_{\text{min}}$ in \cref{eq:kl2} is lesser than the lower limit of measured data. The $t$-dependence of the modulus parametrization given by \cref{eq:ampl_n_mod_param} can be used for extrapolation to the higher values of $|t|$ if the modulus is strongly decreasing with increasing value of $|t|$ in this region. The care needs to be devoted to the allowed fitted values of free parameters specifying the modulus in order to guarantee its vanishing when $|t|$ tends to infinity as required by validity of corresponding dispersion relations.
%The $t$-dependence of the  modulus parametrization given by \cref{eq:ampl_n_mod_param} can be continuously extrapolated to the lower values of $t$ in agreement with \cite{Donnachie1996}.

In \cite{Bailly1987} (and even earlier in \cite{Ayres1976,Bogoljubski1981}) the following parameterization of hadronic phase has been used
\begin{equation} 
  \phase = \arctan{\frac{\rho_0}{1-\frac{t}{t_{\text{dip}}}}}
  \label{eq:ampl_n_stdphase}
\end{equation}
where $t_{\text{dip}}$ is the position of the dip in data and $\rho_0 = \rho(t\!=\!0)$. This parameterization a priori restricts allowed $t$-dependences and reproduces the widely assumed dominance of the imaginary part of \ampl{N} and vanishing of the imaginary part at $t=t_{\text{dip}}$, see \cref{sec:ampl_n}. The parameterization of the phase has been used later in several other analyses of experimental data, including \cite{Kundrat1994_unpolarized} where it has been shown that it leads to central behavior of elastic collisions in impact parameter space. However, this parameterization of the phase is not analytic in $t$; not only due to the pole at $t\!=\!t_{\text{dip}}$ but also due to the fact that the complex function $\arctan(z)$ is not analytic in the points $z=\pm i$ ($i$ being complex unit) \cite{Evgrafov1978}. Moreover, the parameterization of the phase \refp{eq:ampl_n_stdphase} cannot fulfill conclusion of Martin's asymptotic theorem \cite{Martin1997} (derived in 1997) requiring, under certain assumptions, the real part of elastic hadronic amplitude to change sign at some low value of $|t|$. 

In \cite{Kundrat1994_unpolarized} different and more general parameterization of hadronic phase has been also used for analysis of experimental data ($t_0=-1$~GeV$^2$)
\begin{equation} 
  \phase = \zeta_0 + \zeta_1 \left(\frac{t}{t_0}\right)^\kappa \e^{\nu t} %+\zeta_2 \abs{\frac{t}{t_0}}^\lambda
\label{eq:ampl_n_zeta_gen}
\end{equation}
enabling to include a fast increase of $\phase$ with increasing $|t|$ and, consequently, a peripheral behavior of elastic hadronic scattering. 
%In order to preserve the analytic properties of the phase $\phase$ at all physical $t$ values, the phase $\zeta^{\text{N}}(s,t\!=\!0)$ is defined by the limit $\zeta^{\text{N}}(s,t\rightarrow0^{-})$. 

Natural question arises under which conditions both the parameterizations of the modulus given by \cref{eq:ampl_n_mod_param} and of the phase \cref{eq:ampl_n_zeta_gen} represent analytic function of complex variable $t$ as standardly required (see \cite{Epstein1969,Martin1969} and review \cite{Eden1971}). The parameterized modulus in \cref{eq:ampl_n_mod_param} forms the real analytic function and its analytic properties are preserved also in the case of complex variable $t$. However, the same statement is not valid for the phase introduced by \cref{eq:ampl_n_zeta_gen} due to the power $t^\kappa$ in it. For complex variable $t$ this power is analytic at the point $t\!=\!0$ only if parameter $\kappa$ is positive integer. Thus the analyticity of the elastic hadron phase for complex $t$ is guaranteed only for positive integer values of parameter $\kappa$. As the complex goniometric functions $\sin(x)$ and $\cos(x)$ are analytic for complex variable $x$, both the real and imaginary parts of elastic hadron amplitude are analytic, too. It means that the positive integer value of parameter $\kappa$ guarantees that the parameterization of elastic hadronic amplitude given by \cref{eq:ampl_n_mod_param,eq:ampl_n_zeta_gen} is analytic for complex $t$. In \cite{Kundrat1994_unpolarized} this parameter was fitted. \footnote{All the free parameters specifying the elastic hadronic amplitude may be $s$-dependent; their values may be determined at energies corresponding to available measured elastic differential cross sections. This allows to introduce parameterization of these free parameters and, therefore, explicit $s$-dependence of hadronic amplitude which may be required, e.g., to be analytic in $s$. However, taking into account that measured data correspond to discrete energies, it is very common in literature to first apply a model of elastic scattering (hadronic amplitude) at one energy to test some ideas and evaluate the results before trying to establish explicit $s$-dependence.}

The parameterization \refp{eq:ampl_n_zeta_gen} is much more general and flexible than \refp{eq:ampl_n_stdphase} as it may reproduce very broad class of $t$-dependent phases which may all fit measured data and lead to either central or peripheral behavior depending on the values of the free parameters - according to additional assumptions constraining \ampl{N} as it will be explicitly shown in \cref{sec:pp53gev_analysis_central,sec:pp53gev_analysis_peripheral} (at 52.8~GeV) and \cref{sec:pp8000gev_results} (at 8~TeV). %The general parameterizations given by \cref{eq:ampl_n_mod_param,eq:ampl_n_zeta_gen} also allow determination of \ampl{N} on the basis of experimental data which is analytic and which may fulfill the conclusion of the asymptotic theorem of Martin (if required).

When \ampl{N} is not constrained only by the measured differential cross section but also by some other constrains
%All free parameters specifying modulus and phase of elastic hadronic amplitude $\ampl{N}$ may have different values at different collision energies.  
one needs to solve in general the problem of bounded extrema of the $\chi^2$ function, i.e., of the function of the $n$ free parameters $x=(x_1, ..., x_n)$ which may be solved with the help of penalty functions technique. If at the minimum of the $\chi^2$ the values of the free parameters $x$ are limited at point $x_0$ by some condition $g(x\!=\!x_0)$ then one may add to the minimized $\chi^2$ function additional function $[g(x) - g(x\!=\!x_0)]^2 * C_p$, where $C_p$ is some conveniently chosen constant value (weight of the penalty function). In the case of several limiting conditions the resulting penalty function is given by the sum of all individual penalty functions which is added to the original $\chi^2$ during minimization. Performing the minimization procedure one can significantly influence the way how the position of the minimum can be achieved. When performing several successive minimizations one has to decrease successively the values of all the penalty constants $C_p$ in such a way that the position of the minimum is being preserved. Using this approach the added value of total penalty function $\Delta{\chi}^2$ may become finally very small compared to the value of the original $\chi^2$.

\subsection{\label{sec:pp53gev_results}Energy of 52.8~GeV}

\subsubsection{\label{sec:pp53gev_data}Data}

At the energy of 52.8~GeV experimental data in broad region of $|t| \in \langle 0.00126, 7.75 \rangle$ GeV$^2$ taken from \cite{Bystricky1980} have been used, see the data points in \cref{fig:dsdt_data_pp53gev_pp8000gev,fig:pp53gev_1b_dsdt}. %The determined hadronic quantities characterizing pp collisions are only slightly or negligibly changed when derived with the help of the proton effective electric or the effective electromagnetic form factors.

\subsubsection{\label{sec:pp53gev_analysis_central} Elastic hadronic amplitude as constrained in many contemporary models and leading to central behavior of elastic collisions}

\begin{figure*}%[!htb]
\centering
\begin{subfigure}[b]{0.46\textwidth}
\includegraphics*[width=\textwidth]{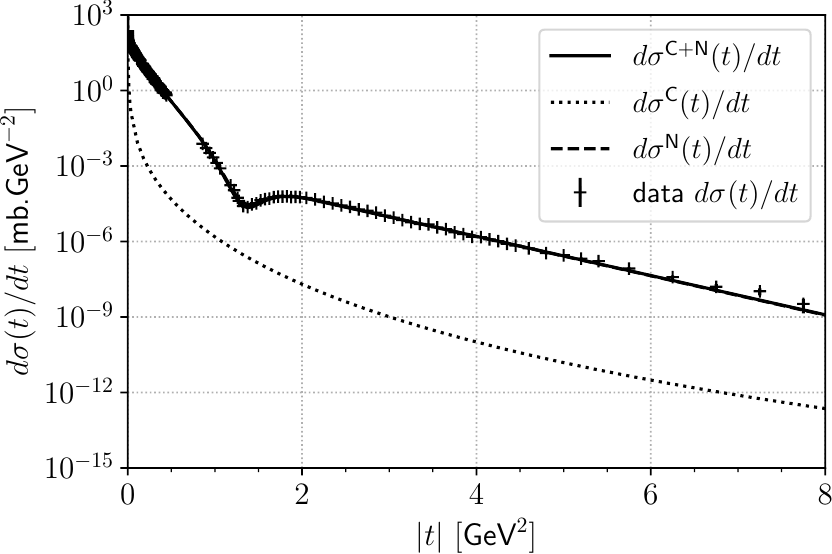}
\caption{\label{fig:pp53gev_1b_dsdt_full}full fitted $|t|$-range of measured data}
\end{subfigure}
\quad%add desired spacing between images, e. g. ~, \quad, \qquad etc.
\begin{subfigure}[b]{0.44\textwidth}
\includegraphics*[width=\textwidth]{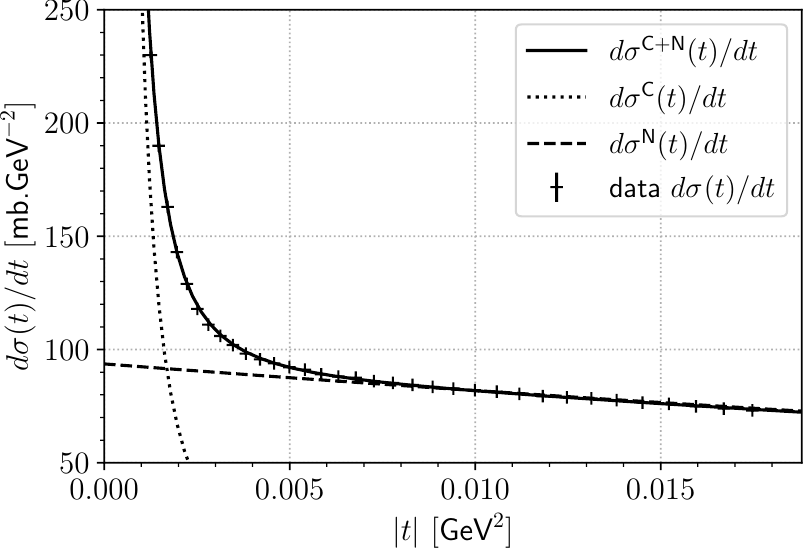}
%\vspace{0.05cm}
\caption{\label{fig:pp53gev_1b_dsdt_zoom}region of the lowest measured values of $|t|$}
\end{subfigure}
\begin{minipage}[t]{.97\textwidth}
		\caption{\label{fig:pp53gev_1b_dsdt} Eikonal model of Coulomb-hadronic interaction fitted to measured elastic pp differential cross section at energy of 52.8~GeV in the interval $|t| \in \langle 0.00126, 7.75 \rangle$ GeV$^2$ corresponding to Fit~1, i.e., central picture of elastic pp scattering. Fit~2 leading to peripheral picture of elastic scattering gives similar graphs.}
\end{minipage}
\end{figure*}

\begin{table*}%[!ht]
\centering
%\resizebox{\textwidth}{!}{
\begin{tabular}{lccccc}
\hline \hline
Particle types                  &              & pp                        & pp                        & pp                   & pp                    \\
$\sqrt{s}$                      & [GeV]        & 52.8                      & 52.8                      & 8000                 & 8000                  \\
Fit                             &              & 1                         & 2                         & 1                    & 2                     \\
Case                            &              & central                   & peripheral                & central              & peripheral            \\
\multirow{2}{*}{Form factor}    &              & effective                 & effective                 & effective            & effective             \\[-1mm]
                                &              & electromagnetic           & electromagnetic           & electromagnetic      & electromagnetic       \\
%program                         &              & pp53sfjmr3.list (11157)  & pp53wjx3.list3 (12427)    &pp8sfjrw3.list10(9475)&pp8mrapexb3.list1(9311)\\
\hline                                                                                                                                                
%$t_{\text{dip}}$                &              & -1.375                   & -                         & -0.53                & -                     \\
%$\rho_{0}$                      &              & 0.0761    $\pm$ 0.0017   & -                         &  0.123  $\pm$ 0.015  & -                     \\
$\zeta_{0}$                     &              & 0.0762     $\pm$ 0.0017   & 0.0825     $\pm$ 0.0017   & 0.121  $\pm$ 0.018   & 0.148  $\pm$ 0.016    \\
$\zeta_{1}$                     &              & -2.605                    & 1974       $\pm$ 37       & -12.02               & 281    $\pm$ 11       \\
$\kappa$                        &              & 3                         & 3                         & 2                    & 2                     \\
$\nu$                           & [GeV$^{-2}$] & 1.028                     & 8.23       $\pm$ 0.14     & 1.304                & 5.68   $\pm$ 0.20     \\
$a_{1}$                         &              & 12149.8   $\pm$ 9.2       & 12202.3    $\pm$ 9.3      & 66.58   $\pm$ 0.12   & 66.79  $\pm$ 0.11     \\
$a_{2}$                         & [GeV$^{-2}$] & 10705     $\pm$ 29        & 10767      $\pm$ 33       & 163.06  $\pm$ 0.73   & 170.39 $\pm$ 0.39     \\
$b_{1}$                         & [GeV$^{-2}$] & 5.905     $\pm$ 0.017     & 5.868      $\pm$ 0.017    & 8.291   $\pm$ 0.038  & 8.137  $\pm$ 0.026    \\
$b_{2}$                         & [GeV$^{-4}$] & 3.677     $\pm$ 0.063     & 3.445      $\pm$ 0.060    & 9.27    $\pm$ 0.23   & 7.58   $\pm$ 0.16     \\
$b_{3}$                         & [GeV$^{-6}$] & 1.678     $\pm$ 0.041     & 1.520      $\pm$ 0.038    & 14.85   $\pm$ 0.34   & 12.15  $\pm$ 0.25     \\
$c_{1}$                         &              & 58.8      $\pm$ 1.4       & 60.4       $\pm$ 1.9      & 1.57    $\pm$ 0.14   & 2.047  $\pm$ 0.067    \\
$c_{2}$                         & [GeV$^{-2}$] & -5.4e-6   $\pm$ 2.9       & -6.3e-8    $\pm$ 2.3      & -3.14   $\pm$ 0.33   & -2.46  $\pm$ 0.14     \\
$d_{1}$                         & [GeV$^{-2}$] & 0.901     $\pm$ 0.050     & 0.907      $\pm$ 0.041    & 2.75    $\pm$ 0.077  & 2.688  $\pm$ 0.019    \\
%$d_{2}$                         & [GeV$^{-4}$] & 0                        & 0                         & 0                    & 0                     \\
%$d_{3}$                         & [GeV$^{-6}$] & 0                        & 0                         & 0                    & 0                     \\
\hline                                                                                                                                                
${\chi}^2/$ndf                  &              & 345/206                   & 303/204                   & 234 / 131            & 368 / 129             \\
$\Delta{\chi}^2$                &              & 0                         & 4.0                       & 0                    & 16                    \\
\hline                                                                                                                                                
$\rho(t\!=\!0)$                 &              & 0.0763    $\pm$ 0.0017    & 0.0827     $\pm$ 0.0016   & 0.122  $\pm$ 0.018   & 0.149  $\pm$ 0.016    \\
$B(t\!=\!0)$                    & [GeV$^{-2}$] & 13.515    $\pm$ 0.035     & 13.444     $\pm$ 0.036    & 21.021 $\pm$ 0.085   & 20.829 $\pm$ 0.055    \\
$\CS[etype={tot,N}]$            & [mb]         & 42.694    $\pm$ 0.033     & 42.861     $\pm$ 0.034    & 103.44 $\pm$ 0.35    & 104.12 $\pm$ 0.31     \\ 
$\CS[etype={el,N}]$             & [mb]         & 7.469                     & 7.539                     & 27.6                 & 28.0                  \\
$\CS[etype={inel}]$             & [mb]         & 35.22                     & 35.32                     & 75.9                 & 76.1                  \\
$\CS[etype={el,N}]/\CS[etype={tot,N}]$  &      & 0.1750                    & 0.1759                    & 0.267                & 0.269                 \\
$\text{d}\sigma^{\text{N}}/                                                                                                                           
\text{d}t(t\!=\!0)$          & [mb.GeV$^{-2}$] & 93.67                     & 94.51                     & 555                  & 566                   \\
\hline                                                                                                                                                
$\sqrt{\meanb[n=2,etype=tot]} $ & [fm]         & 1.026                     & 1.023                     &  1.28                & 1.27                  \\
$\sqrt{\meanb[n=2,etype=el]}  $ & [fm]         & 0.6778                    & 1.959                     &  0.896               & 1.86                  \\
$\sqrt{\meanb[n=2,etype=inel]}$ & [fm]         & 1.085                     & 0.671                     &  1.39                & 0.970                 \\
$\PROF{tot}(b\!=\!0)$           &              & 1.29                      & 1.30                      &  2.01                & 2.04                  \\
$\PROF{el}(b\!=\!0)$            &              & 0.530                     & 0.0342                    &  0.980               & 0.205                 \\
$\PROF{inel}(b\!=\!0)$          &              & 0.762                     & 1.27                      &  1.03                & 1.84                  \\
\hline \hline
\end{tabular}\\ %}
\begin{minipage}[t]{1.\textwidth}
\caption[eikonal model, pp 52.8~GeV and 8~TeV]{\label{tab:pp53gev_pp8000gev_ff_eff} Comparison of several hadronic quantities characterizing pp elastic scattering at energy of 52.8~GeV and 8 TeV. Values of free parameters specifying elastic hadronic amplitude \ampl{N} have been obtained by fitting experimental data under different assumptions using the eikonal model approach.}
\end{minipage}
\end{table*}

\begin{figure*}
\centering
\begin{minipage}[t]{.49\textwidth}
%\begin{minipage}[t]{1.\columnwidth}
\includegraphics[width=\textwidth,keepaspectratio]{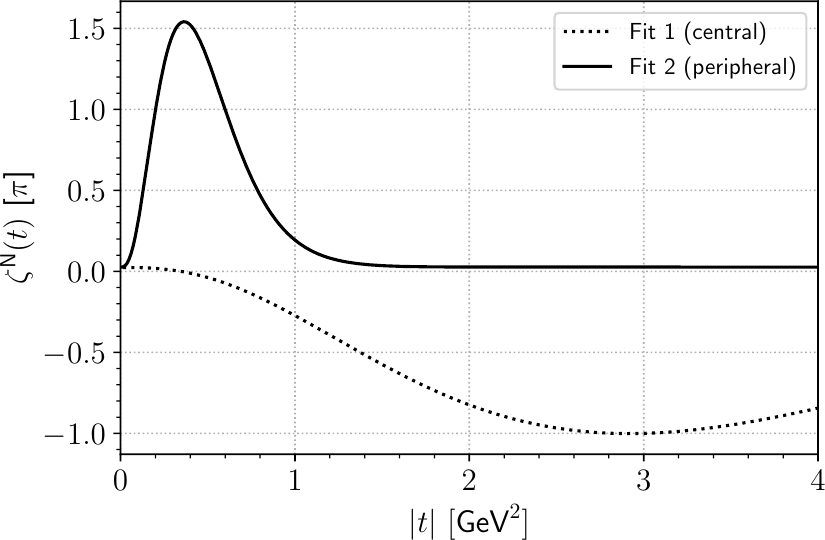}
\caption{\label{fig:pp53gev_zeta_multi}Elastic hadronic phases $\phase$ for central and peripheral pictures of elastic pp scattering (Fits 1 and 2) at energy of 52.8~GeV.}
\end{minipage}%
\hfill
\begin{minipage}[t]{.49\textwidth}
%\begin{minipage}[t]{1.\columnwidth}
  \centering
\includegraphics[width=\textwidth,keepaspectratio]{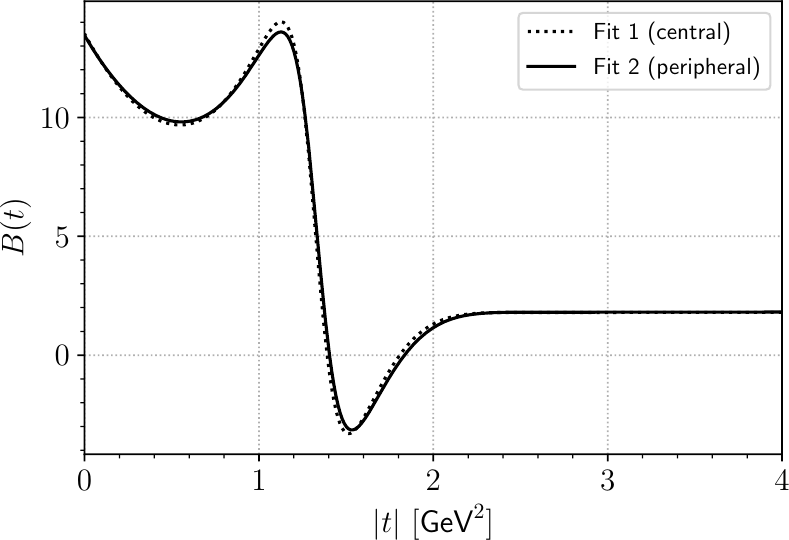}
\caption{\label{fig:pp53gev_diffractive_slope_multi}$t$-dependence of elastic hadronic diffractive slopes $B(t)$ calculated with the help of \cref{eq:slope} and corresponding to Fits~1~and~2 at energy of 52.8~GeV.}
\end{minipage}%
\end{figure*}

\begin{figure*}%[!htb]
\centering
\begin{subfigure}[b]{0.48\textwidth}
\includegraphics*[width=\textwidth]{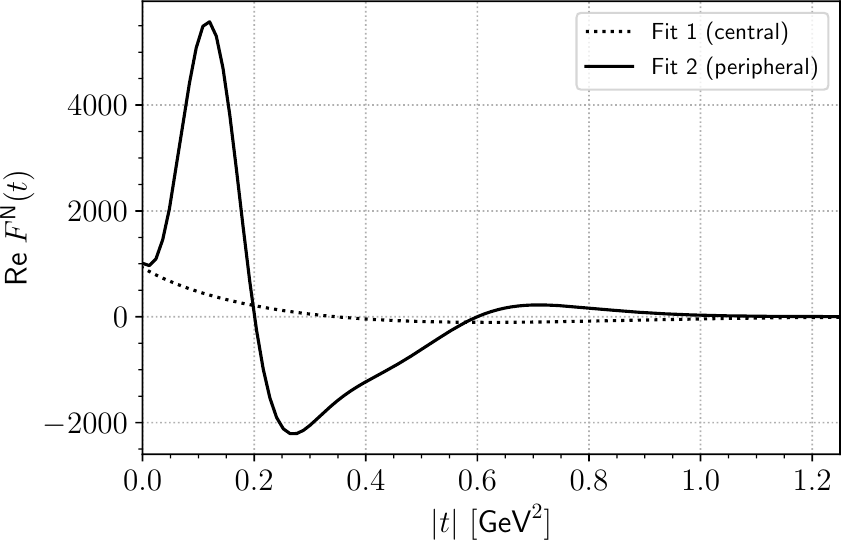}
\caption{\label{fig:pp53gev_ffeff_re} real part }
\end{subfigure}
\quad%add desired spacing between images, e. g. ~, \quad, \qquad etc.
\begin{subfigure}[b]{0.48\textwidth}
\includegraphics*[width=\textwidth]{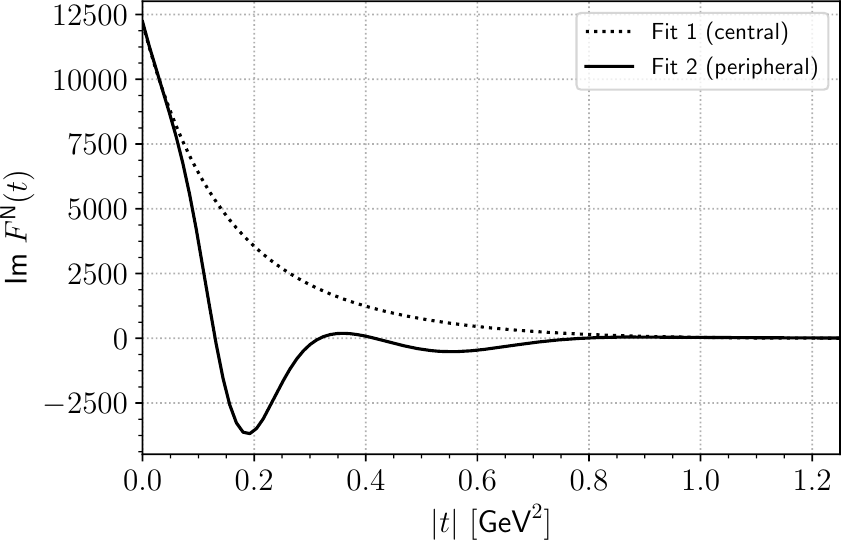}
%\vspace{0.05cm}
\caption{\label{fig:pp53gev_ffeff_im} imaginary part}
\end{subfigure}
\begin{minipage}[t]{.95\textwidth}
\caption{\label{fig:pp53gev_ffeff_reim} The real and imaginary parts of elastic hadron scattering amplitude corresponding to Fits~1~and~2 at 52.8~GeV.}
\end{minipage}
\end{figure*}

%\newpage

The first Fit~1 of data at 52.8~GeV has been performed with the help of parameterization of \ampl{N} given by \cref{eq:ampl_n_mod_param,eq:ampl_n_zeta_gen}. The parameter $\kappa=3$ has been taken to fit data at 52.8~GeV to keep analyticity of elastic hadronic amplitude at all kinematically allowed values of $t$, see \cref{sec:analysis_parameterization}. To obtain $t$-dependence of hadronic amplitude roughly corresponding to many contemporary hadronic models of elastic scattering we have required (as it is often assumed without sufficient reasoning, see \cref{sec:ampl_n} and also fig.~14 in \cite{totem2015}):
\begin{enumerate}
\item{dominance of the imaginary part of \ampl{N} at broad interval of $t$ values in the forward region}
\item{vanishing of the imaginary part of \ampl{N} at (or around) $t=t_{\text{dip}}$}
\item{change of sign of the real part of \ampl{N} at $|t|<|t_{\text{dip}}|$ (motivated by the asymptotic theorem of Martin \cite{Martin1997})}
\end{enumerate}

These constrains of hadronic amplitude may be fulfilled if the phase \phase\ given by \refp{eq:ampl_n_zeta_gen} passes through, e.g., the two following points 
\begin{align}
 [t_1&=t_{\text{dip}}, y_1=-\pi/2] \label{eq:pp53gev_ty1} \\
 [t_2&=-3 \text{ GeV}^2, y_2=-\pi] \label{eq:pp53gev_ty2} \; . 
\end{align}
In this case values of $\nu$ and $\zeta_1$ in \cref{eq:ampl_n_zeta_gen} may be calculated as follows
\begin{align}
\nu     =& \ln\left[\frac{y_2 - \zeta_0}{y_1 - \zeta_0} \left(\frac{t_1}{t_2}\right)^\kappa\right] \frac{1}{t_2-t_1} \label{eq:nu_constrain}\\
\zeta_1 =& \frac{y_1-\zeta_0}{\left(\frac{t_1}{t_0}\right)^\kappa \e^{\nu t_1}} \, . \label{eq:zeta1_constrain}
\end{align}
It means that the hadronic phase given by \cref{eq:ampl_n_zeta_gen} is strongly constrained under the given conditions and it has only one free parameter $\zeta_0$ which may be fitted to experimental data together with other free parameters specifying the modulus of \ampl{N}. 

Fit~1 has been quite straight forward (due to the fact that $t$-dependence of the ``standard'' phase is strongly constrained). Fitted values of all the free parameters specifying the elastic hadronic amplitude at 52.8~GeV are in \cref{tab:pp53gev_pp8000gev_ff_eff}. \Cref{fig:pp53gev_1b_dsdt} shows fitted elastic pp differential cross section $\dcs{\text{C+N}}$ together with corresponding Coulomb $\dcs{\text{C}}$ and hadronic $\dcs{\text{N}}$ differential cross sections. The phase $\phase$ corresponding to the amplitude is pictured in \cref{fig:pp53gev_zeta_multi} (dotted lines). The diffractive slope $B(t)$ calculated with the help of \cref{eq:slope} is shown in \cref{fig:pp53gev_diffractive_slope_multi} (dotted lines).

The corresponding elastic hadronic amplitudes for both the fits have dominant imaginary parts in the large region of $t$ around forward direction which decrease with increasing $|t|$ and vanish in the diffraction dip (as commonly assumed), see Fit~1 in \cref{fig:pp53gev_ffeff_im}. 
The corresponding real parts of \ampl{N} change sign at $|t| \approx 0.35$~GeV$^2$ as motivated by the asymptotic theorem of Martin, see Fit~1 in \cref{fig:pp53gev_ffeff_re}.

Determined values of several physically interesting quantities calculated from the fitted hadronic amplitude corresponding to Fit~1 may be found in \cref{tab:pp53gev_pp8000gev_ff_eff}. The total hadronic cross section $\CS[etype={tot,N}]$ has been calculated using the optical theorem~\refp{eq:optical_theorem}, integrated elastic hadron cross section $\CS[etype={el,N}]$ using the first equation in \cref{eq:cs_el_integ_b} and inelastic $\CS[etype=inel]$ as their difference. 

Root-mean-squares of impact parameter values $\sqrt{\meanb[n=2,etype=tot]}$, $\sqrt{\meanb[n=2,etype=el]}$, $\sqrt{\meanb[n=2,etype=inel]}$ determined with the help of \cref{eq:mstot,eq:msel,eq:msinel} (see appendix~\ref{sec:ampl_b}) may be found also in \cref{tab:pp53gev_pp8000gev_ff_eff}. It holds $\sqrt{\meanb[n=2,etype=el]} < \sqrt{\meanb[n=2,etype=inel]}$, i.e., elastic collisions according to this description should correspond in average to lower impact parameters than average impact parameter corresponding to inelastic collisions ($\sim 0.68$~fm against $\sim 1.09$~fm). Fit~1 will be, therefore, labeled as "central". This centrality of elastic collisions may be further seen from the profile functions $\PROF{X}(b)$  (X=tot, el, inel) calculated at \emph{finite} collision energy $\sqrt{s}$ as explained at the end of appendix~\ref{sec:ampl_b}, see \cref{fig:pp53gev_profiles_1b_profiles} corresponding to Fit~1. The elastic profile function $\PROF{el}(b)$ has Gaussian shape with a maximum at $b=0$. Some other $b$-dependent functions corresponding to Fit~1 and characterizing hadron collisions in $b$-space, see appendix~\ref{sec:ampl_b}, are shown in \cref{fig:pp53gev_profiles_1b_noprofiles}.

%%%%%%%%%%%%%%%%%%%%%%%%%%%%%%%%%%%%%%%%%%%%%%%%%%%%%%%%%%%%%%%%%%%

%\newpage

%\FloatBarrier

%\FloatBarrier
\subsubsection{\label{sec:pp53gev_analysis_peripheral} Possibility of peripheral behavior of elastic scattering}
%%%%%%%%%%%%

The second Fit~2 of the same differential cross section data have been performed similarly as in \cref{sec:pp53gev_analysis_central}, with the help of the same parameterization of \ampl{N} given by \cref{eq:ampl_n_mod_param,eq:ampl_n_zeta_gen} but without the additional, unjustified and widely used constrains on hadronic amplitude expressed by conditions \refp{eq:nu_constrain} and \refp{eq:zeta1_constrain} (and leading to central behavior of elastic collisions).

To obtain peripheral behavior of elastic collisions, to demonstrate this possibility, it has been required for the corresponding root-mean-square impact parameter values to hold $\sqrt{\meanb[n=2,etype=el]} > \sqrt{\meanb[n=2,etype=inel]}$ and $\PROF{el}(b)$ to have its maximum at some non-zero impact parameter $b$. However, the fit has not been unique. We have, therefore, further required value of %parameter $\zeta_1$ to be around 2000 and 
$\sqrt{\meanb[n=2,etype=el]}$ to be around 1.95~fm.
%\mbox{1.85, 1.95 and 2.05~fm} and looked for the values of all the free parameters separately in these 3 cases. 
 If all these additional conditions bounding the values of fitted free parameters have been added then unambiguous fit has been obtained. In this case it has been necessary to solve non-trivial problem of bounded extrema as explained at the end of \cref{sec:analysis_parameterization}. \Cref{tab:pp53gev_pp8000gev_ff_eff} contains the results of Fit 2. It holds $\sqrt{\meanb[n=2,etype=el]} > \sqrt{\meanb[n=2,etype=inel]}$ as required. The table contains also the final values of penalty functions $\Delta{\chi}^2$ which are small compared to the $\chi^2$ values.

Differential cross sections \dcs{\text{N}}, \dcs{\text{C}} and \dcs{\text{C+N}} corresponding to the peripheral Fit~2 are very similar to those shown in \cref{fig:pp53gev_1b_dsdt}. The diffractive slope $B(t)$ for the Fit~2 calculated with the help of \cref{eq:slope} is shown in \cref{fig:pp53gev_diffractive_slope_multi}; its $t$-dependence is very similar to diffractive slope corresponding to Fit~1 (i.e., central pictures of elastic pp scattering discussed). However, $t$-dependence of the phase $\phase$ obtained in the Fit~2 is very different from the dependence corresponding to Fit~1 already at very small values of $|t|$, see \cref{fig:pp53gev_zeta_multi}.

It may be interesting to note that the peripheral Fit~2 fulfill conclusion of Martin's asymptotic theorem \cite{Martin1997}, even if it has not been required. \Cref{fig:pp53gev_ffeff_reim} contains the $t$-dependences of fitted real and imaginary parts of elastic hadronic amplitudes corresponding to Fits~1 and 2. In the peripheral case the corresponding real part changes its sign at $|t| \approx 0.2$~GeV$^2$ and the imaginary parts at $|t| \approx 0.1$~GeV$^2$.
%In the peripheral fits the real part of the hadron amplitude changes its sign at $|t|$ values from approximately $0.19$ to $0.23$~GeV$^2$.
%The $t$-dependence of central hadronic phase \refp{eq:ampl_n_stdphase} is not analytic function in complex $t$-plane due to its pole at $t=t_\text{dip}$; therefore Martin's theorem cannot be applied. 

\begin{figure*}%[!ht]
\centering
\begin{subfigure}[t]{0.48\textwidth}
\includegraphics*[width=\textwidth]{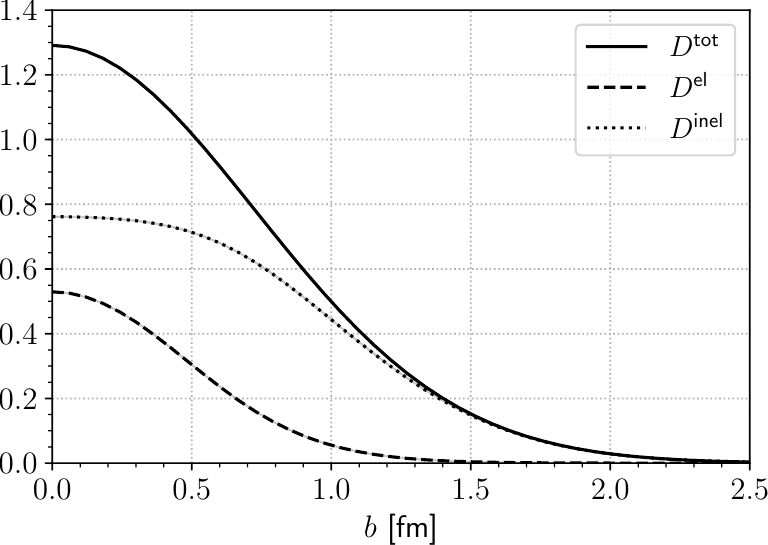}
		\caption{\label{fig:pp53gev_profiles_1b_profiles}Fit~1 (central)}
\end{subfigure}
\quad%add desired spacing between images, e. g. ~, \quad, \qquad etc.
%\begin{subfigure}[t]{0.48\textwidth}
%\includegraphics*[width=\textwidth]{model_ampl_xix_pp53gev_xix_eikonal_xix_calc_errors_xix_profiles_2b_profiles.pdf}
%\caption{\label{fig:pp53gev_profiles_2b_profiles}peripheral case - Fit~2b}
%\end{subfigure}
\begin{subfigure}[t]{0.48\textwidth}
\includegraphics*[width=\textwidth]{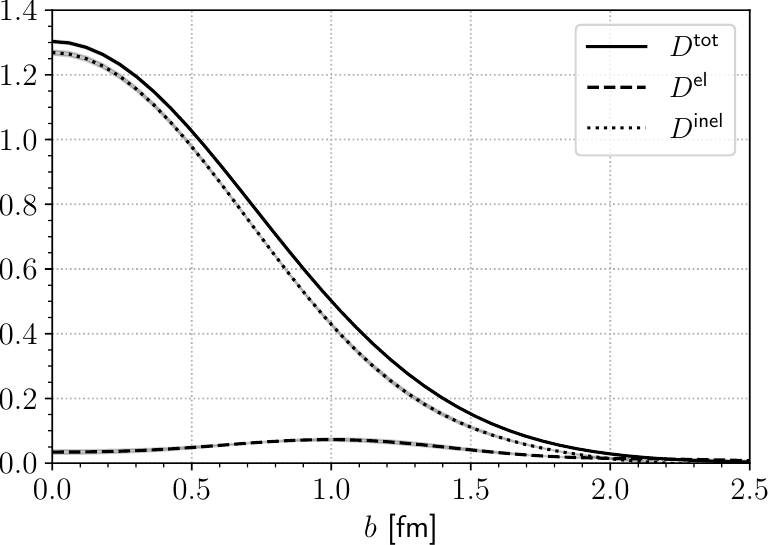}
\caption{\label{fig:pp53gev_profiles_3b_profiles}Fit~2 (peripheral)}
\end{subfigure}
%\quad%add desired spacing between images, e. g. ~, \quad, \qquad etc.
%%(or a blank line to force the subfigure onto a new line)
%\begin{subfigure}[t]{0.48\textwidth}
%\includegraphics*[width=\textwidth]{model_ampl_xix_pp53gev_xix_eikonal_xix_calc_errors_xix_profiles_4b_profiles.pdf}
%\caption{\label{fig:pp53gev_profiles_4b_profiles}peripheral case - Fit~4b}
%\end{subfigure}
\begin{minipage}[t]{.95\textwidth}
%\begin{minipage}[t]{1.\columnwidth}
\caption{\label{fig:pp53gev_profiles}Proton-proton profile functions $\PROF{}(b)$ at energy of 52.8~GeV determined on the basis of \cref{eq:unitarity2,eq:ampl_prof_el_new,eq:prof_tot_gauss_simple}. Full line corresponds to total profile function, dashed line to elastic one and dotted line to inelastic one.} %\\[1cm]} 
\end{minipage}
\end{figure*}
%\begin{figure*}%[!ht]
%\centering
%\begin{subfigure}[t]{0.48\textwidth}
%\includegraphics*[width=\textwidth]{model_ampl_xix_pp53gev_xix_eikonal_xix_calc_xix_ffelc_xix_profile_el_multi.pdf}
%\caption{\label{fig:pp53gev_profile_el_multi_ffelc}}
%\end{subfigure}
%\quad%add desired spacing between images, e. g. ~, \quad, \qquad etc.
%%(or a blank line to force the subfigure onto a new line)
%\begin{subfigure}[t]{0.48\textwidth}
%\includegraphics*[width=\textwidth]{model_ampl_xix_pp53gev_xix_eikonal_xix_calc_xix_ffeff_xix_profile_el_multi.pdf}
%\caption{\label{fig:pp53gev_profile_el_multi_ffeff}}
%\end{subfigure}
%\begin{minipage}[t]{.95\textwidth}
%%\begin{minipage}[t]{1.\columnwidth}
%\caption{\label{fig:pp53gev_profiles_el_multi}Comparison of elastic pp profile functions at energy of 52.8~GeV. Peripheral Fits~2a-4a in (\subref{fig:pp53gev_profile_el_multi_ffelc}) correspond to effective electric form factors while peripheral Fits~2b-4b in (\subref{fig:pp53gev_profile_el_multi_ffeff}) to effective electromagnetic form factors.
%}
%\end{minipage}
%\end{figure*}

\begin{figure*}%[!ht]
\centering
\begin{subfigure}[t]{0.48\textwidth}
\includegraphics*[width=\textwidth]{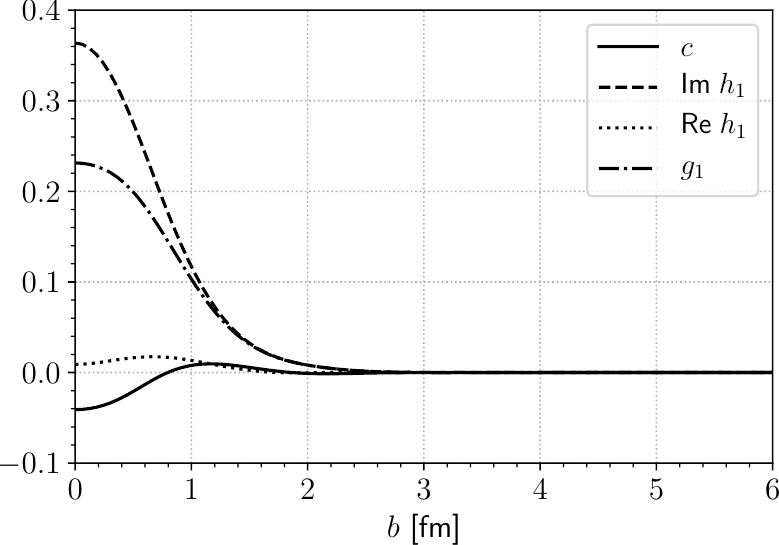}
		\caption{\label{fig:pp53gev_profiles_1b_noprofiles}Fit~1 (central)}
\end{subfigure}
\quad%add desired spacing between images, e. g. ~, \quad, \qquad etc.
%\begin{subfigure}[t]{0.48\textwidth}
%\includegraphics*[width=\textwidth]{model_ampl_xix_pp53gev_xix_eikonal_xix_calc_xix_ffeff_xix_profiles_2b_noprofiles.pdf}
%\caption{\label{fig:pp53gev_profiles_2b_noprofiles}peripheral case - Fit~2b}
%\end{subfigure}
\begin{subfigure}[t]{0.48\textwidth}
\includegraphics*[width=\textwidth]{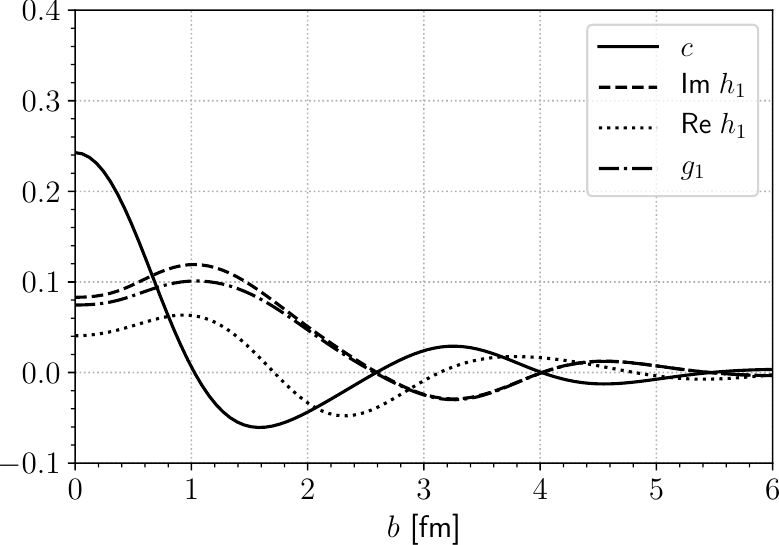}
		\caption{\label{fig:pp53gev_profiles_3b_noprofiles}Fit~2 (peripheral)}
\end{subfigure}
%\quad%add desired spacing between images, e. g. ~, \quad, \qquad etc.
%\begin{subfigure}[t]{0.48\textwidth}
%\includegraphics*[width=\textwidth]{model_ampl_xix_pp53gev_xix_eikonal_xix_calc_xix_ffeff_xix_profiles_4b_noprofiles.pdf}
%\caption{\label{fig:pp53gev_profiles_4b_noprofiles}peripheral case - Fit~4b}
%\end{subfigure}
\begin{minipage}[t]{.95\textwidth}
%\begin{minipage}[t]{1.\columnwidth}
\caption{\label{fig:pp53gev_noprofiles}Function $c(s,b)$ and several other functions characterizing pp collisions in dependence on impact parameter at the energy of 52.8~GeV corresponding to the central and peripheral fits.}
\end{minipage}
\end{figure*}

% (higher value of $\sqrt{\meanb[n=2,etype=el]}$, higher value of $|t|$).
%It may be interesting to note that the central phase \refp{eq:ampl_n_stdphase} does not fulfill Martin's theorem \cite{Martin1997} (derived later in 1997) requiring the real part of elastic hadronic amplitude to be zero at smaller value of $|t|$. The theorem is fulfilled, however, in all the discussed peripheral cases.

For the total mean impact parameter $\sqrt{\meanb[n=2,etype=tot]}$ value of \mbox{$\sim1.02$~fm} has been obtained. As to the numerically greater value 
%($\sim 1.85 \div 2.05$~fm) 
 $\sim1.96$~fm of $\sqrt{\meanb[n=2,etype=el]}$ in the peripheral case it is given by the second term in \cref{eq:msel} representing the influence of the phase; inelastic $\sqrt{\meanb[n=2,etype=inel]}$ being correspondingly lower. %Peripheral fits corresponding to values of $\sqrt{\meanb[n=2,etype=el]}$ significantly bellow $1.6$~fm or above $1.9$~fm lead to much higher values of $\chi^2$ than those shown in \cref{tab:pp53gev_ff_elc,tab:pp53gev_ff_eff}.

The profile functions $\PROF{X}(b)$ for the peripheral Fit 2 is shown in \cref{fig:pp53gev_profiles_3b_profiles}. Additional $b$-dependent functions corresponding to the Fit~2 and further characterizing hadron collisions in dependence on impact parameter are shown in \cref{fig:pp53gev_profiles_3b_noprofiles}. 
%The other peripheral Fits~2a-4a with effective electric form factors give to very similar . 
It may look like that functions $\Im h_1(b)=0$ and $g_1(b)=0$ at the same $b$-values around 2.5~fm, 4~fm and at even higher values, see \cref{fig:pp53gev_profiles_3b_noprofiles}. This would lead to violation of the unitarity given by \cref{eq:unitarity1} (if function $K(s,b)$ is neglected). The two functions are equal to zero in these regions but \emph{not} at the same value of $b$; the unitarity is conserved at all values of $b$. Given that the function $g_1(b)$ at any value of $b$ in any of the performed fits is calculated on the basis of \cref{eq:unitarity1} there is no reason to violate the unitarity.

It may be seen from \cref{tab:pp53gev_pp8000gev_ff_eff} and \cref{fig:pp53gev_profiles,fig:pp53gev_noprofiles} that even if data may be fitted in the central and peripheral cases equally well in terms of ${\chi}^2/$ndf value the corresponding behavior of proton collisions in impact parameter space is completely different. In any of the discussed peripheral cases one may obtain elastic profile function $\PROF{el}(b)$ having its maximum at some $b>0$. The non-zero function $c(s,b)$ discussed in details in appendix~\ref{sec:ampl_b} and shown in \cref{fig:pp53gev_noprofiles} in any of the peripheral cases enables to define non-oscillating and non-negative profile functions. In the central case the function $c(s,b)$ plays much less significant role.

\subsubsection{Comparison to the simplified model of WY}

The results obtained in \cref{sec:pp53gev_analysis_central,sec:pp53gev_analysis_peripheral} may be now compared to the results obtained earlier on the basis of the simplified WY formula~\refp{eq:simplifiedWY} also at the energy of 52.8~GeV. The values of quantities $\CS[etype={tot,N}]$, $\rho(t\!=\!0)$ and $B(t\!=\!0)$ in \cref{tab:pp53gev_pp8000gev_ff_eff} may be compared to similar values
%%
%\begin{align}
\begin{equation}
\begin{aligned}
\CS[etype={tot,N}] =& \; (42.38 \pm 0.27) \; \text{mb}, \\
\rho(t\!=\!0)  =& \; (0.078 \pm 0.010),   \\
B(t\!=\!0)     =& \; (13.1  \pm 0.2) \; \text{GeV}^{-2};
\label{eq:wy_parameters_amaldi1978}
\end{aligned}
\end{equation}
%\end{align}
%%
determined in \cite{Amaldi1978} (see also \cite{Amaldi1977}). However, the simplified WY complete amplitude~\refp{eq:simplifiedWY} has been applied to only in the very narrow region $|t| \in \langle 0.00126, 0.01)\;\; \text{GeV}^{-2}$, while the Fits~1 and 2 have been performed in much broader measured region of $|t| \in \langle 0.00126, 7.75 \rangle$ GeV$^2$ including also the dip-bump structure. While in~\cref{eq:simplifiedWY} it has been assumed that $\phase$ and $B(t)$ are $t$-independent these quantities are $t$-dependent in the Fits~1 and 2, see the graps in \cref{fig:pp53gev_zeta_multi,fig:pp53gev_diffractive_slope_multi}.  %\Cref{fig:pp53gev_ampl_n_fit3b_vs_WY} shows $t$-dependence of elastic hadronic phase $\phase$ and diffractive slope $B(t)$ corresponding to one of the peripheral fit (Fit~3b) and compared to the $t$-independent values~\refp{eq:wy_parameters_amaldi1978} determined on the basis of the simplified approach of WY. 

\Cref{fig:pp53gev_diffractive_slope_multi} clearly shows that diffractive slope is not constant in the analyzed region of $t$; therefore one of the assumptions used in derivation of simplified WY complete amplitude \refp{eq:simplifiedWY} is not fulfilled, see \cref{sec:WY}. It may be interesting to note that in the case of elastic hadronic amplitude in the model of WY with $t$-independent hadronic phase the real part of \ampl{N} does not change sign at any value of $t$; the conclusion of the asymptotic theorem of Martin is \emph{not} fulfilled. %, see \cref{fig:pp53gev_wy_reim_log}. 

%The simplified formula of WY~\refp{eq:simplifiedWY} for description of elastic Coulomb-hadronic scattering of two charged hadrons has been derived without taking into account value of impact parameter which influence value of scattering angle. Theoretical framework used by WY \cite{WY1968} does no provide relevant way how to determine any characteristics of pp collisions in dependence on impact parameter. One may take the hadronic amplitude specified by parameters \refp{eq:wy_parameters_amaldi1978} and calculate $b$-dependent profile functions given by \cref{eq:ampl_prof_el_new,eq:ampl_prof_tot_new,eq:ampl_prof_inel_new} (in the same way as it has been done in \cref{sec:pp53gev_analysis_central,sec:pp53gev_analysis_peripheral} for different elastic hadronic amplitudes) and obtain central behavior of elastic collisions. However, these results have very little or no value taking into account that the hadronic amplitude in the WY approach cannot describe measured differential cross section at higher values of $|t|$ and that the eikonal model framework is based on very different assumptions than those used to derive \refp{eq:simplifiedWY}.

The simplified WY approach can be hardly used for the correct analysis of experimental $\frac{\text{d} \sigma}{\text{d}t}$ data and studying $t$-dependence of elastic hadronic amplitude and corresponding $b$-dependent characteristics of hadrons on the basis of experimental data, see additional calculations in appendix~\refp{sec:wy_study_pp53gev}.

\FloatBarrier
\subsection{\label{sec:pp8000gev_results}Energy of 8~TeV}

\subsubsection{\label{sec:pp8000gev_data}Data}

\begin{figure*}%[!htb]
\centering
\begin{subfigure}[t]{0.48\textwidth}
\includegraphics*[width=\textwidth]{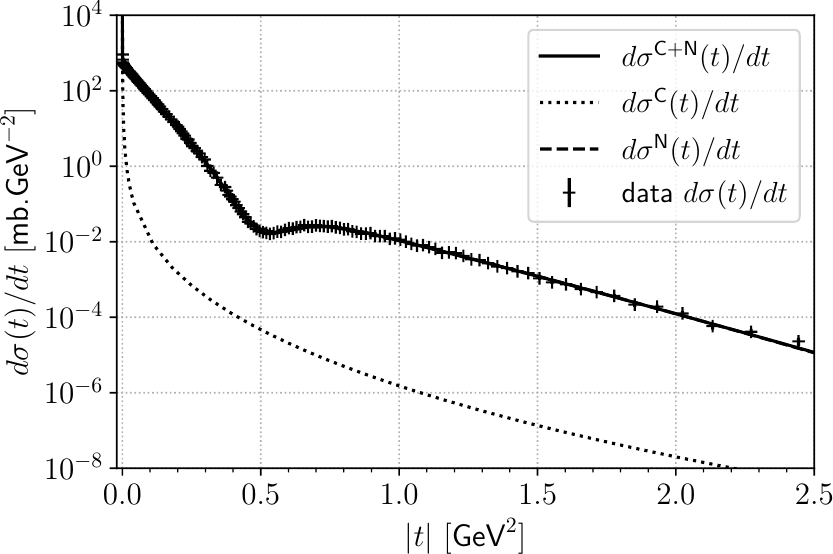}
\caption{\label{fig:pp8000gev_1b_dsdt_full}full fitted $|t|$-range of measured data}
\end{subfigure}
\quad%add desired spacing between images, e. g. ~, \quad, \qquad etc.
\begin{subfigure}[t]{0.48\textwidth}
\includegraphics*[width=\textwidth]{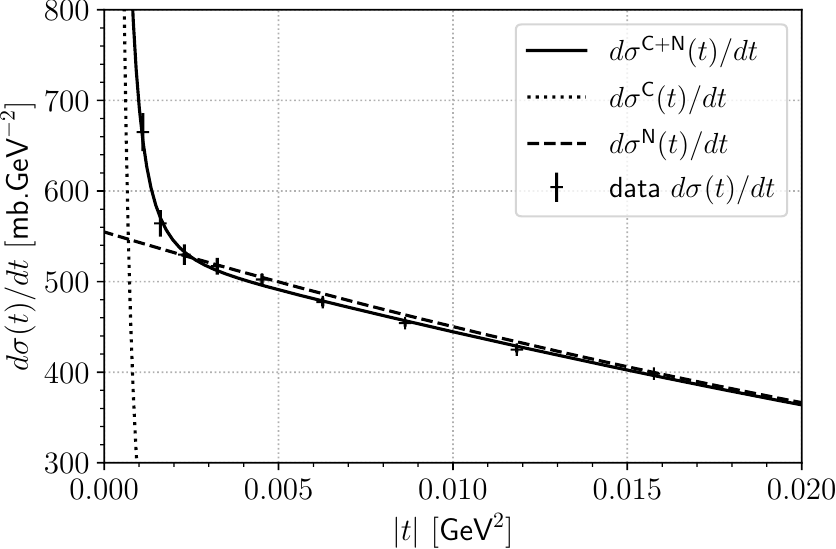}
%\vspace{0.05cm}
\caption{\label{fig:pp8000gev_1b_dsdt_zoom}region of the lowest measured values of $|t|$}
\end{subfigure}
\begin{minipage}[t]{.95\textwidth}
\caption{\label{fig:pp8000gev_1b_dsdt} Eikonal model of Coulomb-hadronic interaction fitted to measured elastic pp differential cross section at energy of 8~TeV in the interval $|t| \in \langle 0.000741, 2.5 \rangle$ GeV$^2$ corresponding to Fit~1, i.e., central picture of elastic pp scattering. Fit~2 leading to peripheral picture of elastic scattering gives similar graphs.}
\end{minipage}
\end{figure*}

Elastic pp differential cross section has been recently measured at the LHC by TOTEM at 8~TeV in the region $0.000741 \leq |t| \leq 0.2010$~GeV$^2$ \cite{totem2015} which contains Coulomb-hadronic interference region. Nearly exponential elastic pp differential cross section at the same energy has been measured by TOTEM \cite{totem9} in the region $0.027 < |t| < 0.2$~GeV$^2$. These two data sets may be combined and continuously extended by renormalized 7~TeV TOTEM data corresponding to the region $0.2 < |t| < 2.5$~GeV$^2$ \cite{totem4} which contains dip-bump structure. %Some more preliminary TOTEM data at 8 TeV corresponding to $|t|<0.027$  will be used. 
This compilation of data will be denoted as ``8 TeV data'' in the following (only statistical errors will be taken into account), see \cref{fig:pp8000gev_1b_dsdt}. %\cref{fig:data_pp53gev_pp8tev}.
The extension by the renormalized 7~TeV data is only an approximation (may be improved when measured data at 8~TeV in this region are available). This is also one of the reason why we have been more interested in overall character of an elastic collision model fitted to data rather than in discussion of ``precise'' numerical values of some quantities. Comparison of measured elastic pp differential cross section at 52.8~GeV and 8~TeV may be found in \cref{fig:dsdt_data_pp53gev_pp8000gev}.

\subsubsection{The eikonal model}

\begin{figure*}
\centering
\begin{minipage}[t]{.48\textwidth}
%\begin{minipage}[t]{1.\columnwidth}
\includegraphics[width=\textwidth,keepaspectratio]{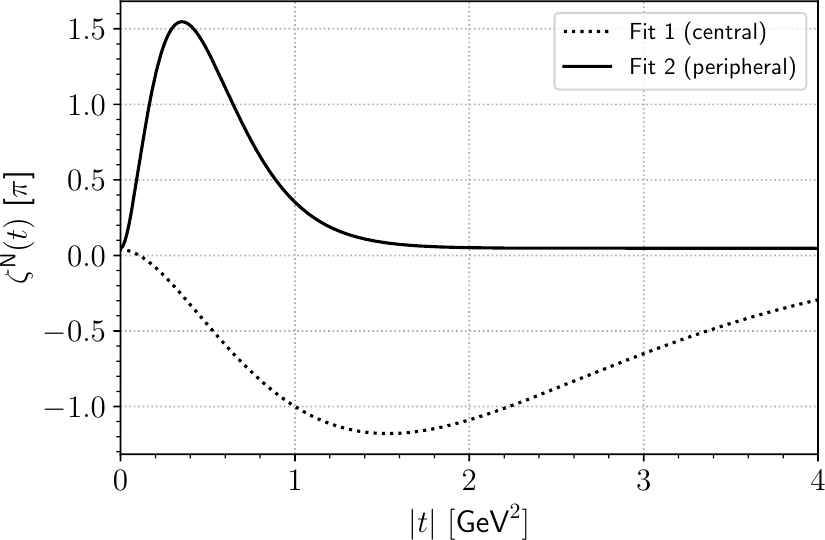}
\caption{\label{fig:pp8000gev_zeta_multi}Elastic hadronic phases $\phase$ for central and peripheral pictures of elastic pp scattering (Fits 1 and 2) at energy of 8~TeV.}
\end{minipage}%
\hfill
\begin{minipage}[t]{.48\textwidth}
%\begin{minipage}[t]{1.\columnwidth}
  \centering
\includegraphics[width=\textwidth,keepaspectratio]{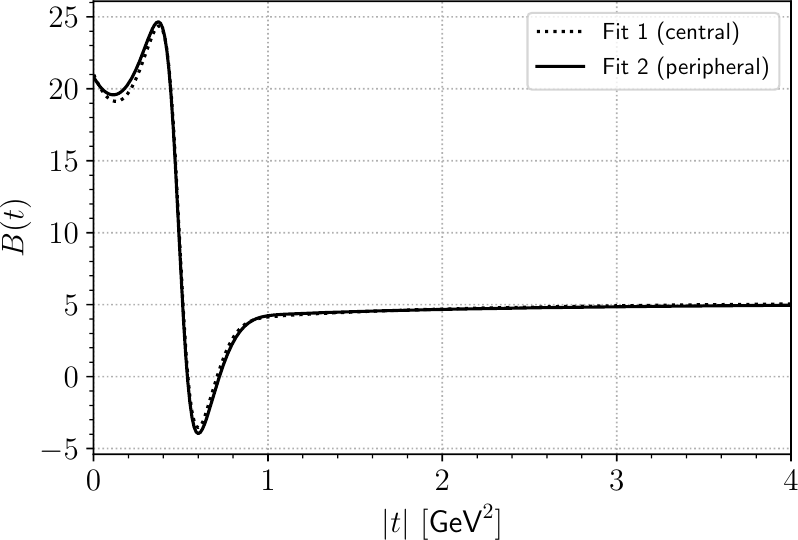}
\caption{\label{fig:pp8000gev_diffractive_slope_multi}$t$-dependence of elastic hadronic diffractive slopes $B(t)$ calculated with the help of \cref{eq:slope} and corresponding to Fits~1~and~2 at energy of 8~TeV.}
\end{minipage}%
\end{figure*}
\begin{figure*}%[!htb]
\centering
\begin{subfigure}[b]{0.46\textwidth}
\includegraphics*[width=\textwidth]{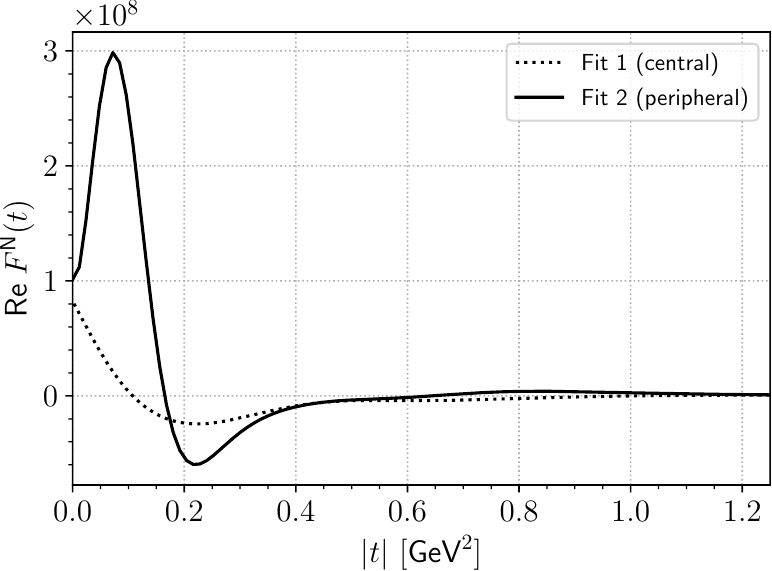}
\caption{\label{fig:pp8000gev_ffeff_re} real part}
\end{subfigure}
\quad%add desired spacing between images, e. g. ~, \quad, \qquad etc.
\begin{subfigure}[b]{0.46\textwidth}
\includegraphics*[width=\textwidth]{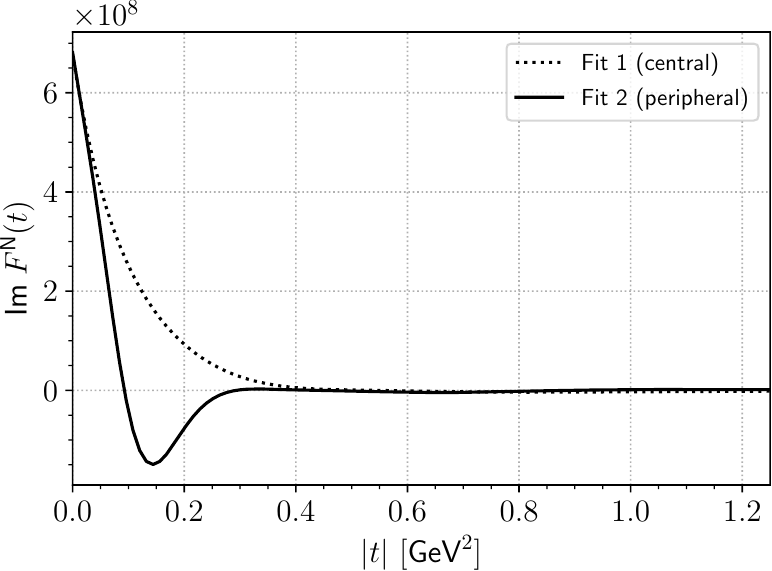}
\caption{\label{fig:pp8000gev_ffeff_im} imaginary part}
\end{subfigure}
\begin{minipage}[t]{.95\textwidth}
\vspace{-2mm}
\caption{\label{fig:pp8000gev_ffeff_reim} The real and imaginary parts of elastic hadron scattering amplitude corresponding to Fits 1 and 2 at 8~TeV.}
\end{minipage}
\end{figure*}

\begin{figure*}%[!ht]
\centering
\begin{subfigure}[t]{0.48\textwidth}
\includegraphics*[width=\textwidth]{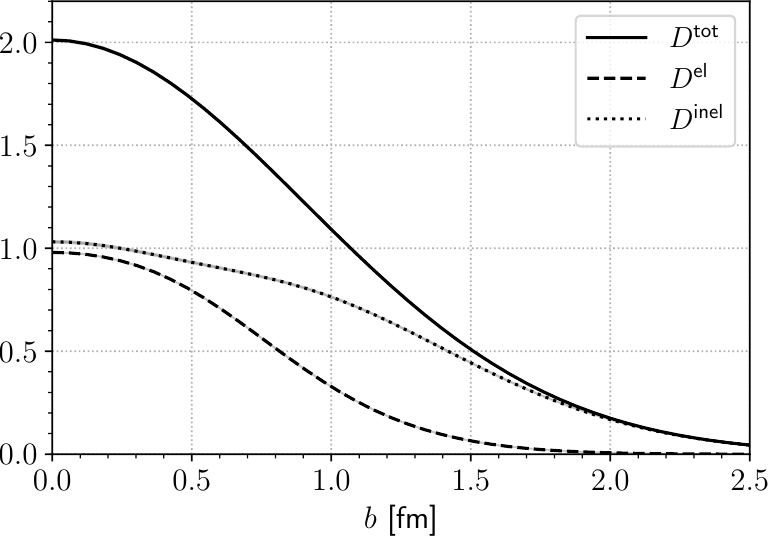}
		\caption{\label{fig:pp8000gev_profiles_1b_profiles}Fit~1 (central)}
\end{subfigure}
\quad%add desired spacing between images, e. g. ~, \quad, \qquad etc.
%\begin{subfigure}[t]{0.48\textwidth}
%\includegraphics*[width=\textwidth]{model_ampl_xix_pp8000gev_xix_eikonal_xix_calc_errors_xix_profiles_2b_profiles.pdf}
%\caption{\label{fig:pp8000gev_profiles_2b_profiles}peripheral case - Fit~2b}
%\end{subfigure}
\begin{subfigure}[t]{0.48\textwidth}
\includegraphics*[width=\textwidth]{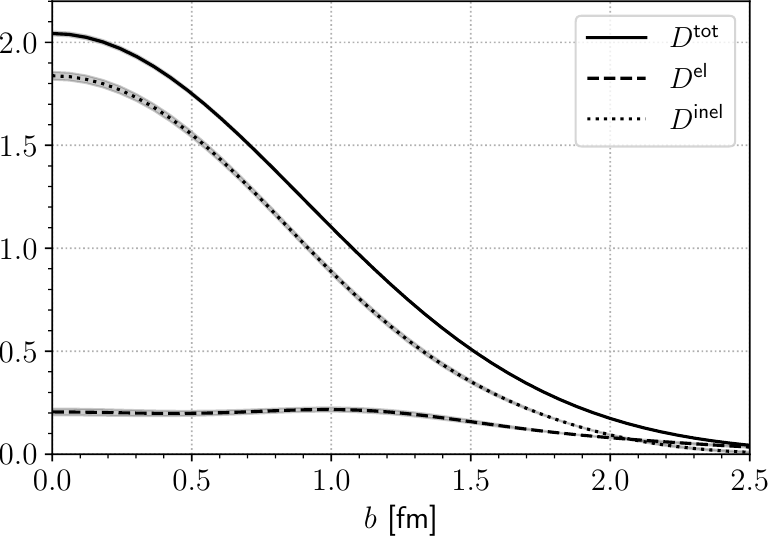}
		\caption{\label{fig:pp8000gev_profiles_3b_profiles}Fit~2 (peripheral)}
\end{subfigure}
%\quad%add desired spacing between images, e. g. ~, \quad, \qquad etc.
%%(or a blank line to force the subfigure onto a new line)
%\begin{subfigure}[t]{0.48\textwidth}
%\includegraphics*[width=\textwidth]{model_ampl_xix_pp8000gev_xix_eikonal_xix_calc_errors_xix_profiles_4b_profiles.pdf}
%\caption{\label{fig:pp8000gev_profiles_4b_profiles}peripheral case - Fit~4b}
%\end{subfigure}
\begin{minipage}[t]{.95\textwidth}
%\begin{minipage}[t]{1.\columnwidth}
\caption{\label{fig:pp8000gev_profiles}Proton-proton profile functions $\PROF{}(b)$ at energy of 8~TeV corresponding to Fits 1 and 2 and determined on the basis of \cref{eq:unitarity2,eq:ampl_prof_el_new,eq:prof_tot_gauss_simple}. Full line corresponds to total profile function, dashed line to elastic one and dotted line to inelastic one.} %\\[1cm]} 
\end{minipage}
\end{figure*}
\begin{figure*}%[!ht]
\centering
\begin{subfigure}[t]{0.48\textwidth}
\includegraphics*[width=\textwidth]{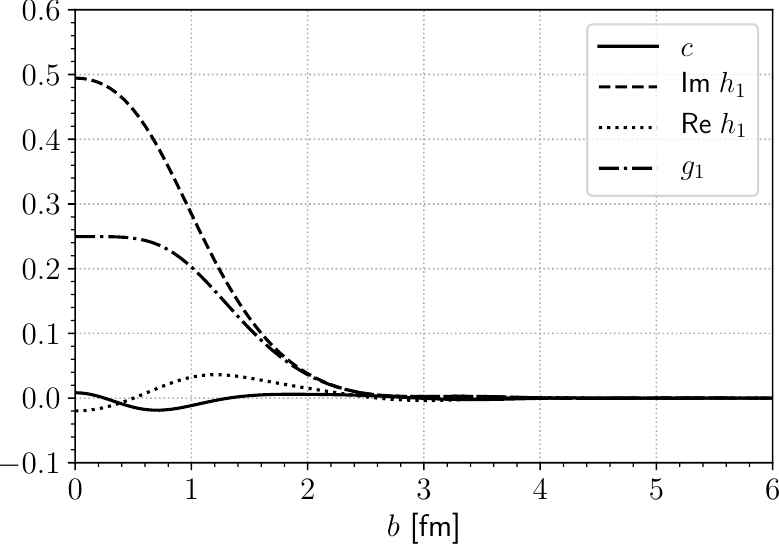}
\caption{\label{fig:pp8000gev_profiles_1b_noprofiles}Fit~1 (central)}
\end{subfigure}
\quad%add desired spacing between images, e. g. ~, \quad, \qquad etc.
%\begin{subfigure}[t]{0.48\textwidth}
%\includegraphics*[width=\textwidth]{model_ampl_xix_pp8000gev_xix_eikonal_xix_calc_xix_ffeff_xix_profiles_2b_noprofiles.pdf}
%\caption{\label{fig:pp8000gev_profiles_2b_noprofiles}peripheral case - Fit~2b}
%\end{subfigure}
\begin{subfigure}[t]{0.48\textwidth}
\includegraphics*[width=\textwidth]{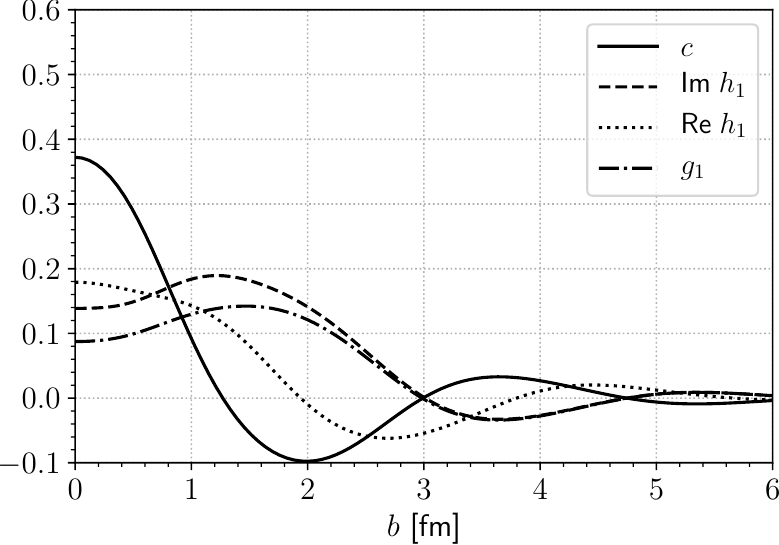}
		\caption{\label{fig:pp8000gev_profiles_3b_noprofiles}Fit~2 (peripheral)}
\end{subfigure}
%\quad%add desired spacing between images, e. g. ~, \quad, \qquad etc.
%\begin{subfigure}[t]{0.48\textwidth}
%\includegraphics*[width=\textwidth]{model_ampl_xix_pp8000gev_xix_eikonal_xix_calc_xix_ffeff_xix_profiles_4b_noprofiles.pdf}
%\caption{\label{fig:pp8000gev_profiles_4b_noprofiles}peripheral case - Fit~4b}
%\end{subfigure}
\begin{minipage}[t]{.95\textwidth}
%\begin{minipage}[t]{1.\columnwidth}
\caption{\label{fig:pp8000gev_noprofiles}Function $c(s,b)$ and several other functions characterizing pp collisions in dependence on impact parameter at energy of 8~GeV corresponding to the central (Fit 1) and the peripheral (Fit 2) picture of elastic scattering.}
\end{minipage}
\end{figure*}

%\subsubsection{\label{sec:pp8000gev_analysis_central} Elastic hadronic amplitude as constrained in many contemporary models and leading to central behavior of elastic collisions}
%\subsubsection{\label{sec:pp8000gev_analysis_peripheral} Possibility of peripheral behavior of elastic scattering}
%\subsubsection{Comparison to the simplified model of WY}

The pp data at 8~TeV have been analysed in very similar way, using the eikonal model, as it has been done in \cref{sec:pp53gev_results} in the case of pp data at 52.8~GeV. Two fits of data under different constrains have been performed.

In Fit~1 at 8~TeV reproducing similar $t$-dependence of elastic hadronic amplitude \ampl{N} as it is assumed in many contemporary models of elastic hadronic scattering, the phase $\phase$ has been required to pass through point
\begin{align}
 [t_2&=-1 \text{ GeV}^2, y_2=-\pi]  
\end{align}
instead of \refp{eq:pp53gev_ty2}. 
To obtain a stable fit at 8~TeV leading to peripherality of elastic collisions (Fit 2) we have required $\sqrt{\meanb[n=2,etype=el]}$ to be around 1.85. The parameter $\kappa=2$ has been used at 8~TeV to keep analyticity of elastic hadronic amplitude, see \cref{sec:analysis_parameterization}. Values of free fitted parameters, together with corresponding values of several hadronic quantities, may be found in \cref{tab:pp53gev_pp8000gev_ff_eff} for both the Fits~1 and 2. 

As one can see from the table total, elastic and inelastic cross sections at 8~TeV are approximately 104~mb, 28~mb and 76~mb differing only slightly in different fit alternatives. These numbers may be compared to values around 43~mb, 7.5~mb and 35~mb at 52.8~GeV in \cref{tab:pp53gev_pp8000gev_ff_eff}. It means that all the three integrated hadronic cross sections increase very significantly with change of collision energy from 52.8~GeV to 8~TeV. It is also interesting to note that ratio $\CS[etype={el,N}]/\CS[etype={tot,N}]$ is around 0.18 at 52.8~GeV and 0.27 at 8~TeV, i.e., significantly higher. 

Hadronic phase $\phase$ and diffractive slope $B(t)$ fitted to experimental data under different conditions are shown in \cref{fig:pp8000gev_zeta_multi,fig:pp8000gev_diffractive_slope_multi}. It may be seen from \cref{tab:pp53gev_pp8000gev_ff_eff} that quantities $B(t\!=\!0)$ and $\rho(t\!=\!0)$ at 8~TeV are higher than at 52.8~GeV.

The real and imaginary parts of elastic hadronic amplitude determined in each fit may be found in \cref{fig:pp8000gev_ffeff_reim}. The peripheral fit at 8~TeV fulfills conclusion of the Martin's asymptotic theorem \cite{Martin1997}, similarly as at lower energy of 52.8~GeV. 

$b$-dependent profile functions corresponding to effective electromagnetic form factors at 8~TeV are shown in \cref{fig:pp8000gev_profiles}; other $b$-dependent functions further characterizing pp collisions in dependence on impact parameter are pictured in \cref{fig:pp8000gev_noprofiles}. One may see big differences between the central case and the peripheral case shown in \cref{fig:pp8000gev_profiles,fig:pp8000gev_noprofiles}. %Comparison of elastic profile functions corresponding to the performed fits at 8~TeV is shown in \cref{fig:pp8000gev_profiles_el_multi}.

%In \cref{fig:data_pp53gev_pp8000gev_models_zoom} one may found comparison of interference regions at very low values of $|t|$ at 52.8~GeV and 8~TeV. One may see that at 52.8~GeV measured $t$-region where $\text{d}\sigma^{\text{C+N}}/\text{d}t$ is significantly different from $\text{d}\sigma^{\text{N}}/\text{d}t$ is much broader than at 8~TeV. It is experimentally more difficult to measure elastic scattering in the Coulomb-hadronic domain at higher energies as it is necessary to detect scattered protons at significantly lower scattering angles. %, see \cref{eq:t_angle_approx} and \cref{sec:TOTEM}.

The eikonal model analysis of experimental data explained and performed in this paper has been prepared and already used for analysis of pp elastic scattering data by the whole TOTEM collaboration, see the very first results of similar analysis of 8~TeV data in the Coulomb-hadronic interference region measured by TOTEM in \cite{totem2015}. Numerical values of some quantities such as $\CS[etype={tot,N}]$, $\CS[etype={el,N}]$, $\CS[etype={inel}]$, $B(t\!=\!0)$ or $\rho(t\!=\!0)$ determined in this section are slightly different from those published in \cite{totem2015}. There are several subtle differences between both the analyses. First of all we have fitted (approximate) data in broad region of $|t|$ values including the dip-bump structure as our aim was to determine, at least approximately, $t$-dependences of elastic hadronic amplitude in the widest possible $t$-range and corresponding $b$-dependences of profile functions - full physical picture under given set of assumptions. In \cite{totem2015} the data were fitted in much narrower $t$-region without the dip-bump structure with focus on determination of only some quantities (see \cite{totem2015} for details). Our parameterization of hadronic modulus differs, therefore, from the one used in \cite{totem2015}. As to $t$-dependence of hadronic phase in peripheral case we have used technique of penalty functions in order to find a few different solutions under given constrains while in \cite{totem2015} the $t$-dependence was chosen in quite fixed form (an ansatz) and then it was demonstrated that the corresponding elastic hadronic amplitude has given properties (leads to peripheral interpretation of elastic collisions in $b$-space). In our analysis only statistical errors of data points were taken into account while in \cite{totem2015} also systematical errors were considered.  

It is evident that describing in one model the Coulomb-hadronic interference region together with the dip-bump structure in measured data (i.e., non-trivial $t$-dependence) is more difficult than fitting, e.g., only (quasi)exponential part of data. Fit of data in broader region of $t$-values typically leads to higher values of reduced $\chi^2$. One may expect that lower values of reduced $\chi^2$ then shown in \cref{tab:pp53gev_pp8000gev_ff_eff} may be obtained by using more flexible parameterization of mainly the modulus of hadronic amplitude. The purpose of the fits performed in this paper has been to study mainly conceptual questions related to different interpretation possibilities of (elastic) pp collisions and corresponding assumptions, see next section.

\subsubsection{Comparison to the simplified model of WY}

The values of quantities $\CS[etype={tot,N}]$, $\rho(t\!=\!0)$, and $B(t\!=\!0)$ at 8~TeV in \cref{tab:pp53gev_pp8000gev_ff_eff} determined on the basis of the eikonal model (under different assumptions) may be compared to the values obtained on the basis of the simplified WY formula~\refp{eq:simplifiedWY} 
\begin{equation}
\begin{aligned}
\CS[etype={tot,N}] =& \; (102.0 \pm 2.2) \; \text{mb}, \\
\rho(t\!=\!0)  =& \; (0.05  \pm 0.02),   \\
B(t\!=\!0)     =& \; (19.42 \pm 0.05) \; \text{GeV}^{-2};
\label{eq:wy_parameters_totem2016}
\end{aligned}
\end{equation}
determined in \cite{totem2015}. %\Cref{fig:pp8000gev_ampl_n_fit3b_vs_WY} shows comparison of $t$-independent values of hadronic phase $\phase$ and diffractive slope $B(t)$ in the simplified approach of WY to strongly $t$-dependent quantities in one of the performed fit (Fit~3b). 
Hadronic phase $\phase$ and diffractive slope $B(t)$ in the simplified approach of WY are $t$-independent at all values of $t$. In the performed fits they are strongly $t$-dependent, see \cref{fig:pp8000gev_zeta_multi,fig:pp8000gev_diffractive_slope_multi}. In the simplified model of WY the real and imaginary parts of elastic hadronic amplitude are purely exponential in $t$ while in the eikonal model describing full measured region of $t$-values they have different $t$-dependence, see \cref{fig:pp8000gev_ffeff_reim}.
%As to the simplified approach of WY the real and imaginary parts of corresponding elastic hadronic amplitude determined by the values~\refp{eq:wy_parameters_totem2016} are shown in \cref{fig:pp8000gev_wy_reim_log}. 

Explicit additional calculations at 8~TeV showing that the WY approach is not suitable for general studies of $t$-dependence of hadronic amplitude determined on the basis of experimental data may be found in appendix~\refp{sec:wy_study_pp8000gev}.

\FloatBarrier
%\clearpage
\section{\label{sec:analysis_summary}Summary and open questions}
In the case of the simplified approach of WY the hadronic phase was, without any justification, assumed to be $t$-independent (see \cref{sec:introduction}). For this reason, there is no ambiguity on determination on the phase from experimental data (in the region of the lowest $|t|$ values). More recent models of hadronic amplitude \ampl{N} do not assume $t$-independent hadronic phase but typically assume dominance of imaginary part of the amplitude in the forward region and vanishing of the imaginary part in the region of diffractive minimum, see \cref{sec:ampl_n}. However, these a priori constrains have never been (up to our knowledge) sufficiently justified in published papers. It may be concluded from our analysis that mainly the dominance of the imaginary part of \ampl{N} in the forward region leads to centrality of elastic collisions in the impact parameter space. %Corresponding structure of colliding particles has never been sufficiently explained in the literature. 

Moreover, these widely used models of elastic hadronic amplitude have been very often fitted to data without taking into account Coulomb-hadronic interference. In this case the elastic hadronic phase has not been constrained by experimental data at all (only the modulus of the amplitude can be determined in the fitted $t$-range, see \cref{eq:difamp_gen}). It is not good sign that choice of $t$-dependence of the phase has not been very often discussed in the corresponding papers at all (usually no plot of the phase has been shown). Many papers devoted to models of elastic hadron scattering have often been focused only on some quantities with unclear physical meaning (such as quantities $\rho$ and $B$ which have only very indirect relation to particle structure or interaction). Not enough attention have been devoted to whole physical picture of particle collisions. $t$ and $b$ dependent characteristics corresponding to given model have often not been discussed together with corresponding assumptions involved in the given description of experimental data. %Transition of initial to final states has not been, therefore, sufficiently studied in these approaches.

We have tried to open the problem and to show some possibilities of description of elastic scattering allowed by the eikonal model approach. Several fits of the same data, including both the peak at very low values of $|t|$ and the dip-bump structure, at energy of 52.8~GeV and 8~TeV have been performed under different assumptions. $t$-dependence of hadronic amplitude \ampl{N} in the whole kinematically allowed region of $t$ has been determined for each of the possibilities. Corresponding physically interesting quantities and $t$ and $b$ dependent functions characterizing hadronic collisions have been calculated and the results compared. The 2 fits at each energy performed in this paper correspond to analytic elastic hadronic amplitude in $t$, satisfy unitarity and also conclusion of the asymptotic theorem of Martin. It is the result of our analysis, that the choice of form factor (effective electric vs.~effective electromagnetic, i.e., inclusion of magnetic moment) has very small or negligible impact on the determined hadronic quantities in the eikonal model approach discussed in this paper. However, the choice of $t$-dependence of hadronic phase fundamentally changes behavior of the collisions in dependence on impact parameters (or vice versa). In this case (rarely mentioned in the literature) there is, therefore, corresponding ambiguity in the description which should be further studied. Our results show that elastic collisions may be interpreted as a peripheral process in agreement to standard ideas corresponding to collisions of two matter objects. Further comments concerning dependence of elastic hadron collisions on impact parameter may be found in \cite{Prochazka2015_bdependence}.

In \cite{Petrov2018_cni} one may find a recent review of calculations concerning Coulomb-nuclear interference at high energies within the eikonal model framework. It confirmed that the Bethe's formula \cref{eq:FCNbethe} and the simplified description of Coulomb-hadronic interference proposed by West and Yennie (see \cref{sec:WY}) can be now hardly used for reliable analysis of contemporary experimental data. A formula very similar to \cref{eq:kl1} (also for arbitrary $t$-dependence of \ampl{N}) describing \ampl{C+N} was derived in slightly modified way. The formula in \cite{Petrov2018_cni} does not contain the first term from \cref{eq:kl2} which was attributed to slightly different way of taking into account electromagnetic form factors in the derivation than it was done in the past, see references in \cref{sec:ampl_cn}. The approach in \cite{Petrov2018_cni}, however, did not sufficiently distinguish between kinematically allowed and forbidden values of $t$ variable in the calculations. Extrapolations outside measured regions (and especially to kinematically forbidden regions) should be treated with care; see bellow a corresponding open question.

\Cref{fig:pp53gev_1b_dsdt_zoom,fig:pp8000gev_1b_dsdt_zoom} clearly show that there is actually no measured $t$ range at 52.8~GeV and 8~TeV where measured differential cross section would be described only by the Coulomb interaction; measured differential cross section at the lowest measured values of $|t|$ is commonly described with the help of \emph{both} the Coulomb \emph{and} hadronic interactions. Coulomb differential cross section for point-like charged particles is in all hitherto descriptions assumed to be "well known" from QED and then modified by some form factors in the case of pp scattering, see \cref{eq:dcs_c_qed}. Influence of the different proton form factors, see \cref{fig:ff_elc,fig:ff_eff}, on $t$-dependence of the Coulomb differential cross section is strongly suppressed with decreasing value of $|t|$ towards $t\!=\!0$. The formula \refp{eq:dcs_c_qed} has, therefore, more and more fixed $t$-dependence (not modified by any free parameter) with decreasing value of $|t|$ and diverging at $t\!=\!0$. It would be, therefore, very useful to test experimentally such $t$-dependence of the ``pure'' Coulomb differential cross section of two protons at very low values of $|t|$. This would require to detect at 52.8~GeV or 8~TeV elastically scattered protons at even lower values of $|t|$ (scattering angles) than those shown in \cref{fig:dsdt_data_pp53gev_pp8000gev}. Taking into account that the widely used description of the Coulomb differential cross section in the high energy limit is $s$-independent, see \cref{eq:dcs_ampl_c_high_energy_limit}, it would be extremely useful to analyze also data at very low values of $|t|$ corresponding to ``pure'' Coulomb scattering at lower energies than those at the ISR. One should be aware, too, that the electromagnetic form factors in ep and pp processes, entering into any contemporary used Coulomb-hadronic interference formula, need not be the same - which should be tested in the future. 

The question of "subtracting" of all Coulomb effects from measured elastic pp scattering to obtain only hadronic scattering represent quite delicate problem. It has been already identified as an open question in sect.~6 in \cite{Prochazka2015_bdependence} where one can find several other open problems. The list of open questions, concerning \emph{all} contemporary models of elastic scattering widely discussed in the literature, may be further extended:

\begin{enumerate}
\item{\textit{Increase of integrated total, elastic and inelastic cross sections and dimensions of colliding particles in dependence on collision energy}\\

Values of cross sections $\CS[etype={tot,N}]$, $\CS[etype={el,N}]$, $\CS[etype={inel}]$ and ratio $\CS[etype={el,N}]/\CS[etype={tot,N}]$ determined in analyses of experimental pp data strongly increase with increasing collision energy (see, e.g., figs.~4 and 6 in \cite{TOTEM2019}). Various contemporary models of elastic scattering applied to experimental data have led to very similar values of cross sections. The increase of the cross sections is clearly visible from the results of our analysis, too. In \cref{tab:sizes_and_dimensions} one may find values of the cross sections at 52.8~GeV and 8~TeV corresponding to two fits performed in this paper and selected here as examples for comparison. The values of total and inelastic cross sections may be used to estimate corresponding average radius of colliding proton (interaction range) according to formula $r_X \approx \sqrt{\CS[etype={X}]/\pi}$ (for spherical particles, in the case of some changeable sizes of colliding particles similar average values of dimensions may be expected). The very significant increase of the cross sections and corresponding dimension values of colliding particles with energy has never be explained. 

More attention should be devoted also to the ratio $\CS[etype={el,N}]/\CS[etype={tot,N}]$ which significantly increases with energy, see \cref{tab:sizes_and_dimensions}. According to this result relatively more and more collisions should be elastic instead of inelastic when energy increases; and one may ask how the trend continues at even higher energies (or even at $s \to \infty$). It is especially difficult to understand this result, if one considers that the speeds of colliding protons at both the 52.8~GeV and 8~TeV energies are (according to the theory of relativity) nearly the same and equal the speed of light. In sect.~3.1.~in \cite{Petrov2016_problems} understanding of the increase of the cross sections with energy has been denoted as a pressing problem. 

\indent
\begin{table}%[!ht]
\centering
%\resizebox{\textwidth}{!}{
%\resizebox*{!}{\dimexpr\textheight-2\baselineskip\relax}{%
\begin{tabular}{lccccc}
\hline \hline
$\sqrt{s}$                     & [GeV] & 52.8  & 8000  \\
Fit                            &       & 2     & 2     \\
\hline                                        
$\CS[etype=tot]$               & [mb]  & 42.9  & 104.1 \\ 
$\CS[etype=el]$                & [mb]  & 7.54  & 28.0  \\
$\CS[etype=inel]$              & [mb]  & 35.3  & 76.1  \\
$\CS[etype=el]/\CS[etype=tot]$ &       & 0.176 & 0.269 \\
\hline
$r_{\text{tot}}$               & [fm]  & 1.17  & 1.82  \\
$r_{\text{inel}}$              & [fm]  & 1.06  & 1.56  \\
\hline \hline
\end{tabular} %} %\\ %}
\begin{minipage}[t]{0.4\textwidth}
\caption{\label{tab:sizes_and_dimensions}Comparison of values of cross sections and estimated radii of protons corresponding to Fit~2 at 52.8 GeV and Fit~2 at 8~TeV.}
\end{minipage}
\end{table}

It is interesting to compare, too, values of the root-mean squares of impact parameter corresponding to total, elastic and inelastic hadronic collisions ($\sqrt{\meanb[n=2,etype=tot]}$, $\sqrt{\meanb[n=2,etype=el]}$ and $\sqrt{\meanb[n=2,etype=inel]}$) in \cref{tab:pp53gev_pp8000gev_ff_eff}. One may see clear differences between a central and a peripheral case at given energy and also significant energy dependence. It further illustrates that the question of the sizes of colliding protons (or interaction regions) represents an important open question for the future. The fact that protons cannot be taken as point-like particles during high energy collisions at small values of impact parameter comparable to the sizes of the particles should be taken into account in derivations of all formulas describing the given collision processes.\\
}

\item{\textit{Extrapolations outside measured regions}\\

In physics in general extrapolations of quantities outside measured regions are very delicate. In the case of models of elastic pp scattering more attention should be devoted mainly to extrapolation of $\text{d}\sigma^{\text{N}}/\text{d}t$ to $t=0$ (\modulus{N} to $t=0$). In all contemporary models a (quasi)exponential dependence (having maximal value at $t=0$) has been assumed. Small value of parameter $\rho(t=0)$, determined from Coulomb-hadronic interference using the WY or the eikonal model approach, has been in one way or another assumed in all widely used models of elastic hadron scattering not considering any Coulomb effect. All the models fitted to experimental data (represented by measured elastic differential cross section), therefore, led to very similar values of $\CS[etype={tot,N}]$, $\CS[etype={el,N}]$ and $\CS[etype={inel}]$ when determined using optical theorem (see \cref{eq:optical_theorem,eq:cs_el_integ_b,eq:cs_inel}). 

Kinematics of two particles limits allowed values of $t$ variable: $t\in\langle t_{\text{min}},0 \rangle$. This should be respected in derivations of relations used for description of elastic collisions measured at finite energies (see, for example, impact parameter representation of elastic scattering amplitude at \emph{finite} energies in appendix~\ref{sec:ampl_b}). However, many descriptions of elastic scattering have not distinguished between physically allowed and forbidden region of $t$ values. More attention should be devoted also to extrapolations of $s$ dependent quantities to energy regions where no elastic differential cross section was measured (especially in asymptotic region $s \to \infty$). These extrapolations of $t$ or $s$ dependent quantities typically enter into calculations whenever it is required to integrate an expression over "all" values of $t$ or $s$.

Extrapolations which are done only on mathematical level may unpredictably fail. Instead of using phenomenological models for extrapolations of quantities outside measured regions one should develop causal ontological models which may provide insight into the studied physical process and, therefore, provide arguments for given extrapolation which may be further analyzed and tested, see \cite{Lokajicek2007_models}. %This has not yet been sufficiently done in the case of elastic hadron collisions.
}
\end{enumerate}

%Future analyses of elastic collisions should focus on all the assumptions involved in a given model applied to experimental data and testing of their consequences. and discussion of quantities which have clear physical meaning and relation to properties of the colliding particles \cite{Prochazka2015_bdependence}.
%Better understanding of elastic collisions should focus on further analysis of assumptions involved in the models applied to experimental data and on discussion on of quantities which have clear physical meaning and relation to properties of the colliding particles \cite{Prochazka2015_bdependence}.

%% TODO PETROV
%open questions and problems (see also, e.g., \cite{Petrov2016_problems,Petrov2016_sizes})

\FloatBarrier

%\FloatBarrier
%\newpage
%%%%%%%%%%%%%%%%%%%%
\section{\label{sec:conclusion}Conclusion}
%%%%%%%%%%%%%%%%%%%%
The measurement of elastic pp differential cross section $\frac{\text{d}\sigma}{\text{d}t}$ represents main source of experimental data for the analysis of elastic processes of protons. The goal of contemporary theoretical description consists in separation of the Coulomb effect from data to determine elastic hadronic amplitude $\ampl{N}$ from which conclusions concerning structure and interactions of colliding hadrons should be derived and further tested. There has not been any actual theory until now which would consistently determine its corresponding $t$-dependence on the basis of measured elastic differential cross section at \emph{all} measured values of $|t|$ (including both the Coulomb-hadronic region at very low values of $|t|$ \emph{and} the dip-bump region at higher values of $|t|$) - except the eikonal model approach. 

In the past the simplified approach of West and Yennie has been made use of for separation of Coulomb and hadron interactions. However, this method is not theoretically consistent and is not in sufficient agreement with the measured data. It contains many limitations as it has been discussed in \cref{sec:introduction}. It has been applied to the analysis of the data only in a very narrow region of momentum transfers in forward direction and the influence of Coulomb scattering at higher values of momentum transfers has been always neglected by definition. The elastic scattering at higher values of momentum transfer has been always described phenomenologically as purely hadronic scattering on the basis of assumptions not consistent with the ones used in the approach of WY. Such an inconsistent dual description of data in the description of elastic hadron collisions can be hardly justified. It has been argued in \cite{KL2007} that already the integral formula \refp{eq:phaseWY} for relative phase $\alpha\phi(s,t)$ is limited to $t$-independent quantity $\rho(s,t)$. The WY approach cannot be, therefore, used for analysis of experimental data with arbitrary $t$-dependence of hadronic phase, i.e., one cannot study $b$-dependent characteristics of pp collisions in this approach, see also appendix~\ref{sec:wy_study}.

%It has been shown explicitly in \cref{sec:wy_study} that already the integral formula \refp{eq:phaseWY} for relative phase $\alpha\phi(s,t)$ is limited to $t$-independent quantity $\rho(s,t)$ (see also \cite{KL2007}); the WY approach cannot be used for analysis of experimental data with arbitrary $t$-dependence of hadronic phase, i.e., one cannot study $b$-dependent characteristics of pp collisions in this approach.
%It has been shown explicitly in \cref{sec:wy_study} that already the integral formula \refp{eq:phaseWY} for relative phase $\alpha\phi(s,t)$ is limited to $t$-independent quantity $\rho(s,t)$ (see also \cite{KL2007}); the WY approach cannot be used for analysis of experimental data with arbitrary $t$-dependence of hadronic phase, i.e., one cannot study $b$-dependent characteristics of pp collisions in this approach.

The eikonal model approach, based on the complete elastic scattering amplitude $\ampl{C+N}$ fulfilling \cref{eq:kl1,eq:kl2,eq:KLampI}, provides more reliable basis for analysis of elastic collisions of (charged) hadrons. In principle it is established on the fact that the common influence of the Coulomb and hadronic elastic scattering can be reliably described by the sum of the Coulomb and elastic hadronic potentials (eikonals) and without any a priori limitation on $t$-dependence of the elastic hadronic amplitude. However, analyses of experimental data have shown that the complex hadronic component $\ampl{N}$ cannot be uniquely established. Only its modulus is strongly constrained on the basis of measured elastic differential cross section. The $t$-dependence of its phase has been only partially constrained when Coulomb-hadronic interference (the region of very small $|t|$) has been taken into account. 

In the majority of published analyses of experimental data the corresponding freedom has been, however, strongly limited by the choice of amplitude parameterization. The imaginary part has been usually assumed to be dominant in a great interval of $t$ and vanishing in the region around diffractive minimum; with the real part determining the non-zero value of differential cross section in the diffractive minimum; see, e.g., the earlier papers \cite{Miettinen1975,Groot1973,Amaldi1976_385,Amaldi1980,Castaldi1985,Barone2002,Ayres1976,Bailly1987}, \cite{Heckman1972,Henyey1974,Henzi1974,Henzi1979,Henzi1983,Henzi1984,Henzi1985_plb160,Henzi1985_zpc27,Henzi1976} and also recent papers \cite{Carvalho1997,Carvalho2005,Menon2005,Menon2007,Avila2006,Avila2007,Avila2008,Silva2007,Ferreira1997,Ferreira1998,Campos2010,Fagundes2011_epjc71,Fagundes2011_ijmpa26,Shoshi2002,Ryskin2007,Lipari2013}. Description of elastic hadronic amplitude corresponding to these widely used assumptions has been fitted to experimental data at energy of 52.8~GeV and 8~TeV in \cref{sec:pp53gev_analysis_central,sec:pp8000gev_results}. The so-called central behavior in impact parameter space has been then obtained in such a constrained case; elastic processes being more central (i.e., existing for very small $b$ even at $b=0$) than inelastic ones. Transparent protons during elastic processes may be, however, hardly brought to agreement with the existence of inelastic processes when hundreds of particles have been formed at the same collision energy. 

Much more general parameterization of the hadronic amplitude $\ampl{N}$ has been used in \cref{sec:pp53gev_analysis_peripheral,sec:pp8000gev_results}. A rather steep rise of phase $\phase$ with increasing $|t|$ already at very small values of $|t|$ has been allowed. It has been possible to obtain strongly peripheral impact parameter profile for elastic processes; the imaginary part (dominant at $t=0$) going quickly to zero with rising $|t|$ ($\Im \ampl{N} = 0$ at $t \simeq -0.1$~GeV$^2$, at 52.8~GeV and 8~TeV).

Similar analysis of experimental data with the help of the eikonal model (see \cref{eq:kl1,eq:kl2,eq:KLampI}) has been done already earlier in \cite{Kundrat1994_unpolarized}. %In this paper different alternatives corresponding to the peripheral behavior have been newly shown at 52.8~GeV and compared to similar analysis at much higher energy of 8~TeV. 
In \cite{Kundrat1994_unpolarized} only electric form factors have been taken into account. It has been shown in this paper that addition of magnetic form factors does not lead to significant change of determined amplitude \ampl{N}. For the purpose of this analysis integral $I(t,t')$ defined by \cref{eq:KLampI} has been calculated also analytically for one suitable parameterization of the form factors, see \cref{sec:integral_I}. In \cref{sec:data_analysis} we have determined $t$-dependences of elastic hadronic amplitude under different constrains and showed that all the solutions may be constructed as analytic, while in \cite{Kundrat1994_unpolarized} used parameterizations have not been analytic. All the performed fits at older energy of 52.8~GeV and newer energy of 8~TeV discussed in detail in \cref{sec:data_analysis} are analytic, satisfy condition of unitarity and the real parts of all elastic hadronic amplitudes change sign at low value of $|t|$ as motivated by the asymptotic theorem of Martin \cite{Martin1997}.

Analysis of data presented in this paper with the help of the eikonal model has been prepared and already used for analysis of pp elastic scattering at the LHC energies, see the very first results of similar analysis of 8~TeV data in the Coulomb-hadronic interference region measured by TOTEM in \cite{totem2015}. Similar analysis of experimental data under different assumptions may be performed at any other (high) energy and the results further studied. However, it is possible to say (against earlier conviction) that there is not any reason against more realistic interpretation of elastic processes when protons are regarded as compact (non-transparent) objects.

In \cref{sec:analysis_summary} we have pointed out to several difficulties in description of elastic pp collisions. Several other fundamental open problems concerning description of elastic hadron collisions may be found in \cite{Prochazka2015_bdependence}. %, see also \cite{Lokajicel_physics_status} for further comments. 

\section*{Acknowledgement}
%We would like to thank to profs.~A.~Martin, O.~Nachtmann, V.~A.~Petrov and M.~M.~Islam for stimulating discussions concerning analyticity of elastic hadronic amplitude. 
We would like to thank to profs.~M.~M.~Islam, A.~Martin, O.~Nachtmann and V.~A.~Petrov for stimulating discussions concerning various aspects of elastic pp scattering. 

\begin{appendices}
%%%%%%%%%%%%%%%%%%%%%%%%%%%%%%%%%%%%%%%%%%%%%%%%%%%%%%%%%%%%%%%%
\section{\label{sec:ampl_b}Impact parameter representation of elastic scattering amplitude at finite energies}
%%%%%%%%%%%%%%%%%%%%%%%%%%%%%%%%%%%%%%%%%%%%%%%%%%%%%%%%%%%%%%%%
In this section we will summarize the needed formulae for the mathematical consistent formalism of the elastic scattering amplitude in the impact parameter representation at finite energies. Similarly, the formulas enabling determination of the values of root-mean-squares of impact parameters characterizing total, elastic and inelastic collisions will be mentioned, too. 

The function $h_{\text{el}}(s,b)$ defined by \cref{eq:hel_standard} determines the impact parameter profile in the limit of $s$ going to infinity as the FB transform introducing the impact parameter representation of elastic scattering amplitude requires the amplitude to be defined for all values of $t$ from the interval $(-\infty, 0 \rangle$ \cite{PTPS.E65.316,PTP.35.463,PTP.35.485,PTPS.37.297,PTP.39.430,PTP.39.785,Islam1968}. For finite energy values the function $\ampl{N}$ may be specified, however, in the kinematically allowed interval only: {\mbox{$t \in \langle t_{\text{min}}, 0 \rangle$}}. %It should hold $F^{\text{N}}(s,t_{\text{min}}) = 0$.

In the following we shall follow the approach proposed in \cite{Islam1968}. One may write
\begin{equation}
\sqrt{-t} \; = \; 2p\,y\,,  \;\;\;\;\;    y\;=\; \sin {\frac{\theta}{2}},
\label{eq:tt1}
\end{equation}
where $\theta$ stands for elastic scattering angle in the center-of-mass system. Let us define then the function $A(s,y)$ by relation
\begin{equation}
A(s,y) =
\begin{cases}
F^{\text{N}}(s,y)                                           & 0 \leq y < 1    \\
\lambda (s,y) \equiv \lambda_R (s,y) + \text{i} \; \lambda_I (s,y) & 1 < y < \infty 
\end{cases}
\label{eq:A_sy}
\end{equation} 
where $\lambda(s,y)$ is unknown complex function the real and imaginary parts of which are supposed to have following properties: 
\begin{itemize}
\item{ $\int\limits_1^{\infty} y^{1/2} \lambda_{R,I} (s,y) \text{d}y\;\;$ are absolutely convergent,}

\item{ $\lambda_{R,I} (s,y) \;\;$ are of bounded variation for  $1 < y < \infty$.} 
\end{itemize}
Then according to Hankel theorem \cite{Watson1962} the amplitude $A(s,y)$ has FB transform for $0 < y < \infty$
\begin{equation}
A(s,y) = \frac{\sqrt{s}}{2 p} 
\int \limits_{0}^{\infty} \beta \text{d}\beta \; J_{0}(\beta y) \; h(s, \beta),
\label{eq:imp3}
\end{equation}
\begin{equation}
h_{\text{el}}(s, \beta ) = \frac{2 p}{\sqrt{s}}
\int \limits_{0}^{\infty} y \text{d}y \; J_{0}(\beta y)\; A(s,y);
%\label{eq:imp4}
\end{equation}
here we have introduced a new variable $\beta = 2 p b$. 

More detailed insight to inverse FB transform offers MacRobert's theorem \cite{Macrobert1931,Macrobert1947,Sneddon1951} which may be formulated as follows. Let the function $F(s,y)$ is holomorphic in the interval $p < y < q$ and let the function $a(s,\beta)$ can be expressed by the integral
\begin{equation}
a(s, \beta) = \frac{2 p}{\sqrt{s}}
\int \limits_{p}^{q} y \text{d}y \; J_{\nu}(\beta y) \; F(s, y) \\
\label{eq:mat1}
\end{equation}
for $\;0 \leq p < q \leq \infty \; $ and $\; \Re \nu > -1$ then  \\
\begin{equation}
\begin{split}
A(s,y) &= \frac{\sqrt{s}}{2p} \int \limits_{0}^{\infty} \beta \text{d} \beta \;
J_{\nu}(\beta y) \;
a(s, \beta) \\
&= \begin{cases}
{F(s,y)}& \text{for $\;p<y<q$}, \\
0\, & \text{for $\;0<y<p\;$ or $\;y>q$}.
\end{cases}
\end{split}
\label{eq:mat2}
\end{equation}
MacRobert's theorem may be used to the FB transform of function $F(s,y)$ as the elastic hadronic amplitude $F^{\text{N}}(s,y)$ is the holomorphic function inside Lehman's ellipse with foci -1 and 1 (see, e.g., \cite{Eden1971}). 

Then the original elastic scattering amplitude $\ampl{N}$ is given by relation
\begin{equation}
\ampl{N} = 2 p \sqrt{s}
\int \limits_{0}^{\infty} b \text{d}b \; J_{0}(b\sqrt{-t})\; 
h_{\text{el}}(s,b),
\label{eq:imp5}
\end{equation}
which is the representation of elastic scattering amplitude in the impact parameter space. With the help of relation \refp{eq:A_sy} the inverse relation to the relation \refp{eq:imp5} has a form
\begin{equation}
h_{\text{el}}(s,\beta) = h_{1} (s, \beta) + h_{2} (s, \beta),
\label{eq:imp6}
\end{equation}
where
\begin{equation}
h_{1}(s,\beta)=  \frac{2 p}{\sqrt{s}}
\int \limits_{0}^{1} y \text{d}y \; J_{0}(\beta y) \; F^{\text{N}}(s,y)
\label{eq:imp7}
\end{equation}
and
\begin{equation}
h_{2}(s,\beta)= \frac{2 p}{\sqrt{s}}
\int \limits_{1}^{\infty} y \text{d}y \; J_{0}(\beta y) \; \lambda (s,y).
\label{eq:imp8}
\end{equation}
The function $\lambda (s,y)$ as well as its FB image $h_2(s,\beta)$ are in general complex functions. 

Similar relations may be derived also for the inelastic processes. Starting from the unitarity condition \refp{eq:preunitarity} in $t$ variable (expressed now in $y$ variable) and performing the FB transform of the real inelastic overlap function $G_{\text{inel}}(s,y)$ one may obtain
\begin{equation}
g_{\text{inel}}(s,\beta) = g_{1}(s,\beta) + g_{2}(s,\beta),
\label{eq:imp12}
\end{equation}
where
\begin{equation}
g_{1}(s,\beta) =  \frac{2 p}{\sqrt{s}}
\int \limits_{0}^{1} y \text{d}y \; J_{0}(\beta y) \; G_{\text{inel}}(s,y)
\label{eq:imp13}
\end{equation}
and 
\begin{equation}
g_{2}(s,\beta) = \frac{2 p}{\sqrt{s}}
\int \limits_{1}^{\infty} y \text{d}y \; J_{0}(\beta y) \; \mu(s, y)
\label{eq:imp14}
\end{equation}
where the real function $\mu(s, y)$ must fulfill the same conditions as real and imaginary parts of function $\lambda(s,y)$.

The representation of elastic hadronic amplitude in the impact parameter space, i.e., $h_1(s,\beta)$ in the physical region, should contain the same amount of physical information as the original amplitude $\ampl{N}$. With the help of the optical theorem \refp{eq:optical_theorem} the total cross section may be expressed using the $b$-dependent function $h_1(s,b)$ as \cite{PTPS.E65.316,PTP.35.463,PTP.35.485,PTPS.37.297,PTP.39.430,PTP.39.785,Islam1968}    
\begin{equation}
\begin{split}
\CS[etype={tot,N}](s) &= \frac{4 \pi}{p \sqrt{s}} \; \Im F^{\text{N}}(s, t=0) 
%= \frac{2 \pi}{p^2} \int \limits_{0}^{\infty} \beta \text{d}\beta \; \Im h_{1}(s,\beta) \\
= 2 \pi \int \limits_{0}^{\infty} b \text{d}b \; 4 \Im h_{1}(s, b);
\end{split}
\label{eq:cs_tot_integ_b}
\end{equation}
and also the integrated elastic cross section may be written as
\begin{equation}
\begin{split}
\CS[etype={el,N}](s) &= \frac{ 8 \pi}{s} \int \limits_{0}^{1} y \text{d}y \; |F^{\text{N}}(s,y)|^2 
%= \frac{2 \pi}{p^2} \int \limits_{0}^{\infty} \beta \text{d}\beta \; |h_{1}(s, \beta)|^2 \\
= 2 \pi \int \limits_{0}^{\infty} b \text{d}b \; 4 |h_{1}(s,b)|^2.
\end{split}
\label{eq:cs_el_integ_b}
\end{equation}
The integrated inelastic cross section defined as 
\begin{equation}
\CS[etype={inel}](s) = \CS[etype={tot,N}](s) - \CS[etype={el,N}](s)
\label{eq:cs_inel}
\end{equation}
is then given by relation \cite{PTPS.E65.316,PTP.35.463,PTP.35.485,PTPS.37.297,PTP.39.430,PTP.39.785,Islam1968} 
\begin{equation}
\CS[etype={inel}](s) = 
%\frac{2 \pi}{p^2} \int \limits_{0}^{\infty} \beta \text{d}\beta \; g_{1}(s,\beta)= 
2 \pi \int \limits_{0}^{\infty} b \text{d}b \; 4 g_{1}(s,b).
\label{eq:cs_inel_integ_b}
\end{equation}
The unitarity equation in the impact parameter space can be written in a generalized form as \cite{PTPS.E65.316,PTP.35.463,PTP.35.485,PTPS.37.297,PTP.39.430,PTP.39.785,Islam1968}
\begin{equation}
%\Im h_{1}^{\text{N}}(s,\beta) = |h_{1}^{\text{N}}(s,\beta)|^2 + g_{1}(s,\beta) + K(s,\beta) 
\Im h_1(s,\beta)  =  |h_1(s,\beta)|^2+g_1(s,\beta) + K(s,\beta) 
\label{eq:imp16}
\end{equation}
where the correction function $K(s,\beta)$ is limited by a condition
\begin{equation}
\int \limits_{0}^{\infty} \beta \text{d}\beta \; K(s, \beta) = 0.
\label{eq:imp17}
\end{equation}
Also the functions $h_2(s,\beta)$ and $g_2(s,\beta)$ are limited by the similar conditions \cite{PTPS.E65.316,PTP.35.463,PTP.35.485,PTPS.37.297,PTP.39.430,PTP.39.785,Islam1968}, i.e.,
\begin{equation}
\int \limits_{0}^{\infty} \beta \text{d}\beta \; h_2(s, \beta) = 0,
\hspace*{1cm}
\int \limits_{0}^{\infty} \beta \text{d}\beta \; g_2(s, \beta) = 0.
\label{eq:imp19}
\end{equation}
The function $K(s,\beta)$ equals \cite{PTPS.E65.316,PTP.35.463,PTP.35.485,PTPS.37.297,PTP.39.430,PTP.39.785} 
\begin{equation}
K(s,\beta) = \frac{1}{16 \pi^2 s}\int \limits_{t_{\text{min}}}^{0} \! \text{d}t_1 \!
 \int \limits_{t_{\text{min}}}^{0}\! \text{d}t_2 \;
F^{N*}(s,t_2) \; F^{\text{N}}(s, t_1) \; M(\beta; t_1, t_2)
\label{eq:cf1}
\end{equation}
where
\begin{equation}
\begin{split}
M(\beta; t_1, t_2)  =& 
 J_{0} \Bigg ( \beta \sqrt {t_1 \bigg (\frac{t_2}{t_{\text{min}}}-1 \bigg )} \Bigg )
 J_{0} \Bigg ( \beta \sqrt {t_2 \bigg (\frac{t_1}{t_{\text{min}}}-1 \bigg )} \Bigg ) \\
         &- J_{0} (\beta \sqrt{-t_1}) J_{0} (\beta \sqrt{-t_2}).  
\end{split}  
\raisetag{1.18\baselineskip}
\label{eq:cf2}
\end{equation}
The function $K(s,\beta)$ vanishes at $\beta=0$ and $b \rightarrow \infty$; and also at asymptotic energies $(t_{\text{min}} \rightarrow - \infty)$ \cite{PTPS.E65.316,PTP.35.463,PTP.35.485,PTPS.37.297,PTP.39.430,PTP.39.785,Islam1968}. Detailed analysis of high energy elastic pp scattering \cite{Kaspar2011} has showed that the value of function $K(s,\beta)$ has very small impact on determination of the value of function $g_1(s,\beta)$ calculated on the basis of \cref{eq:imp16}.

The shape of elastic amplitude in the impact parameter space $h_{\text{el}}(s,\beta)$ determined by \cref{eq:imp6} depends on the $t$-dependence of elastic hadronic amplitude $\ampl{N}$ in the unphysical region of $t$. As shown by Islam \cite{Islam1976} the uniqueness of the $\ampl{N}$ can be achieved if two $t$-dependent parts of the amplitude $\ampl{N}$ in the physical and unphysical regions are bounded by the Sommerfeld-Watson transform. The elastic amplitude $h_{\text{el}}(s,\beta)$ in the impact parameter space oscillates at larger $\beta$ values; the oscillations disappear at infinite energies only. 

The representation of the scattering amplitude in the impact parameter space has been defined in \cite{PTPS.E65.316,PTP.35.463,PTP.35.485,PTPS.37.297,PTP.39.430,PTP.39.785,Islam1968} as an analogy to partial wave analysis. From the requirement of equivalence of both these representations the question arises which conditions imposed on the elastic hadronic amplitude $\ampl{N}$ guarantee the existence of its impact parameter representation. It is shown in \cite{PTPS.E65.316,PTP.35.463,PTP.35.485,PTPS.37.297,PTP.39.430,PTP.39.785,Islam1968} that the finiteness of the integrated elastic cross section \refp{eq:cs_el_integ_b} at finite energies guarantees its existence.

It has been shown in \cite{Kundrat1994_unpolarized} that the $4 \Im h_{1}(s,b)$ and $4 g_{1}(s,b)$ obtained with the help of FB transforms oscillate at larger values of impact parameter $b$ due to the fact that the region of kinematically allowed values of momentum transfers $t$ at finite energies is limited and the region for $t < t_{\text{min}}$ is not taken into account. The oscillations appear not only in the case of peripheral behavior of elastic hadron scattering where they are very significant but also in the case of central behavior. 
The physical meaning may be, therefore, hardly attributed to the functions $4 \Im h_{1}(s,b)$ and $4 g_{1}(s,b)$ in \cref{eq:cs_tot_integ_b,eq:cs_inel_integ_b,eq:imp16}, even if their integrals represent corresponding cross sections, see also \cite{Islam1976}. Only the non-negative function $4 |h_{1}(s,b)|^2$ has been denoted as elastic profile function. According to \cite{Kundrat2000,Kundrat2001,Deile:2010mv_Kundrat,elba2010_klkp,Kundrat2002} non-negative (non-oscillating) total and inelastic profile functions at finite energies may be defined if a convenient real function $c(s,b)$ is added to both the sides of the unitarity equation \refp{eq:imp16} 
\begin{align}
  \Im h_1(s,b) + c(s,b) &=  |h_1(s,b)|^2+g_1(s,b) + K(s,b) + c(s,b).
  \label{eq:unitarity1} 
\end {align}
It is then possible to define at finite energies total, elastic and inelastic profile functions $\PROF{X}(s,b)$ 
\begin{align}
  \PROF{el}(s,b)   &\equiv 4\,|h_1(s,b)|^2,               \label{eq:ampl_prof_el_new}\\
  \PROF{tot}(s,b)  &\equiv 4\,(\Im h_1(s,b) + c(s,b)),    \label{eq:ampl_prof_tot_new}\\
  \PROF{inel}(s,b) &\equiv 4\,(g_1(s,b) + K(s,b)+c(s,b)). \label{eq:ampl_prof_inel_new}
%  \PROF{inel}(s,b) &\equiv 4\, \Tilde{g}_{\text{inel}}(s,b). \label{eq:ampl_prof_inel} 
\end{align}
and rewrite the unitarity condition in $b$-space in the form
\begin{equation}
  \PROF{tot}(s,b) = \PROF{el}(s,b) + \PROF{inel}(s,b).
  \label{eq:unitarity2}
\end{equation}
The shape of $\PROF{tot}(s,b)$ and $\PROF{inel}(s,b)$ might be then modified to become non-negative; the shape of elastic profile remains the same. The function $c(s,b)$ should, however, fulfill some additional conditions. The total and inelastic cross section given by  
\begin{equation}
\CS[etype=X](s) = 2\pi\int\limits_0^{\infty} b \text{d}b\;\PROF{X}(s,b)
\label{eq:integ_cs_prof}
\end{equation}
(see \cref{eq:cs_tot_integ_b,eq:cs_el_integ_b,eq:cs_inel_integ_b}) remains unchanged if
\begin{equation}
\int \limits_{0}^{\infty} b\; \text{d}b \; c(s,b) = 0. 
\label{eq:rs1}
\end{equation}
The other physical quantities which should be preserved are the mean squared values of the total and inelastic impact parameters, i.e., function $c(s,b)$ should not change the quantities $\meanb[n=2,etype=tot]$ and $\meanb[n=2,etype=inel]$ defined as
\begin{equation}
  \meanb[n=2,etype=X] = \frac{\int\limits_0^{\infty} b^2 \; 2 \pi b\PROF{X}(s,b) \text{d}b}{\int\limits_0^{\infty} 2 \pi b \PROF{X}(s,b) \text{d}b}.
\label{eq:integ_meanb}
\end{equation}
These quantities will be preserved if also
\begin{equation}
\int \limits_{0}^{\infty} b^3\; \text{d}b\; c(s,b) \; = \; 0.
\label{eq:rs3}
\end{equation}

By definition all the mentioned processes (total, elastic and inelastic) are realized by strong interactions which are of finite ranges. Therefore both the integrals appearing in \cref{eq:integ_meanb} should be convergent. Condition $\beta^{1/2} h_1(s, \beta) \in L^2(0, \infty)$ guarantees that all three integrals (for total, elastic or inelastic type $X$) in the denominator of \cref{eq:integ_meanb} are convergent; for the inelastic case also on the basis of unitarity condition given by \cref{eq:imp16}. However, this condition does not guarantee the convergence of the integrals in the nominator of \cref{eq:integ_meanb}; in order to achieve this we have to require the validity of stronger condition, i.e., that $\beta^{3/2} h_1(s, \beta) \in L^2(0, \infty)$. Due to the unitarity equation the remaining two integrals corresponding to the elastic and inelastic scattering will be convergent, too.

The function $c(s,b)$ should fulfill, therefore, the following conditions: it must remove the oscillations (provide the non-negative function $\PROF{tot}(s,b)$) and fulfill \cref{eq:rs1} and \cref{eq:rs3}.

It follows then from the Islam's approach \cite{Islam1976} that the two conditions given by \cref{eq:rs1,eq:rs3} are fulfilled when 
\begin{equation}
c(s,b) \; = \; - \; \Im h_2(s,b),
\label{eq:rs4}
\end{equation}
where $h_2(s,b)$ is defined by \cref{eq:imp8} and is based on analytic continuation of complex amplitude $\ampl{N}$.
%that fulfills the condition
%%%
%\begin{equation}
%F^{\text{N}}(s,t_{\text{min}}) \; = \; 0.
%\label{eq:rs5}
%\end{equation}
%%%
It also means that one can hardly determine the function $c(s,b)$ quite exactly on the basis of analyzing experimental data of elastic scattering corresponding always to very limited $t$-region (see also \cite{Deile:2010mv_Kundrat,elba2010_klkp}). 

According to \cite{Kundrat2002} the mean squares of total, elastic and inelastic impact parameter defined by \cref{eq:integ_meanb} may be determined directly from the hadronic amplitude \ampl{N} in $t$ variable without being necessary to know the corresponding profile function or the function $c(s,b)$. It is possible to write for the mean squared value of elastic impact parameters
\newcommand{\msint}[1][]{\ensuremath{\int\limits_{t_{\text{min}}}^0 \text{d} t {#1} \modulus{N}^2}}
\begin{equation}
  \begin{split}
    \meanb[n=2,etype=el] =& \meanb[n=2,etype=mod]+\meanb[n=2,etype=ph] \\
    =& \frac{4 \int\limits_{t_{\text{min}}}^0 \text{d} t |t| \left(\frac{\text{d}}{\text{d} t} \modulus{N} \right)^2}{\msint} \\
     &+ \frac{4 \msint{|t|}\left( \frac{\text{d}}{\text{d} t} \phase \right)^2}{\msint}
  \label{eq:msel}
\end{split}
\end{equation}
and for the total mean squared value  
%%
%\begin{equation}
%{\langle } b^2(s){\rangle }_{\text{tot}}\;\; = 
%{ { {8 \pi \int \limits_{0}}^{\infty} b \;db \; b^2 h_{\text{tot}}(s,b)}
%\over {\sigma_{\text{tot}}(s)}} = 2 B(s,0).  
%\label{eq:mv3}
%\end{equation}
\begin{equation}
  \meanb[n=2,etype=tot] = \left. 4\left( \frac{\frac{\text{d}}{\text{d}t}\modulus{N}}{\modulus{N}} -\tan{\phase}{\frac{\text{d}}{\text{d}t} \phase}\right) \right|_{t=0}.
 \label{eq:mstot}
\end{equation}
The inelastic mean squared value is then given by
\begin{equation}
    \meanb[n=2,etype=inel] = \frac{\CS[etype={tot,N}](s) \meanb[n=2,etype=tot] -  \CS[etype={el,N}](s) \meanb[n=2,etype=el]}{\CS[etype=inel](s)}
  \label{eq:msinel}
\end{equation}
if the cross sections are determined using the optical theorem~\refp{eq:optical_theorem} and the first equation in \refp{eq:cs_el_integ_b}.

The $b$-dependent profile functions may be determined in the following way. We may chose Gaussian shape of total profile function $\PROF{tot}(b)$ corresponding to the commonly assumed one \cite{Deile:2010mv_Kundrat}
\begin{equation}
\PROF{tot}(b) = \Tilde{a}_2 \e^{-\Tilde{a}_1 b^2}
\label{eq:prof_tot_gauss_simple}
\end{equation}
where $\Tilde{a}_1$ and $\Tilde{a}_2$ are some parameters which may be expressed using \cref{eq:integ_cs_prof,eq:integ_meanb} as (see integral formulas 3.461 in \cite{Gradshteyn1980})
\begin{align}
\Tilde{a}_1 &= \frac{1}{\meanb[n=2,etype=tot]},  \\
\Tilde{a}_2 &= \frac{\CS[etype={tot,N}]}{\pi\meanb[n=2,etype=tot]}.
\end{align}
The total profile function $\PROF{tot}$ given by \cref{eq:prof_tot_gauss_simple} may be, therefore, determined from values of \CS[etype={tot,N}] and \meanb[n=2,etype=tot] using optical theorem~\refp{eq:optical_theorem} and \cref{eq:mstot}, i.e., from $t$-dependent elastic amplitude \ampl{N}. It means that using FB transformation \refp{eq:imp7} of \ampl{N} and \cref{eq:unitarity2} the total, elastic and inelastic profile functions (and also the corresponding $c(s,b)$ function) may be determined for a given \ampl{N}. This approach has been used in \cref{sec:data_analysis} where the hadronic amplitude \ampl{N} have been determined on the basis of experimental data using the eikonal model description of Coulomb-hadronic interference discussed in \cref{sec:eikonal}. %The Coulomb interaction of charged hadrons depends on electromagnetic form factors which will be explained first.

\section{\label{sec:wy_study}A priori limitation of $t$-dependence of hadronic phase in the WY approach}
%\section{\label{sec:wy_study}Relative phase $\alpha\phi(s,t)$ in the WY approach}
\subsection{\label{sec:wy_study_pp53gev}Energy of 52.8~GeV}

\begin{figure*}%[!ht]%[!ht]
\centering
\begin{subfigure}[t]{0.48\textwidth}
%\begin{subfigure}[t]{1.\columnwidth}
\includegraphics*[width=\textwidth]{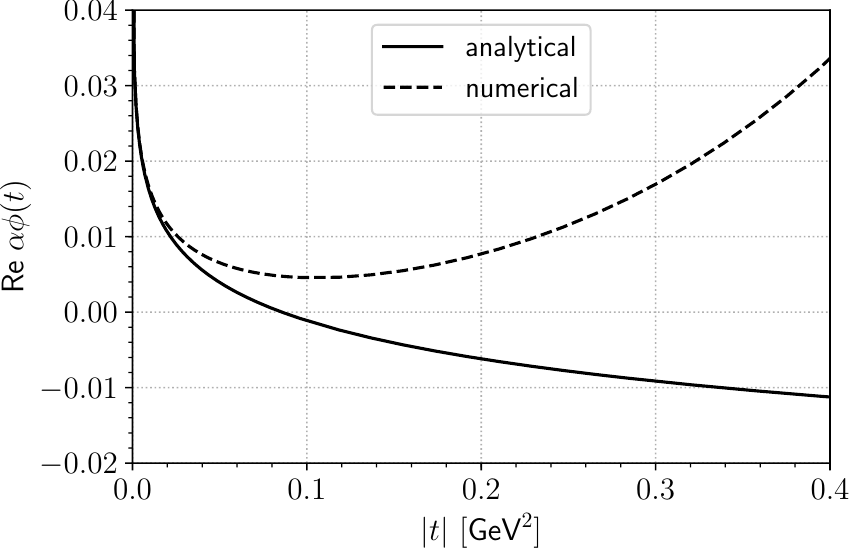}
\caption{\label{fig:pp53gev_wy_anal_vs_num_full}}
\end{subfigure}
%\quad%add desired spacing between images, e. g. ~, \quad, \qquad etc.
%%(or a blank line to force the subfigure onto a new line)
\begin{subfigure}[t]{0.48\textwidth}
%\begin{subfigure}[t]{1.\columnwidth}
\includegraphics*[width=\textwidth]{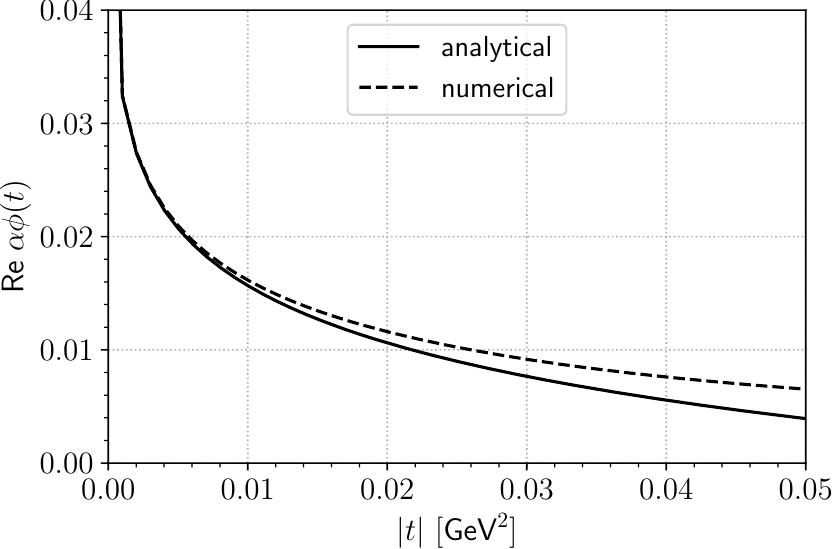}
\caption{\label{fig:pp53gev_wy_anal_vs_num_zoom}}
\end{subfigure}
\begin{minipage}[t]{.95\textwidth}
%\begin{minipage}[t]{1.\columnwidth}
\caption{\label{fig:pp53gev_wy_anal_vs_num}The real part of $\alpha\phi(s,t)$ given by \cref{eq:phaseWY} (denoted as "numerical" calculation) and \cref{eq:phiWY} (denoted as "analytical" calculation) in two different $t$ regions under the assumptions of $t$-independent quantities $\rho(t)$ and $B(t)$ whose values have been taken from \refp{eq:wy_parameters_amaldi1978} corresponding to pp scattering at 52.8~GeV. (\subref{fig:pp53gev_wy_anal_vs_num_zoom}) shows region of very low values of $|t|$.}
\end{minipage}
\end{figure*}

As it has been mentioned in \cref{sec:introduction} the quantities $\rho(t)$ and $B(t)$ in the simplified formula \refp{eq:simplifiedWY} of WY are assumed to be $t$-independent in this approach. In this case the imaginary part of relative phase $\alpha\phi(s,t)$ given by \cref{eq:phaseWY} is equal to zero by definition. Taking numerical values of the free parameters for pp collisions at 52.8~GeV from \refp{eq:wy_parameters_amaldi1978} one may calculate the real part of relative phase $\alpha\phi(s,t)$ according to \cref{eq:phaseWY}. The integral in \cref{eq:phaseWY} may be calculated numerically and the result may be then compared to corresponding analytical calculation given by \cref{eq:phiWY} which has been widely used in the past for analysis of experimental data. \Cref{fig:pp53gev_wy_anal_vs_num} shows that the analytically and numerically calculated $\text{Re}\; \alpha\phi(s,t)$ are compatible at $|t| \lesssim 0.01$~GeV$^2$; significant differences exist at higher values of $|t|$. 

The formula~\refp{eq:phaseWY} might seem to be considered as quite general, i.e., that the $t$-dependence of the relative phase $\alpha \phi(s,t)$ could be uniquely determined for any $t$-dependence of elastic hadronic amplitude $\ampl{N}$. However, it has been shown in \cite{KL2007} that the mentioned relative phase which has to be real by definition (it is defined as imaginary part of another function, see \cite{WY1968}) is real only provided the elastic hadronic phase $\phase$ is $t$-independent in the whole integration region of \cref{eq:phaseWY}. If the elastic hadronic phase is $t$-dependent then the function $\alpha\phi(s,t)$ in \cref{eq:phaseWY} becomes complex and looses its physical sense. 

\begin{figure*}%[!ht]
\centering
\begin{subfigure}[t]{0.48\textwidth}
\includegraphics*[width=\textwidth]{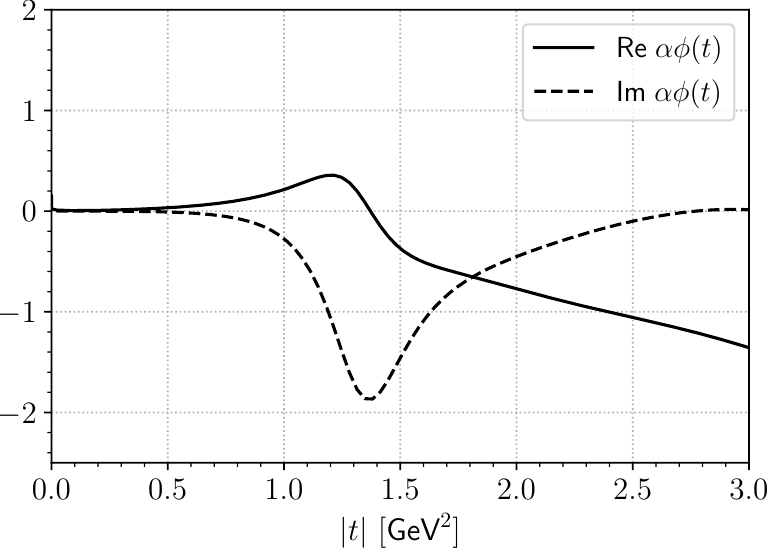}
		\caption{\label{fig:pp53gev_wy_Fit1b_phi_phase_re_im_zoom}}
\end{subfigure}
\quad%add desired spacing between images, e. g. ~, \quad, \qquad etc.
\begin{subfigure}[t]{0.48\textwidth}
\includegraphics*[width=\textwidth]{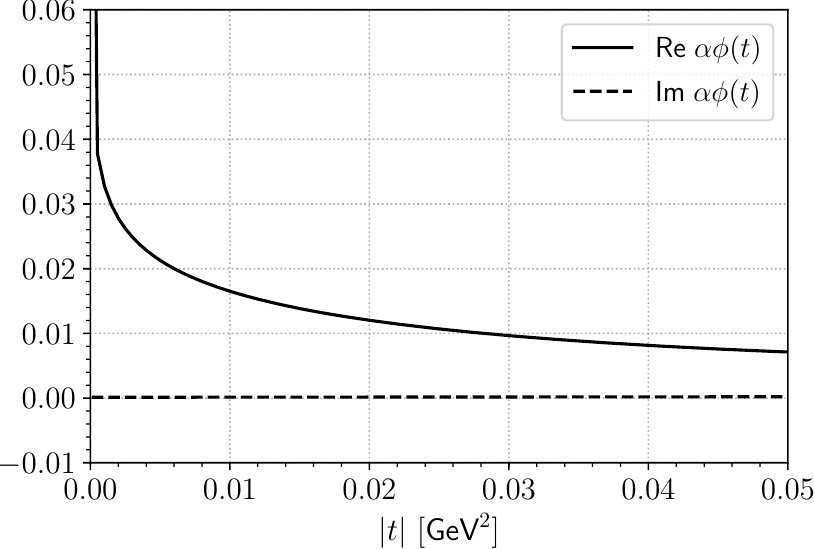}
		\caption{\label{fig:pp53gev_wy_Fit1b_phi_phase_re_im_zoom2}}
\end{subfigure}
\begin{subfigure}[t]{0.48\textwidth}
\includegraphics*[width=\textwidth]{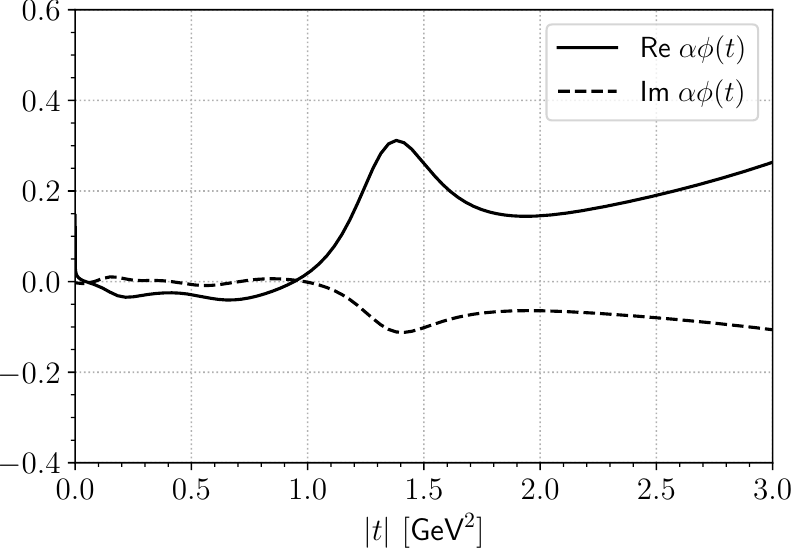}
		\caption{\label{fig:pp53gev_wy_Fit3b_phi_phase_re_im_zoom}}
\end{subfigure}
\quad%add desired spacing between images, e. g. ~, \quad, \qquad etc.
\begin{subfigure}[t]{0.48\textwidth}
\includegraphics*[width=\textwidth]{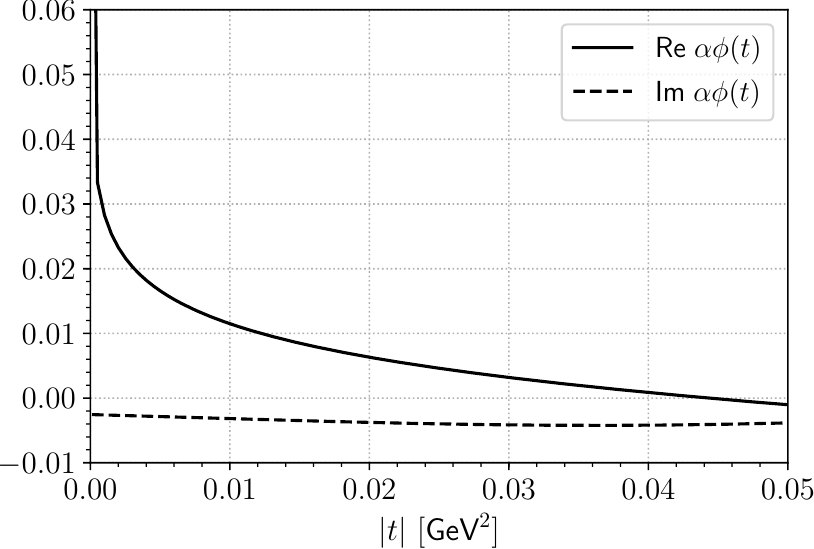}
		\caption{\label{fig:pp53gev_wy_Fit3b_phi_phase_re_im_zoom2}}
\end{subfigure}
\begin{minipage}[t]{.95\textwidth}
%\begin{minipage}[t]{1.\columnwidth}
\caption{\label{fig:pp53gev_wy_phi_phase_re_im}Comparison of the real and imaginary parts of $\alpha\phi(s,t)$ given by \cref{eq:phaseWY} and calculated for elastic pp hadronic amplitude at 52.8~GeV corresponding to Fit~1 ((\protect\subref{fig:pp53gev_wy_Fit1b_phi_phase_re_im_zoom}) and (\protect\subref{fig:pp53gev_wy_Fit1b_phi_phase_re_im_zoom2})) and Fit~2 ((\protect\subref{fig:pp53gev_wy_Fit3b_phi_phase_re_im_zoom}) and (\protect\subref{fig:pp53gev_wy_Fit3b_phi_phase_re_im_zoom2})) in different $t$-regions. (\subref{fig:pp53gev_wy_Fit1b_phi_phase_re_im_zoom2}) and (\subref{fig:pp53gev_wy_Fit3b_phi_phase_re_im_zoom2}) show region of very low values of $|t|$.}
\end{minipage}
\end{figure*}

One may test these aspects by calculating numerically the relative phase $\alpha \phi(s,t)$ given by \cref{eq:phaseWY} for hadronic amplitude \ampl{N} having $t$-dependent hadronic phase. For this purpose one may take elastic hadronic amplitudes determined in the central Fit~1 and the peripheral Fit~2 of pp elastic data at 52.8~GeV which have been performed in \cref{sec:data_analysis}. Both the cases represent very different $t$-dependences of hadronic phases, see \cref{fig:pp53gev_zeta_multi}. \Cref{fig:pp53gev_wy_Fit1b_phi_phase_re_im_zoom,fig:pp53gev_wy_Fit1b_phi_phase_re_im_zoom2} show comparison of $t$-dependence of the real and imaginary parts of $\alpha\phi(s,t)$ corresponding to hadronic amplitude determined in Fit~1. \Cref{fig:pp53gev_wy_Fit3b_phi_phase_re_im_zoom,fig:pp53gev_wy_Fit3b_phi_phase_re_im_zoom2} then correspond to hadronic amplitude determined in Fit~2. As one may see the function $\text{Im}\; \alpha\phi(s,t)$ is zero at $|t| \lesssim 0.9$~GeV$^2$ in the central case given by Fit~1 which reflects the $t$-region where the corresponding hadronic phase $\phase$ is roughly $t$-independent, see \cref{fig:pp53gev_zeta_multi}. At higher values of $|t|$ the function $\text{Im}\; \alpha\phi(s,t)$ is strongly $t$-dependent and significantly non-zero. In the peripheral case corresponding to Fit~2 function $\text{Im}\; \alpha\phi(s,t)$ is significantly non-zero in the whole $t$-range, including very low values of $|t|$. In the peripheral case the hadronic phase $\phase$ has strong $t$-dependence already at very low $|t|$ values, see \cref{fig:pp53gev_zeta_multi}.

These new calculations explicitly show that the approach of WY may be, therefore, hardly suitable for analysis of experimental data with the help of general $t$-dependence of the elastic hadronic phase $\phase$ (whole amplitude $\ampl{N}$), even in the region of very low values of $|t|$.

\subsection{\label{sec:wy_study_pp8000gev}Energy of 8~TeV}
\begin{figure*}%[!ht]
\centering
\begin{subfigure}[t]{0.48\textwidth}
%\begin{subfigure}[t]{1.\columnwidth}
\includegraphics*[width=\textwidth]{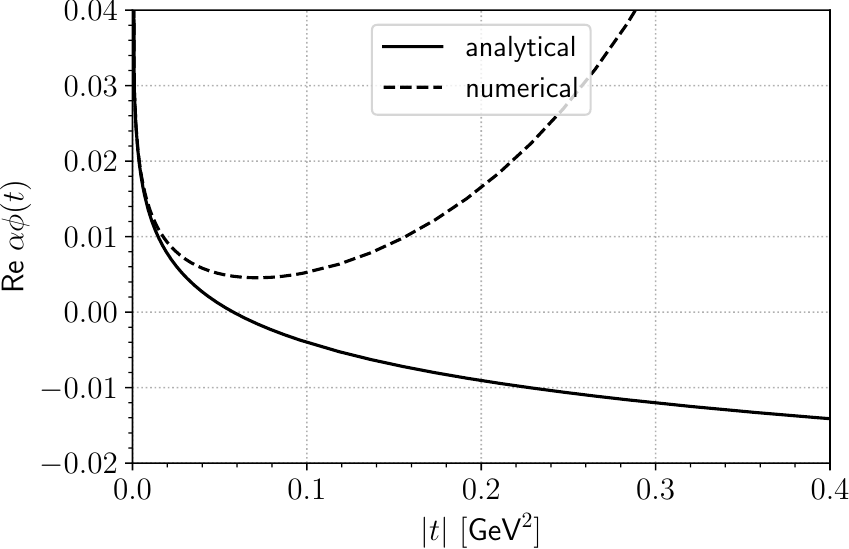}
\caption{\label{fig:pp8000gev_wy_anal_vs_num_full}}
\end{subfigure}
%\quad%add desired spacing between images, e. g. ~, \quad, \qquad etc.
%%(or a blank line to force the subfigure onto a new line)
\begin{subfigure}[t]{0.48\textwidth}
%\begin{subfigure}[t]{1.\columnwidth}
\includegraphics*[width=\textwidth]{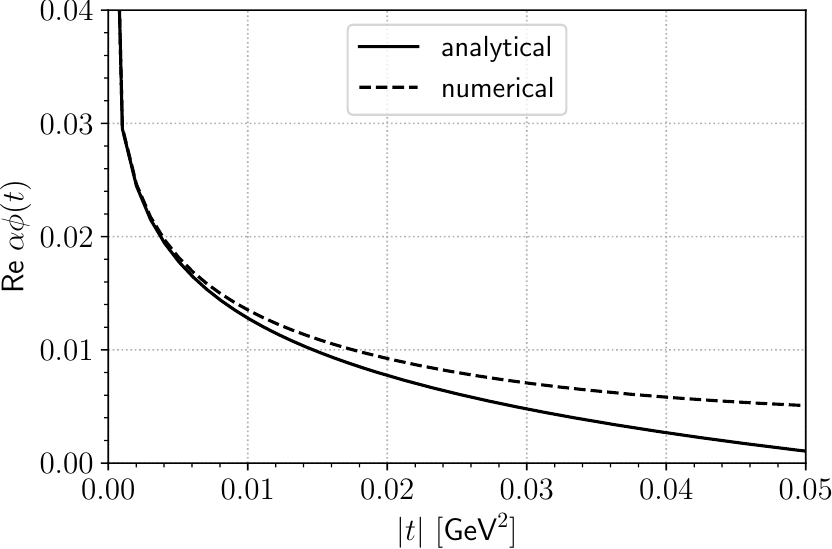}
\caption{\label{fig:pp8000gev_wy_anal_vs_num_zoom}}
\end{subfigure}
\begin{minipage}[t]{.95\textwidth}
%\begin{minipage}[t]{1.\textwidth}
%\begin{minipage}[t]{1.\columnwidth}
\caption{\label{fig:pp8000gev_wy_anal_vs_num}The real part of $\alpha\phi(s,t)$ given by \cref{eq:phaseWY} (denoted as "numerical" calculation) and \cref{eq:phiWY} (denoted as "analytical" calculation) in two different $t$ regions under the assumptions of $t$-independent quantities $\rho(t)$ and $B(t)$ whose values have been taken from \refp{eq:wy_parameters_totem2016} corresponding to pp scattering at 8~TeV. (\subref{fig:pp8000gev_wy_anal_vs_num_zoom}) shows region of very low values of $|t|$.}
\end{minipage}
\end{figure*}

\begin{figure*}%[!ht]
\centering
\begin{subfigure}[t]{0.48\textwidth}
\includegraphics*[width=\textwidth]{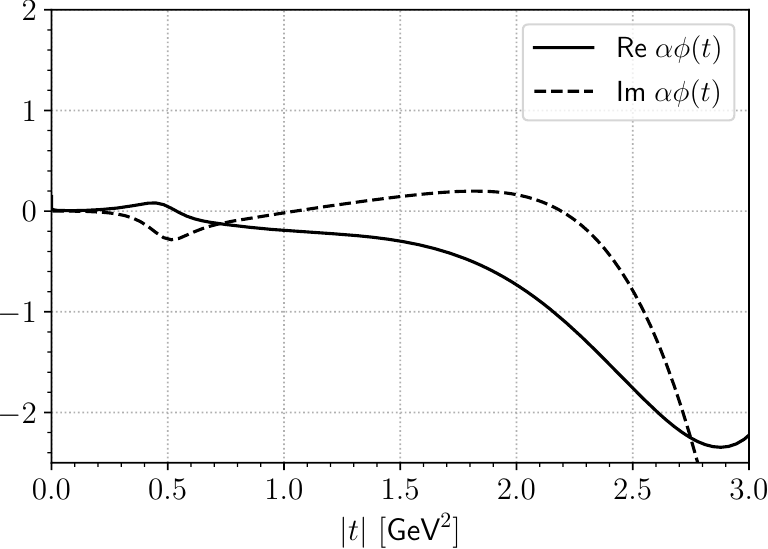}
		\caption{\label{fig:pp8000gev_wy_fit1b_phi_phase_re_im_zoom}}
\end{subfigure}
\quad%add desired spacing between images, e. g. ~, \quad, \qquad etc.
\begin{subfigure}[t]{0.48\textwidth}
\includegraphics*[width=\textwidth]{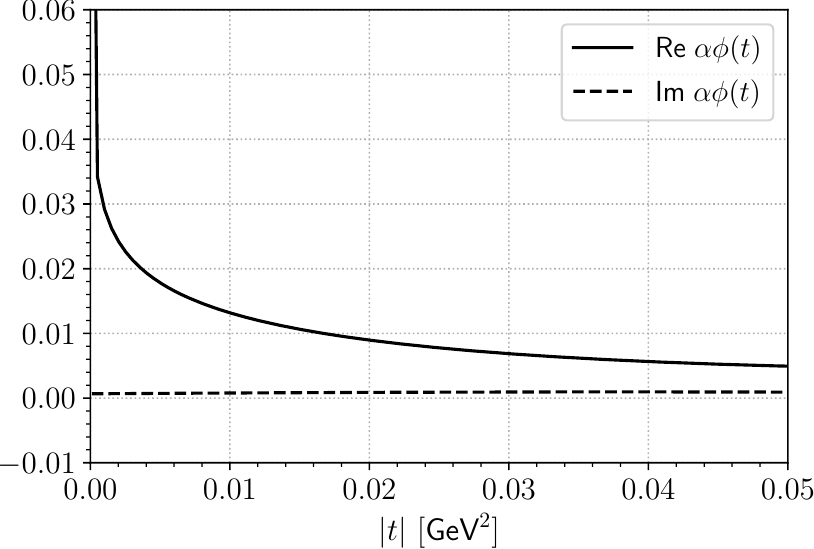}
		\caption{\label{fig:pp8000gev_wy_fit1b_phi_phase_re_im_zoom2}}
\end{subfigure}
\begin{subfigure}[t]{0.48\textwidth}
\includegraphics*[width=\textwidth]{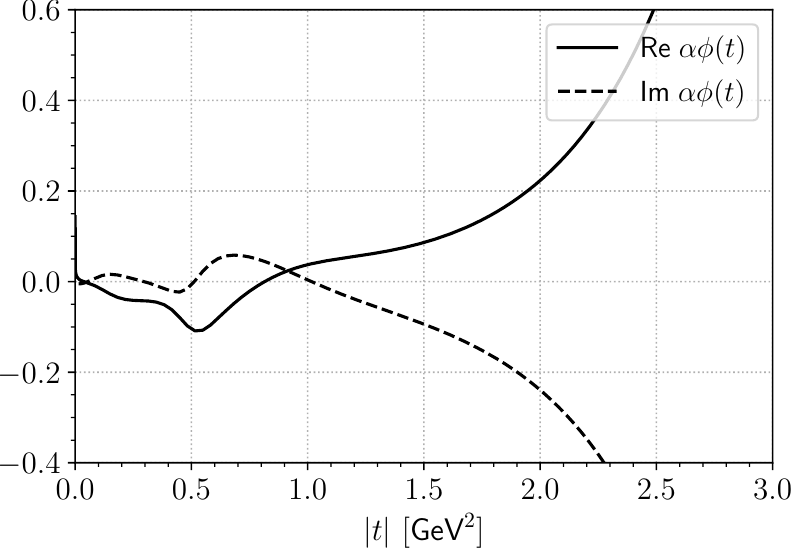}
		\caption{\label{fig:pp8000gev_wy_fit3b_phi_phase_re_im_zoom}}
\end{subfigure}
\quad%add desired spacing between images, e. g. ~, \quad, \qquad etc.
\begin{subfigure}[t]{0.48\textwidth}
\includegraphics*[width=\textwidth]{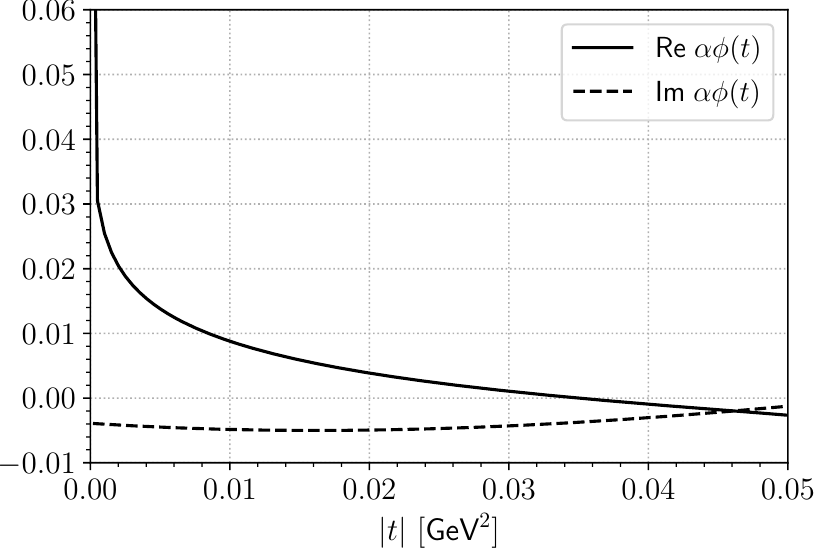}
		\caption{\label{fig:pp8000gev_wy_fit3b_phi_phase_re_im_zoom2}}
\end{subfigure}
\begin{minipage}[t]{.95\textwidth}
%\begin{minipage}[t]{1.\textwidth}
%\begin{minipage}[t]{1.\columnwidth}
\caption{\label{fig:pp8000gev_wy_phi_phase_re_im}Comparison of the real and imaginary parts of $\alpha\phi(s,t)$ given by \cref{eq:phaseWY} and calculated for elastic pp hadronic amplitude at 8~TeV corresponding to Fit~1 ((\protect\subref{fig:pp8000gev_wy_fit1b_phi_phase_re_im_zoom}) and (\protect\subref{fig:pp8000gev_wy_fit1b_phi_phase_re_im_zoom2})) and Fit~2 ((\protect\subref{fig:pp8000gev_wy_fit3b_phi_phase_re_im_zoom}) and (\protect\subref{fig:pp8000gev_wy_fit3b_phi_phase_re_im_zoom2})) in different $t$-regions. (\subref{fig:pp8000gev_wy_fit1b_phi_phase_re_im_zoom2}) and (\subref{fig:pp8000gev_wy_fit3b_phi_phase_re_im_zoom2}) show region of very low values of $|t|$.}
\end{minipage}
\end{figure*}

Calculations at 52.8~GeV discussed in appendix~\refp{sec:wy_study_pp53gev} may be analogically performed also at 8~TeV with similar conclusions. \Cref{fig:pp8000gev_wy_anal_vs_num} shows that the analytically and numerically calculated $\text{Re}\; \alpha\phi(s,t)$ (taking numerical values of the free parameters for pp collisions at 8~TeV from \refp{eq:wy_parameters_totem2016}) are again compatible at $|t| \lesssim 0.01$~GeV$^2$ and that significant differences exist at higher values of $|t|$. \Cref{fig:pp8000gev_wy_phi_phase_re_im} at 8 TeV (where fits at 8 TeV discussed in \cref{sec:pp8000gev_results} have been used of) explicitly shows that the approach of WY is not suitable for general analysis of experimental data and study of $t$-dependence of elastic hadronic amplitude.



\FloatBarrier
\end{appendices}

\clearpage

%\addcontentsline{toc}{section}{References}
%\section*{References}

%%\tableofcontents
%
\end{document}